%% file: main.tex
\documentclass[conference]{IEEEtran}

\usepackage{xspace}
\usepackage{booktabs}
\usepackage{cite}
\usepackage{amsmath,amssymb,amsfonts}
\usepackage{algorithmic}
\usepackage{graphicx}
\usepackage{textcomp}
\usepackage{xcolor}
\usepackage{balance}
\usepackage{tcolorbox}
\usepackage{subcaption}
\usepackage{url}
\usepackage[capitalise]{cleveref}
\usepackage[
    colorinlistoftodos,prependcaption,textsize=tiny
]{todonotes}

\usepackage[ruled,linesnumbered]{algorithm2e}
\SetKwInOut{Input}{input}
\SetKwInOut{Output}{output}
\SetKwProg{Fn}{function}{:}{end}
\SetKw{Break}{break}
\SetKw{Continue}{continue}
\SetFuncSty{textsf}
\newcommand\definetool[2]{\newcommand{#1}{{\textsc{#2}}\xspace}}
\definetool{\Scratch}{Scratch}
\definetool{\Whisker}{Whisker}
\definetool{\Neat}{Neat}
\definetool{\Neatest}{Neatest}
\definetool{\MOSA}{MOSA}
\definetool{\MIO}{MIO}
\definetool{\NEWSD}{NEWS/D}
\definetool{\MOSANeatest}{MOSA-Neatest}
\definetool{\MIONeatest}{MIO-Neatest}
\definetool{\NEWSDNeatest}{NEWS/D-Neatest}

\begin{document}

\input{macros.tex}

\title{Many-Objective Neuroevolution for Testing Games}

\author{\IEEEauthorblockN{Patric Feldmeier}
\IEEEauthorblockA{\textit{University of Passau} \\
Germany}
\and
\IEEEauthorblockN{Katrin Schmelz}
\IEEEauthorblockA{\textit{University of Passau} \\
Germany}
\and
\IEEEauthorblockN{Gordon Fraser}
\IEEEauthorblockA{\textit{University of Passau} \\
Germany}
}

\maketitle

\begin{abstract}
  Games are designed to challenge human players, but this also makes
  it challenging to generate software tests for computer games
  automatically.
  Neural networks have therefore been proposed to serve as dynamic
  test cases trained to reach statements in the underlying code,
  similar to how static test cases consisting of event sequences would
  do in traditional software.
  The \Neatest approach combines search-based software testing
  principles with neuroevolution to generate such dynamic test cases.
  However, it may take long or even be impossible to evolve a network that
  can cover individual program statements, and since \Neatest is a
  single-objective algorithm, it will have to be sequentially invoked
  for a potentially large number of coverage goals.
%
%
  In this paper, we therefore propose to treat the neuroevolution of
  dynamic test cases as a many-objective search problem. By targeting
  all coverage goals at the same time, easy goals are covered quickly,
  and the search can focus on more challenging ones. We extend the
  state-of-the-art many-objective test generation algorithms
  \MIO and \MOSA as well as the state-of-the-art 
  many-objective neuroevolution algorithm \NEWSD to generate dynamic 
  test cases.
  Experiments on 20 \Scratch games show that targeting several
  objectives simultaneously increases \Neatest's average branch
  coverage from \MeanBranchCoverageTotalNeatest\% to
  \MeanBranchCoverageTotalMOSA\% while reducing the search time by
  \SpeedUpBranchMOSA\%.
\end{abstract}

\begin{IEEEkeywords}
Neuroevolution, Many-Objective Search, Software Testing, Games
\end{IEEEkeywords}

\input{sections/1-Introduction}
\input{sections/2-Background}
\input{sections/3-Approach}
\input{sections/4-Evaluation}
\input{sections/5-RelatedWork}
\input{sections/6-Conclusions}

\balance
\bibliographystyle{IEEEtran}
\bibliography{library}

\end{document}

%% file: macros.tex
\newcommand{\WinCountCatchTheDotsCompatMIO}{6}
\newcommand{\WinCountCatchTheDotsNoveltyMIO}{7}
\newcommand{\WinCountCatchTheDotsSpeciesMIO}{8}
\newcommand{\WinCountCityDefenderCompatMIO}{0}
\newcommand{\WinCountCityDefenderNoveltyMIO}{0}
\newcommand{\WinCountCityDefenderSpeciesMIO}{0}
\newcommand{\WinCountCreateYourWorldCompatMIO}{0}
\newcommand{\WinCountCreateYourWorldNoveltyMIO}{3}
\newcommand{\WinCountCreateYourWorldSpeciesMIO}{1}
\newcommand{\WinCountDessertRallyCompatMIO}{0}
\newcommand{\WinCountDessertRallyNoveltyMIO}{0}
\newcommand{\WinCountDessertRallySpeciesMIO}{0}
\newcommand{\WinCountDieZauberlehrlingeCompatMIO}{14}
\newcommand{\WinCountDieZauberlehrlingeNoveltyMIO}{17}
\newcommand{\WinCountDieZauberlehrlingeSpeciesMIO}{18}
\newcommand{\WinCountDodgeballCompatMIO}{0}
\newcommand{\WinCountDodgeballNoveltyMIO}{0}
\newcommand{\WinCountDodgeballSpeciesMIO}{0}
\newcommand{\WinCountDragonsCompatMIO}{27}
\newcommand{\WinCountDragonsNoveltyMIO}{28}
\newcommand{\WinCountDragonsSpeciesMIO}{29}
\newcommand{\WinCountEndlessRunnerCompatMIO}{30}
\newcommand{\WinCountEndlessRunnerNoveltyMIO}{30}
\newcommand{\WinCountEndlessRunnerSpeciesMIO}{30}
\newcommand{\WinCountFallingStarsCompatMIO}{30}
\newcommand{\WinCountFallingStarsNoveltyMIO}{30}
\newcommand{\WinCountFallingStarsSpeciesMIO}{30}
\newcommand{\WinCountFinalFightCompatMIO}{30}
\newcommand{\WinCountFinalFightNoveltyMIO}{30}
\newcommand{\WinCountFinalFightSpeciesMIO}{30}
\newcommand{\WinCountFlappyParrotCompatMIO}{5}
\newcommand{\WinCountFlappyParrotNoveltyMIO}{4}
\newcommand{\WinCountFlappyParrotSpeciesMIO}{7}
\newcommand{\WinCountFroggerCompatMIO}{0}
\newcommand{\WinCountFroggerNoveltyMIO}{0}
\newcommand{\WinCountFroggerSpeciesMIO}{2}
\newcommand{\WinCountFruitCatchingCompatMIO}{4}
\newcommand{\WinCountFruitCatchingNoveltyMIO}{2}
\newcommand{\WinCountFruitCatchingSpeciesMIO}{1}
\newcommand{\WinCountHackAttackCompatMIO}{1}
\newcommand{\WinCountHackAttackNoveltyMIO}{1}
\newcommand{\WinCountHackAttackSpeciesMIO}{1}
\newcommand{\WinCountOceanCleanupCompatMIO}{11}
\newcommand{\WinCountOceanCleanupNoveltyMIO}{12}
\newcommand{\WinCountOceanCleanupSpeciesMIO}{12}
\newcommand{\WinCountPokemonClickerCompatMIO}{0}
\newcommand{\WinCountPokemonClickerNoveltyMIO}{0}
\newcommand{\WinCountPokemonClickerSpeciesMIO}{0}
\newcommand{\WinCountSnakeCompatMIO}{0}
\newcommand{\WinCountSnakeNoveltyMIO}{0}
\newcommand{\WinCountSnakeSpeciesMIO}{0}
\newcommand{\WinCountSnowballFightCompatMIO}{9}
\newcommand{\WinCountSnowballFightNoveltyMIO}{14}
\newcommand{\WinCountSnowballFightSpeciesMIO}{7}
\newcommand{\WinCountSpaceOdysseyCompatMIO}{0}
\newcommand{\WinCountSpaceOdysseyNoveltyMIO}{0}
\newcommand{\WinCountSpaceOdysseySpeciesMIO}{0}
\newcommand{\WinCountWhackAMoleCompatMIO}{27}
\newcommand{\WinCountWhackAMoleNoveltyMIO}{27}
\newcommand{\WinCountWhackAMoleSpeciesMIO}{25}
\newcommand{\MeanWinsCompatMIO}{9.70}
\newcommand{\MeanWinsNoveltyMIO}{10.25}
\newcommand{\MeanWinsSpeciesMIO}{10.05}
\newcommand{\MeanStatementCoverageCatchTheDotsCompatMIO}{96.34}
\newcommand{\MeanStatementCoverageCityDefenderCompatMIO}{74.23}
\newcommand{\MeanStatementCoverageCreateYourWorldCompatMIO}{74.61}
\newcommand{\MeanStatementCoverageDessertRallyCompatMIO}{93.90}
\newcommand{\MeanStatementCoverageDieZauberlehrlingeCompatMIO}{83.29}
\newcommand{\MeanStatementCoverageDodgeballCompatMIO}{96.03}
\newcommand{\MeanStatementCoverageDragonsCompatMIO}{87.21}
\newcommand{\MeanStatementCoverageEndlessRunnerCompatMIO}{82.42}
\newcommand{\MeanStatementCoverageFallingStarsCompatMIO}{99.49}
\newcommand{\MeanStatementCoverageFinalFightCompatMIO}{99.65}
\newcommand{\MeanStatementCoverageFlappyParrotCompatMIO}{93.15}
\newcommand{\MeanStatementCoverageFroggerCompatMIO}{93.85}
\newcommand{\MeanStatementCoverageFruitCatchingCompatMIO}{85.21}
\newcommand{\MeanStatementCoverageHackAttackCompatMIO}{92.72}
\newcommand{\MeanStatementCoverageOceanCleanupCompatMIO}{83.70}
\newcommand{\MeanStatementCoveragePokemonClickerCompatMIO}{26.21}
\newcommand{\MeanStatementCoverageSnakeCompatMIO}{95.00}
\newcommand{\MeanStatementCoverageSnowballFightCompatMIO}{95.64}
\newcommand{\MeanStatementCoverageSpaceOdysseyCompatMIO}{87.07}
\newcommand{\MeanStatementCoverageWhackAMoleCompatMIO}{85.03}
\newcommand{\MeanStatementCoverageTotalCompatMIO}{86.24}
\newcommand{\MeanStatementCoverageCatchTheDotsNoveltyMIO}{96.99}
\newcommand{\MeanStatementCoverageCityDefenderNoveltyMIO}{74.33}
\newcommand{\MeanStatementCoverageCreateYourWorldNoveltyMIO}{75.23}
\newcommand{\MeanStatementCoverageDessertRallyNoveltyMIO}{93.90}
\newcommand{\MeanStatementCoverageDieZauberlehrlingeNoveltyMIO}{84.22}
\newcommand{\MeanStatementCoverageDodgeballNoveltyMIO}{95.98}
\newcommand{\MeanStatementCoverageDragonsNoveltyMIO}{87.46}
\newcommand{\MeanStatementCoverageEndlessRunnerNoveltyMIO}{84.43}
\newcommand{\MeanStatementCoverageFallingStarsNoveltyMIO}{99.30}
\newcommand{\MeanStatementCoverageFinalFightNoveltyMIO}{99.65}
\newcommand{\MeanStatementCoverageFlappyParrotNoveltyMIO}{92.97}
\newcommand{\MeanStatementCoverageFroggerNoveltyMIO}{94.10}
\newcommand{\MeanStatementCoverageFruitCatchingNoveltyMIO}{82.06}
\newcommand{\MeanStatementCoverageHackAttackNoveltyMIO}{92.72}
\newcommand{\MeanStatementCoverageOceanCleanupNoveltyMIO}{84.32}
\newcommand{\MeanStatementCoveragePokemonClickerNoveltyMIO}{26.21}
\newcommand{\MeanStatementCoverageSnakeNoveltyMIO}{95.00}
\newcommand{\MeanStatementCoverageSnowballFightNoveltyMIO}{96.07}
\newcommand{\MeanStatementCoverageSpaceOdysseyNoveltyMIO}{87.27}
\newcommand{\MeanStatementCoverageWhackAMoleNoveltyMIO}{84.61}
\newcommand{\MeanStatementCoverageTotalNoveltyMIO}{86.34}
\newcommand{\MeanStatementCoverageCatchTheDotsSpeciesMIO}{96.50}
\newcommand{\MeanStatementCoverageCityDefenderSpeciesMIO}{74.33}
\newcommand{\MeanStatementCoverageCreateYourWorldSpeciesMIO}{74.81}
\newcommand{\MeanStatementCoverageDessertRallySpeciesMIO}{93.90}
\newcommand{\MeanStatementCoverageDieZauberlehrlingeSpeciesMIO}{84.57}
\newcommand{\MeanStatementCoverageDodgeballSpeciesMIO}{95.98}
\newcommand{\MeanStatementCoverageDragonsSpeciesMIO}{87.31}
\newcommand{\MeanStatementCoverageEndlessRunnerSpeciesMIO}{83.17}
\newcommand{\MeanStatementCoverageFallingStarsSpeciesMIO}{99.41}
\newcommand{\MeanStatementCoverageFinalFightSpeciesMIO}{99.65}
\newcommand{\MeanStatementCoverageFlappyParrotSpeciesMIO}{93.69}
\newcommand{\MeanStatementCoverageFroggerSpeciesMIO}{94.36}
\newcommand{\MeanStatementCoverageFruitCatchingSpeciesMIO}{82.79}
\newcommand{\MeanStatementCoverageHackAttackSpeciesMIO}{92.72}
\newcommand{\MeanStatementCoverageOceanCleanupSpeciesMIO}{83.89}
\newcommand{\MeanStatementCoveragePokemonClickerSpeciesMIO}{26.21}
\newcommand{\MeanStatementCoverageSnakeSpeciesMIO}{94.89}
\newcommand{\MeanStatementCoverageSnowballFightSpeciesMIO}{95.47}
\newcommand{\MeanStatementCoverageSpaceOdysseySpeciesMIO}{87.27}
\newcommand{\MeanStatementCoverageWhackAMoleSpeciesMIO}{84.72}
\newcommand{\MeanStatementCoverageTotalSpeciesMIO}{86.28}
\newcommand{\EffectSizeStatementCoverageCatchTheDotsNoveltyCompatMIO}{0.41}
\newcommand{\PValStatementCoverageCatchTheDotsNoveltyCompatMIO}{0.20}
\newcommand{\EffectSizeStatementCoverageCatchTheDotsSpeciesCompatMIO}{0.49}
\newcommand{\PValStatementCoverageCatchTheDotsSpeciesCompatMIO}{0.85}
\newcommand{\EffectSizeStatementCoverageCatchTheDotsSpeciesNoveltyMIO}{0.57}
\newcommand{\PValStatementCoverageCatchTheDotsSpeciesNoveltyMIO}{0.31}
\newcommand{\EffectSizeStatementCoverageCityDefenderNoveltyCompatMIO}{0.48}
\newcommand{\PValStatementCoverageCityDefenderNoveltyCompatMIO}{0.33}
\newcommand{\EffectSizeStatementCoverageCityDefenderSpeciesCompatMIO}{0.48}
\newcommand{\PValStatementCoverageCityDefenderSpeciesCompatMIO}{0.33}
\newcommand{\EffectSizeStatementCoverageCityDefenderSpeciesNoveltyMIO}{0.50}
\newcommand{\PValStatementCoverageCityDefenderSpeciesNoveltyMIO}{1.00}
\newcommand{\EffectSizeStatementCoverageCreateYourWorldNoveltyCompatMIO}{0.43}
\newcommand{\PValStatementCoverageCreateYourWorldNoveltyCompatMIO}{0.14}
\newcommand{\EffectSizeStatementCoverageCreateYourWorldSpeciesCompatMIO}{0.50}
\newcommand{\PValStatementCoverageCreateYourWorldSpeciesCompatMIO}{0.97}
\newcommand{\EffectSizeStatementCoverageCreateYourWorldSpeciesNoveltyMIO}{0.57}
\newcommand{\PValStatementCoverageCreateYourWorldSpeciesNoveltyMIO}{0.18}
\newcommand{\EffectSizeStatementCoverageDessertRallyNoveltyCompatMIO}{0.50}
\newcommand{\PValStatementCoverageDessertRallyNoveltyCompatMIO}{1.00}
\newcommand{\EffectSizeStatementCoverageDessertRallySpeciesCompatMIO}{0.50}
\newcommand{\PValStatementCoverageDessertRallySpeciesCompatMIO}{1.00}
\newcommand{\EffectSizeStatementCoverageDessertRallySpeciesNoveltyMIO}{0.50}
\newcommand{\PValStatementCoverageDessertRallySpeciesNoveltyMIO}{1.00}
\newcommand{\EffectSizeStatementCoverageDieZauberlehrlingeNoveltyCompatMIO}{0.46}
\newcommand{\PValStatementCoverageDieZauberlehrlingeNoveltyCompatMIO}{0.59}
\newcommand{\EffectSizeStatementCoverageDieZauberlehrlingeSpeciesCompatMIO}{0.44}
\newcommand{\PValStatementCoverageDieZauberlehrlingeSpeciesCompatMIO}{0.40}
\newcommand{\EffectSizeStatementCoverageDieZauberlehrlingeSpeciesNoveltyMIO}{0.47}
\newcommand{\PValStatementCoverageDieZauberlehrlingeSpeciesNoveltyMIO}{0.68}
\newcommand{\EffectSizeStatementCoverageDodgeballNoveltyCompatMIO}{0.52}
\newcommand{\PValStatementCoverageDodgeballNoveltyCompatMIO}{0.70}
\newcommand{\EffectSizeStatementCoverageDodgeballSpeciesCompatMIO}{0.52}
\newcommand{\PValStatementCoverageDodgeballSpeciesCompatMIO}{0.70}
\newcommand{\EffectSizeStatementCoverageDodgeballSpeciesNoveltyMIO}{0.50}
\newcommand{\PValStatementCoverageDodgeballSpeciesNoveltyMIO}{1.00}
\newcommand{\EffectSizeStatementCoverageDragonsNoveltyCompatMIO}{0.44}
\newcommand{\PValStatementCoverageDragonsNoveltyCompatMIO}{0.43}
\newcommand{\EffectSizeStatementCoverageDragonsSpeciesCompatMIO}{0.49}
\newcommand{\PValStatementCoverageDragonsSpeciesCompatMIO}{0.85}
\newcommand{\EffectSizeStatementCoverageDragonsSpeciesNoveltyMIO}{0.56}
\newcommand{\PValStatementCoverageDragonsSpeciesNoveltyMIO}{0.41}
\newcommand{\EffectSizeStatementCoverageEndlessRunnerNoveltyCompatMIO}{\textbf{0.36}}
\newcommand{\PValStatementCoverageEndlessRunnerNoveltyCompatMIO}{\textbf{0.03}}
\newcommand{\EffectSizeStatementCoverageEndlessRunnerSpeciesCompatMIO}{0.44}
\newcommand{\PValStatementCoverageEndlessRunnerSpeciesCompatMIO}{0.33}
\newcommand{\EffectSizeStatementCoverageEndlessRunnerSpeciesNoveltyMIO}{0.57}
\newcommand{\PValStatementCoverageEndlessRunnerSpeciesNoveltyMIO}{0.27}
\newcommand{\EffectSizeStatementCoverageFallingStarsNoveltyCompatMIO}{0.58}
\newcommand{\PValStatementCoverageFallingStarsNoveltyCompatMIO}{0.20}
\newcommand{\EffectSizeStatementCoverageFallingStarsSpeciesCompatMIO}{0.53}
\newcommand{\PValStatementCoverageFallingStarsSpeciesCompatMIO}{0.61}
\newcommand{\EffectSizeStatementCoverageFallingStarsSpeciesNoveltyMIO}{0.45}
\newcommand{\PValStatementCoverageFallingStarsSpeciesNoveltyMIO}{0.44}
\newcommand{\EffectSizeStatementCoverageFinalFightNoveltyCompatMIO}{0.50}
\newcommand{\PValStatementCoverageFinalFightNoveltyCompatMIO}{1.00}
\newcommand{\EffectSizeStatementCoverageFinalFightSpeciesCompatMIO}{0.50}
\newcommand{\PValStatementCoverageFinalFightSpeciesCompatMIO}{1.00}
\newcommand{\EffectSizeStatementCoverageFinalFightSpeciesNoveltyMIO}{0.50}
\newcommand{\PValStatementCoverageFinalFightSpeciesNoveltyMIO}{1.00}
\newcommand{\EffectSizeStatementCoverageFlappyParrotNoveltyCompatMIO}{0.51}
\newcommand{\PValStatementCoverageFlappyParrotNoveltyCompatMIO}{0.77}
\newcommand{\EffectSizeStatementCoverageFlappyParrotSpeciesCompatMIO}{0.47}
\newcommand{\PValStatementCoverageFlappyParrotSpeciesCompatMIO}{0.52}
\newcommand{\EffectSizeStatementCoverageFlappyParrotSpeciesNoveltyMIO}{0.45}
\newcommand{\PValStatementCoverageFlappyParrotSpeciesNoveltyMIO}{0.35}
\newcommand{\EffectSizeStatementCoverageFroggerNoveltyCompatMIO}{0.46}
\newcommand{\PValStatementCoverageFroggerNoveltyCompatMIO}{0.37}
\newcommand{\EffectSizeStatementCoverageFroggerSpeciesCompatMIO}{0.42}
\newcommand{\PValStatementCoverageFroggerSpeciesCompatMIO}{0.06}
\newcommand{\EffectSizeStatementCoverageFroggerSpeciesNoveltyMIO}{0.45}
\newcommand{\PValStatementCoverageFroggerSpeciesNoveltyMIO}{0.19}
\newcommand{\EffectSizeStatementCoverageFruitCatchingNoveltyCompatMIO}{0.60}
\newcommand{\PValStatementCoverageFruitCatchingNoveltyCompatMIO}{0.17}
\newcommand{\EffectSizeStatementCoverageFruitCatchingSpeciesCompatMIO}{0.61}
\newcommand{\PValStatementCoverageFruitCatchingSpeciesCompatMIO}{0.13}
\newcommand{\EffectSizeStatementCoverageFruitCatchingSpeciesNoveltyMIO}{0.51}
\newcommand{\PValStatementCoverageFruitCatchingSpeciesNoveltyMIO}{0.89}
\newcommand{\EffectSizeStatementCoverageHackAttackNoveltyCompatMIO}{0.50}
\newcommand{\PValStatementCoverageHackAttackNoveltyCompatMIO}{1.00}
\newcommand{\EffectSizeStatementCoverageHackAttackSpeciesCompatMIO}{0.50}
\newcommand{\PValStatementCoverageHackAttackSpeciesCompatMIO}{1.00}
\newcommand{\EffectSizeStatementCoverageHackAttackSpeciesNoveltyMIO}{0.50}
\newcommand{\PValStatementCoverageHackAttackSpeciesNoveltyMIO}{1.00}
\newcommand{\EffectSizeStatementCoverageOceanCleanupNoveltyCompatMIO}{0.47}
\newcommand{\PValStatementCoverageOceanCleanupNoveltyCompatMIO}{0.69}
\newcommand{\EffectSizeStatementCoverageOceanCleanupSpeciesCompatMIO}{0.50}
\newcommand{\PValStatementCoverageOceanCleanupSpeciesCompatMIO}{0.98}
\newcommand{\EffectSizeStatementCoverageOceanCleanupSpeciesNoveltyMIO}{0.52}
\newcommand{\PValStatementCoverageOceanCleanupSpeciesNoveltyMIO}{0.79}
\newcommand{\EffectSizeStatementCoveragePokemonClickerNoveltyCompatMIO}{0.50}
\newcommand{\PValStatementCoveragePokemonClickerNoveltyCompatMIO}{1.00}
\newcommand{\EffectSizeStatementCoveragePokemonClickerSpeciesCompatMIO}{0.50}
\newcommand{\PValStatementCoveragePokemonClickerSpeciesCompatMIO}{1.00}
\newcommand{\EffectSizeStatementCoveragePokemonClickerSpeciesNoveltyMIO}{0.50}
\newcommand{\PValStatementCoveragePokemonClickerSpeciesNoveltyMIO}{1.00}
\newcommand{\EffectSizeStatementCoverageSnakeNoveltyCompatMIO}{0.50}
\newcommand{\PValStatementCoverageSnakeNoveltyCompatMIO}{1.00}
\newcommand{\EffectSizeStatementCoverageSnakeSpeciesCompatMIO}{0.52}
\newcommand{\PValStatementCoverageSnakeSpeciesCompatMIO}{0.33}
\newcommand{\EffectSizeStatementCoverageSnakeSpeciesNoveltyMIO}{0.52}
\newcommand{\PValStatementCoverageSnakeSpeciesNoveltyMIO}{0.33}
\newcommand{\EffectSizeStatementCoverageSnowballFightNoveltyCompatMIO}{0.42}
\newcommand{\PValStatementCoverageSnowballFightNoveltyCompatMIO}{0.19}
\newcommand{\EffectSizeStatementCoverageSnowballFightSpeciesCompatMIO}{0.53}
\newcommand{\PValStatementCoverageSnowballFightSpeciesCompatMIO}{0.57}
\newcommand{\EffectSizeStatementCoverageSnowballFightSpeciesNoveltyMIO}{0.62}
\newcommand{\PValStatementCoverageSnowballFightSpeciesNoveltyMIO}{0.06}
\newcommand{\EffectSizeStatementCoverageSpaceOdysseyNoveltyCompatMIO}{0.48}
\newcommand{\PValStatementCoverageSpaceOdysseyNoveltyCompatMIO}{0.33}
\newcommand{\EffectSizeStatementCoverageSpaceOdysseySpeciesCompatMIO}{0.48}
\newcommand{\PValStatementCoverageSpaceOdysseySpeciesCompatMIO}{0.33}
\newcommand{\EffectSizeStatementCoverageSpaceOdysseySpeciesNoveltyMIO}{0.50}
\newcommand{\PValStatementCoverageSpaceOdysseySpeciesNoveltyMIO}{1.00}
\newcommand{\EffectSizeStatementCoverageWhackAMoleNoveltyCompatMIO}{0.57}
\newcommand{\PValStatementCoverageWhackAMoleNoveltyCompatMIO}{0.33}
\newcommand{\EffectSizeStatementCoverageWhackAMoleSpeciesCompatMIO}{0.59}
\newcommand{\PValStatementCoverageWhackAMoleSpeciesCompatMIO}{0.23}
\newcommand{\EffectSizeStatementCoverageWhackAMoleSpeciesNoveltyMIO}{0.50}
\newcommand{\PValStatementCoverageWhackAMoleSpeciesNoveltyMIO}{0.95}
\newcommand{\MeanEffectSizeStatementCoverageNoveltyCompatMIO}{0.49}
\newcommand{\MeanEffectSizeStatementCoverageSpeciesCompatMIO}{0.50}
\newcommand{\MeanEffectSizeStatementCoverageSpeciesNoveltyMIO}{0.51}
\newcommand{\MeanBranchCoverageCatchTheDotsCompatMIO}{87.14}
\newcommand{\MeanBranchCoverageCityDefenderCompatMIO}{67.57}
\newcommand{\MeanBranchCoverageCreateYourWorldCompatMIO}{70.44}
\newcommand{\MeanBranchCoverageDessertRallyCompatMIO}{96.15}
\newcommand{\MeanBranchCoverageDieZauberlehrlingeCompatMIO}{69.57}
\newcommand{\MeanBranchCoverageDodgeballCompatMIO}{92.79}
\newcommand{\MeanBranchCoverageDragonsCompatMIO}{81.26}
\newcommand{\MeanBranchCoverageEndlessRunnerCompatMIO}{69.80}
\newcommand{\MeanBranchCoverageFallingStarsCompatMIO}{94.39}
\newcommand{\MeanBranchCoverageFinalFightCompatMIO}{99.23}
\newcommand{\MeanBranchCoverageFlappyParrotCompatMIO}{84.55}
\newcommand{\MeanBranchCoverageFroggerCompatMIO}{90.00}
\newcommand{\MeanBranchCoverageFruitCatchingCompatMIO}{78.45}
\newcommand{\MeanBranchCoverageHackAttackCompatMIO}{97.46}
\newcommand{\MeanBranchCoverageOceanCleanupCompatMIO}{73.71}
\newcommand{\MeanBranchCoveragePokemonClickerCompatMIO}{17.46}
\newcommand{\MeanBranchCoverageSnakeCompatMIO}{79.05}
\newcommand{\MeanBranchCoverageSnowballFightCompatMIO}{85.22}
\newcommand{\MeanBranchCoverageSpaceOdysseyCompatMIO}{91.23}
\newcommand{\MeanBranchCoverageWhackAMoleCompatMIO}{84.05}
\newcommand{\MeanBranchCoverageTotalCompatMIO}{80.48}
\newcommand{\MeanBranchCoverageCatchTheDotsNoveltyMIO}{87.90}
\newcommand{\MeanBranchCoverageCityDefenderNoveltyMIO}{67.66}
\newcommand{\MeanBranchCoverageCreateYourWorldNoveltyMIO}{71.36}
\newcommand{\MeanBranchCoverageDessertRallyNoveltyMIO}{96.15}
\newcommand{\MeanBranchCoverageDieZauberlehrlingeNoveltyMIO}{70.54}
\newcommand{\MeanBranchCoverageDodgeballNoveltyMIO}{92.71}
\newcommand{\MeanBranchCoverageDragonsNoveltyMIO}{81.79}
\newcommand{\MeanBranchCoverageEndlessRunnerNoveltyMIO}{72.20}
\newcommand{\MeanBranchCoverageFallingStarsNoveltyMIO}{94.02}
\newcommand{\MeanBranchCoverageFinalFightNoveltyMIO}{99.23}
\newcommand{\MeanBranchCoverageFlappyParrotNoveltyMIO}{84.24}
\newcommand{\MeanBranchCoverageFroggerNoveltyMIO}{90.00}
\newcommand{\MeanBranchCoverageFruitCatchingNoveltyMIO}{75.36}
\newcommand{\MeanBranchCoverageHackAttackNoveltyMIO}{97.46}
\newcommand{\MeanBranchCoverageOceanCleanupNoveltyMIO}{74.41}
\newcommand{\MeanBranchCoveragePokemonClickerNoveltyMIO}{17.46}
\newcommand{\MeanBranchCoverageSnakeNoveltyMIO}{78.57}
\newcommand{\MeanBranchCoverageSnowballFightNoveltyMIO}{86.52}
\newcommand{\MeanBranchCoverageSpaceOdysseyNoveltyMIO}{91.40}
\newcommand{\MeanBranchCoverageWhackAMoleNoveltyMIO}{83.47}
\newcommand{\MeanBranchCoverageTotalNoveltyMIO}{80.62}
\newcommand{\MeanBranchCoverageCatchTheDotsSpeciesMIO}{87.33}
\newcommand{\MeanBranchCoverageCityDefenderSpeciesMIO}{67.66}
\newcommand{\MeanBranchCoverageCreateYourWorldSpeciesMIO}{70.73}
\newcommand{\MeanBranchCoverageDessertRallySpeciesMIO}{96.15}
\newcommand{\MeanBranchCoverageDieZauberlehrlingeSpeciesMIO}{70.43}
\newcommand{\MeanBranchCoverageDodgeballSpeciesMIO}{92.71}
\newcommand{\MeanBranchCoverageDragonsSpeciesMIO}{81.19}
\newcommand{\MeanBranchCoverageEndlessRunnerSpeciesMIO}{70.67}
\newcommand{\MeanBranchCoverageFallingStarsSpeciesMIO}{94.24}
\newcommand{\MeanBranchCoverageFinalFightSpeciesMIO}{99.23}
\newcommand{\MeanBranchCoverageFlappyParrotSpeciesMIO}{85.76}
\newcommand{\MeanBranchCoverageFroggerSpeciesMIO}{90.17}
\newcommand{\MeanBranchCoverageFruitCatchingSpeciesMIO}{75.48}
\newcommand{\MeanBranchCoverageHackAttackSpeciesMIO}{97.46}
\newcommand{\MeanBranchCoverageOceanCleanupSpeciesMIO}{73.98}
\newcommand{\MeanBranchCoveragePokemonClickerSpeciesMIO}{17.46}
\newcommand{\MeanBranchCoverageSnakeSpeciesMIO}{78.57}
\newcommand{\MeanBranchCoverageSnowballFightSpeciesMIO}{84.64}
\newcommand{\MeanBranchCoverageSpaceOdysseySpeciesMIO}{91.40}
\newcommand{\MeanBranchCoverageWhackAMoleSpeciesMIO}{83.90}
\newcommand{\MeanBranchCoverageTotalSpeciesMIO}{80.46}
\newcommand{\EffectSizeBranchCoverageCatchTheDotsNoveltyCompatMIO}{0.41}
\newcommand{\PValBranchCoverageCatchTheDotsNoveltyCompatMIO}{0.20}
\newcommand{\EffectSizeBranchCoverageCatchTheDotsSpeciesCompatMIO}{0.49}
\newcommand{\PValBranchCoverageCatchTheDotsSpeciesCompatMIO}{0.85}
\newcommand{\EffectSizeBranchCoverageCatchTheDotsSpeciesNoveltyMIO}{0.57}
\newcommand{\PValBranchCoverageCatchTheDotsSpeciesNoveltyMIO}{0.31}
\newcommand{\EffectSizeBranchCoverageCityDefenderNoveltyCompatMIO}{0.48}
\newcommand{\PValBranchCoverageCityDefenderNoveltyCompatMIO}{0.33}
\newcommand{\EffectSizeBranchCoverageCityDefenderSpeciesCompatMIO}{0.48}
\newcommand{\PValBranchCoverageCityDefenderSpeciesCompatMIO}{0.33}
\newcommand{\EffectSizeBranchCoverageCityDefenderSpeciesNoveltyMIO}{0.50}
\newcommand{\PValBranchCoverageCityDefenderSpeciesNoveltyMIO}{1.00}
\newcommand{\EffectSizeBranchCoverageCreateYourWorldNoveltyCompatMIO}{0.43}
\newcommand{\PValBranchCoverageCreateYourWorldNoveltyCompatMIO}{0.14}
\newcommand{\EffectSizeBranchCoverageCreateYourWorldSpeciesCompatMIO}{0.50}
\newcommand{\PValBranchCoverageCreateYourWorldSpeciesCompatMIO}{0.97}
\newcommand{\EffectSizeBranchCoverageCreateYourWorldSpeciesNoveltyMIO}{0.57}
\newcommand{\PValBranchCoverageCreateYourWorldSpeciesNoveltyMIO}{0.18}
\newcommand{\EffectSizeBranchCoverageDessertRallyNoveltyCompatMIO}{0.50}
\newcommand{\PValBranchCoverageDessertRallyNoveltyCompatMIO}{1.00}
\newcommand{\EffectSizeBranchCoverageDessertRallySpeciesCompatMIO}{0.50}
\newcommand{\PValBranchCoverageDessertRallySpeciesCompatMIO}{1.00}
\newcommand{\EffectSizeBranchCoverageDessertRallySpeciesNoveltyMIO}{0.50}
\newcommand{\PValBranchCoverageDessertRallySpeciesNoveltyMIO}{1.00}
\newcommand{\EffectSizeBranchCoverageDieZauberlehrlingeNoveltyCompatMIO}{0.44}
\newcommand{\PValBranchCoverageDieZauberlehrlingeNoveltyCompatMIO}{0.43}
\newcommand{\EffectSizeBranchCoverageDieZauberlehrlingeSpeciesCompatMIO}{0.45}
\newcommand{\PValBranchCoverageDieZauberlehrlingeSpeciesCompatMIO}{0.52}
\newcommand{\EffectSizeBranchCoverageDieZauberlehrlingeSpeciesNoveltyMIO}{0.51}
\newcommand{\PValBranchCoverageDieZauberlehrlingeSpeciesNoveltyMIO}{0.87}
\newcommand{\EffectSizeBranchCoverageDodgeballNoveltyCompatMIO}{0.52}
\newcommand{\PValBranchCoverageDodgeballNoveltyCompatMIO}{0.70}
\newcommand{\EffectSizeBranchCoverageDodgeballSpeciesCompatMIO}{0.52}
\newcommand{\PValBranchCoverageDodgeballSpeciesCompatMIO}{0.70}
\newcommand{\EffectSizeBranchCoverageDodgeballSpeciesNoveltyMIO}{0.50}
\newcommand{\PValBranchCoverageDodgeballSpeciesNoveltyMIO}{1.00}
\newcommand{\EffectSizeBranchCoverageDragonsNoveltyCompatMIO}{0.44}
\newcommand{\PValBranchCoverageDragonsNoveltyCompatMIO}{0.40}
\newcommand{\EffectSizeBranchCoverageDragonsSpeciesCompatMIO}{0.50}
\newcommand{\PValBranchCoverageDragonsSpeciesCompatMIO}{0.95}
\newcommand{\EffectSizeBranchCoverageDragonsSpeciesNoveltyMIO}{0.58}
\newcommand{\PValBranchCoverageDragonsSpeciesNoveltyMIO}{0.31}
\newcommand{\EffectSizeBranchCoverageEndlessRunnerNoveltyCompatMIO}{\textbf{0.35}}
\newcommand{\PValBranchCoverageEndlessRunnerNoveltyCompatMIO}{\textbf{0.02}}
\newcommand{\EffectSizeBranchCoverageEndlessRunnerSpeciesCompatMIO}{0.44}
\newcommand{\PValBranchCoverageEndlessRunnerSpeciesCompatMIO}{0.40}
\newcommand{\EffectSizeBranchCoverageEndlessRunnerSpeciesNoveltyMIO}{0.59}
\newcommand{\PValBranchCoverageEndlessRunnerSpeciesNoveltyMIO}{0.19}
\newcommand{\EffectSizeBranchCoverageFallingStarsNoveltyCompatMIO}{0.58}
\newcommand{\PValBranchCoverageFallingStarsNoveltyCompatMIO}{0.20}
\newcommand{\EffectSizeBranchCoverageFallingStarsSpeciesCompatMIO}{0.53}
\newcommand{\PValBranchCoverageFallingStarsSpeciesCompatMIO}{0.61}
\newcommand{\EffectSizeBranchCoverageFallingStarsSpeciesNoveltyMIO}{0.45}
\newcommand{\PValBranchCoverageFallingStarsSpeciesNoveltyMIO}{0.44}
\newcommand{\EffectSizeBranchCoverageFinalFightNoveltyCompatMIO}{0.50}
\newcommand{\PValBranchCoverageFinalFightNoveltyCompatMIO}{1.00}
\newcommand{\EffectSizeBranchCoverageFinalFightSpeciesCompatMIO}{0.50}
\newcommand{\PValBranchCoverageFinalFightSpeciesCompatMIO}{1.00}
\newcommand{\EffectSizeBranchCoverageFinalFightSpeciesNoveltyMIO}{0.50}
\newcommand{\PValBranchCoverageFinalFightSpeciesNoveltyMIO}{1.00}
\newcommand{\EffectSizeBranchCoverageFlappyParrotNoveltyCompatMIO}{0.51}
\newcommand{\PValBranchCoverageFlappyParrotNoveltyCompatMIO}{0.77}
\newcommand{\EffectSizeBranchCoverageFlappyParrotSpeciesCompatMIO}{0.47}
\newcommand{\PValBranchCoverageFlappyParrotSpeciesCompatMIO}{0.52}
\newcommand{\EffectSizeBranchCoverageFlappyParrotSpeciesNoveltyMIO}{0.45}
\newcommand{\PValBranchCoverageFlappyParrotSpeciesNoveltyMIO}{0.35}
\newcommand{\EffectSizeBranchCoverageFroggerNoveltyCompatMIO}{0.50}
\newcommand{\PValBranchCoverageFroggerNoveltyCompatMIO}{1.00}
\newcommand{\EffectSizeBranchCoverageFroggerSpeciesCompatMIO}{0.47}
\newcommand{\PValBranchCoverageFroggerSpeciesCompatMIO}{0.16}
\newcommand{\EffectSizeBranchCoverageFroggerSpeciesNoveltyMIO}{0.47}
\newcommand{\PValBranchCoverageFroggerSpeciesNoveltyMIO}{0.16}
\newcommand{\EffectSizeBranchCoverageFruitCatchingNoveltyCompatMIO}{0.62}
\newcommand{\PValBranchCoverageFruitCatchingNoveltyCompatMIO}{0.10}
\newcommand{\EffectSizeBranchCoverageFruitCatchingSpeciesCompatMIO}{0.62}
\newcommand{\PValBranchCoverageFruitCatchingSpeciesCompatMIO}{0.10}
\newcommand{\EffectSizeBranchCoverageFruitCatchingSpeciesNoveltyMIO}{0.50}
\newcommand{\PValBranchCoverageFruitCatchingSpeciesNoveltyMIO}{1.00}
\newcommand{\EffectSizeBranchCoverageHackAttackNoveltyCompatMIO}{0.50}
\newcommand{\PValBranchCoverageHackAttackNoveltyCompatMIO}{1.00}
\newcommand{\EffectSizeBranchCoverageHackAttackSpeciesCompatMIO}{0.50}
\newcommand{\PValBranchCoverageHackAttackSpeciesCompatMIO}{1.00}
\newcommand{\EffectSizeBranchCoverageHackAttackSpeciesNoveltyMIO}{0.50}
\newcommand{\PValBranchCoverageHackAttackSpeciesNoveltyMIO}{1.00}
\newcommand{\EffectSizeBranchCoverageOceanCleanupNoveltyCompatMIO}{0.47}
\newcommand{\PValBranchCoverageOceanCleanupNoveltyCompatMIO}{0.72}
\newcommand{\EffectSizeBranchCoverageOceanCleanupSpeciesCompatMIO}{0.50}
\newcommand{\PValBranchCoverageOceanCleanupSpeciesCompatMIO}{0.99}
\newcommand{\EffectSizeBranchCoverageOceanCleanupSpeciesNoveltyMIO}{0.53}
\newcommand{\PValBranchCoverageOceanCleanupSpeciesNoveltyMIO}{0.73}
\newcommand{\EffectSizeBranchCoveragePokemonClickerNoveltyCompatMIO}{0.50}
\newcommand{\PValBranchCoveragePokemonClickerNoveltyCompatMIO}{1.00}
\newcommand{\EffectSizeBranchCoveragePokemonClickerSpeciesCompatMIO}{0.50}
\newcommand{\PValBranchCoveragePokemonClickerSpeciesCompatMIO}{1.00}
\newcommand{\EffectSizeBranchCoveragePokemonClickerSpeciesNoveltyMIO}{0.50}
\newcommand{\PValBranchCoveragePokemonClickerSpeciesNoveltyMIO}{1.00}
\newcommand{\EffectSizeBranchCoverageSnakeNoveltyCompatMIO}{0.53}
\newcommand{\PValBranchCoverageSnakeNoveltyCompatMIO}{0.16}
\newcommand{\EffectSizeBranchCoverageSnakeSpeciesCompatMIO}{0.53}
\newcommand{\PValBranchCoverageSnakeSpeciesCompatMIO}{0.16}
\newcommand{\EffectSizeBranchCoverageSnakeSpeciesNoveltyMIO}{0.50}
\newcommand{\PValBranchCoverageSnakeSpeciesNoveltyMIO}{1.00}
\newcommand{\EffectSizeBranchCoverageSnowballFightNoveltyCompatMIO}{0.42}
\newcommand{\PValBranchCoverageSnowballFightNoveltyCompatMIO}{0.22}
\newcommand{\EffectSizeBranchCoverageSnowballFightSpeciesCompatMIO}{0.53}
\newcommand{\PValBranchCoverageSnowballFightSpeciesCompatMIO}{0.57}
\newcommand{\EffectSizeBranchCoverageSnowballFightSpeciesNoveltyMIO}{0.61}
\newcommand{\PValBranchCoverageSnowballFightSpeciesNoveltyMIO}{0.07}
\newcommand{\EffectSizeBranchCoverageSpaceOdysseyNoveltyCompatMIO}{0.48}
\newcommand{\PValBranchCoverageSpaceOdysseyNoveltyCompatMIO}{0.33}
\newcommand{\EffectSizeBranchCoverageSpaceOdysseySpeciesCompatMIO}{0.48}
\newcommand{\PValBranchCoverageSpaceOdysseySpeciesCompatMIO}{0.33}
\newcommand{\EffectSizeBranchCoverageSpaceOdysseySpeciesNoveltyMIO}{0.50}
\newcommand{\PValBranchCoverageSpaceOdysseySpeciesNoveltyMIO}{1.00}
\newcommand{\EffectSizeBranchCoverageWhackAMoleNoveltyCompatMIO}{0.57}
\newcommand{\PValBranchCoverageWhackAMoleNoveltyCompatMIO}{0.37}
\newcommand{\EffectSizeBranchCoverageWhackAMoleSpeciesCompatMIO}{0.53}
\newcommand{\PValBranchCoverageWhackAMoleSpeciesCompatMIO}{0.70}
\newcommand{\EffectSizeBranchCoverageWhackAMoleSpeciesNoveltyMIO}{0.45}
\newcommand{\PValBranchCoverageWhackAMoleSpeciesNoveltyMIO}{0.53}
\newcommand{\MeanEffectSizeBranchCoverageNoveltyCompatMIO}{0.49}
\newcommand{\MeanEffectSizeBranchCoverageSpeciesCompatMIO}{0.50}
\newcommand{\MeanEffectSizeBranchCoverageSpeciesNoveltyMIO}{0.51}
\newcommand{\MeanFitnessEvaluationsCatchTheDotsCompatMIO}{687.87}
\newcommand{\MeanFitnessEvaluationsCityDefenderCompatMIO}{108.90}
\newcommand{\MeanFitnessEvaluationsCreateYourWorldCompatMIO}{618.97}
\newcommand{\MeanFitnessEvaluationsDessertRallyCompatMIO}{827.50}
\newcommand{\MeanFitnessEvaluationsDieZauberlehrlingeCompatMIO}{76.53}
\newcommand{\MeanFitnessEvaluationsDodgeballCompatMIO}{585.67}
\newcommand{\MeanFitnessEvaluationsDragonsCompatMIO}{79.70}
\newcommand{\MeanFitnessEvaluationsEndlessRunnerCompatMIO}{87.83}
\newcommand{\MeanFitnessEvaluationsFallingStarsCompatMIO}{630.30}
\newcommand{\MeanFitnessEvaluationsFinalFightCompatMIO}{803.27}
\newcommand{\MeanFitnessEvaluationsFlappyParrotCompatMIO}{349.40}
\newcommand{\MeanFitnessEvaluationsFroggerCompatMIO}{585.60}
\newcommand{\MeanFitnessEvaluationsFruitCatchingCompatMIO}{4219.57}
\newcommand{\MeanFitnessEvaluationsHackAttackCompatMIO}{646.93}
\newcommand{\MeanFitnessEvaluationsOceanCleanupCompatMIO}{3527.80}
\newcommand{\MeanFitnessEvaluationsPokemonClickerCompatMIO}{444.83}
\newcommand{\MeanFitnessEvaluationsSnakeCompatMIO}{23414.20}
\newcommand{\MeanFitnessEvaluationsSnowballFightCompatMIO}{660.53}
\newcommand{\MeanFitnessEvaluationsSpaceOdysseyCompatMIO}{281.23}
\newcommand{\MeanFitnessEvaluationsWhackAMoleCompatMIO}{1683.70}
\newcommand{\MeanFitnessEvaluationsTotalCompatMIO}{2016.02}
\newcommand{\MeanFitnessEvaluationsCatchTheDotsNoveltyMIO}{669.67}
\newcommand{\MeanFitnessEvaluationsCityDefenderNoveltyMIO}{114.73}
\newcommand{\MeanFitnessEvaluationsCreateYourWorldNoveltyMIO}{620.93}
\newcommand{\MeanFitnessEvaluationsDessertRallyNoveltyMIO}{824.90}
\newcommand{\MeanFitnessEvaluationsDieZauberlehrlingeNoveltyMIO}{77.27}
\newcommand{\MeanFitnessEvaluationsDodgeballNoveltyMIO}{591.40}
\newcommand{\MeanFitnessEvaluationsDragonsNoveltyMIO}{77.40}
\newcommand{\MeanFitnessEvaluationsEndlessRunnerNoveltyMIO}{87.03}
\newcommand{\MeanFitnessEvaluationsFallingStarsNoveltyMIO}{629.60}
\newcommand{\MeanFitnessEvaluationsFinalFightNoveltyMIO}{867.87}
\newcommand{\MeanFitnessEvaluationsFlappyParrotNoveltyMIO}{346.37}
\newcommand{\MeanFitnessEvaluationsFroggerNoveltyMIO}{587.40}
\newcommand{\MeanFitnessEvaluationsFruitCatchingNoveltyMIO}{3872.90}
\newcommand{\MeanFitnessEvaluationsHackAttackNoveltyMIO}{632.73}
\newcommand{\MeanFitnessEvaluationsOceanCleanupNoveltyMIO}{3812.63}
\newcommand{\MeanFitnessEvaluationsPokemonClickerNoveltyMIO}{633.10}
\newcommand{\MeanFitnessEvaluationsSnakeNoveltyMIO}{23749.33}
\newcommand{\MeanFitnessEvaluationsSnowballFightNoveltyMIO}{654.63}
\newcommand{\MeanFitnessEvaluationsSpaceOdysseyNoveltyMIO}{269.90}
\newcommand{\MeanFitnessEvaluationsWhackAMoleNoveltyMIO}{1765.03}
\newcommand{\MeanFitnessEvaluationsTotalNoveltyMIO}{2044.24}
\newcommand{\MeanFitnessEvaluationsCatchTheDotsSpeciesMIO}{724.97}
\newcommand{\MeanFitnessEvaluationsCityDefenderSpeciesMIO}{113.03}
\newcommand{\MeanFitnessEvaluationsCreateYourWorldSpeciesMIO}{619.17}
\newcommand{\MeanFitnessEvaluationsDessertRallySpeciesMIO}{843.60}
\newcommand{\MeanFitnessEvaluationsDieZauberlehrlingeSpeciesMIO}{75.03}
\newcommand{\MeanFitnessEvaluationsDodgeballSpeciesMIO}{597.73}
\newcommand{\MeanFitnessEvaluationsDragonsSpeciesMIO}{76.93}
\newcommand{\MeanFitnessEvaluationsEndlessRunnerSpeciesMIO}{93.07}
\newcommand{\MeanFitnessEvaluationsFallingStarsSpeciesMIO}{627.80}
\newcommand{\MeanFitnessEvaluationsFinalFightSpeciesMIO}{794.83}
\newcommand{\MeanFitnessEvaluationsFlappyParrotSpeciesMIO}{348}
\newcommand{\MeanFitnessEvaluationsFroggerSpeciesMIO}{586.87}
\newcommand{\MeanFitnessEvaluationsFruitCatchingSpeciesMIO}{4771.87}
\newcommand{\MeanFitnessEvaluationsHackAttackSpeciesMIO}{647.23}
\newcommand{\MeanFitnessEvaluationsOceanCleanupSpeciesMIO}{3692.40}
\newcommand{\MeanFitnessEvaluationsPokemonClickerSpeciesMIO}{672.83}
\newcommand{\MeanFitnessEvaluationsSnakeSpeciesMIO}{24944.93}
\newcommand{\MeanFitnessEvaluationsSnowballFightSpeciesMIO}{665.70}
\newcommand{\MeanFitnessEvaluationsSpaceOdysseySpeciesMIO}{298.20}
\newcommand{\MeanFitnessEvaluationsWhackAMoleSpeciesMIO}{1773.43}
\newcommand{\MeanFitnessEvaluationsTotalSpeciesMIO}{2148.38}
\newcommand{\EffectSizeFitnessEvaluationsCatchTheDotsNoveltyCompatMIO}{0.44}
\newcommand{\PValFitnessEvaluationsCatchTheDotsNoveltyCompatMIO}{0.45}
\newcommand{\EffectSizeFitnessEvaluationsCatchTheDotsSpeciesCompatMIO}{0.44}
\newcommand{\PValFitnessEvaluationsCatchTheDotsSpeciesCompatMIO}{0.42}
\newcommand{\EffectSizeFitnessEvaluationsCatchTheDotsSpeciesNoveltyMIO}{0.50}
\newcommand{\PValFitnessEvaluationsCatchTheDotsSpeciesNoveltyMIO}{0.99}
\newcommand{\EffectSizeFitnessEvaluationsCityDefenderNoveltyCompatMIO}{0.57}
\newcommand{\PValFitnessEvaluationsCityDefenderNoveltyCompatMIO}{0.38}
\newcommand{\EffectSizeFitnessEvaluationsCityDefenderSpeciesCompatMIO}{0.55}
\newcommand{\PValFitnessEvaluationsCityDefenderSpeciesCompatMIO}{0.54}
\newcommand{\EffectSizeFitnessEvaluationsCityDefenderSpeciesNoveltyMIO}{0.47}
\newcommand{\PValFitnessEvaluationsCityDefenderSpeciesNoveltyMIO}{0.73}
\newcommand{\EffectSizeFitnessEvaluationsCreateYourWorldNoveltyCompatMIO}{0.48}
\newcommand{\PValFitnessEvaluationsCreateYourWorldNoveltyCompatMIO}{0.77}
\newcommand{\EffectSizeFitnessEvaluationsCreateYourWorldSpeciesCompatMIO}{0.47}
\newcommand{\PValFitnessEvaluationsCreateYourWorldSpeciesCompatMIO}{0.70}
\newcommand{\EffectSizeFitnessEvaluationsCreateYourWorldSpeciesNoveltyMIO}{0.49}
\newcommand{\PValFitnessEvaluationsCreateYourWorldSpeciesNoveltyMIO}{0.94}
\newcommand{\EffectSizeFitnessEvaluationsDessertRallyNoveltyCompatMIO}{0.47}
\newcommand{\PValFitnessEvaluationsDessertRallyNoveltyCompatMIO}{0.66}
\newcommand{\EffectSizeFitnessEvaluationsDessertRallySpeciesCompatMIO}{0.38}
\newcommand{\PValFitnessEvaluationsDessertRallySpeciesCompatMIO}{0.10}
\newcommand{\EffectSizeFitnessEvaluationsDessertRallySpeciesNoveltyMIO}{0.37}
\newcommand{\PValFitnessEvaluationsDessertRallySpeciesNoveltyMIO}{0.09}
\newcommand{\EffectSizeFitnessEvaluationsDieZauberlehrlingeNoveltyCompatMIO}{0.53}
\newcommand{\PValFitnessEvaluationsDieZauberlehrlingeNoveltyCompatMIO}{0.34}
\newcommand{\EffectSizeFitnessEvaluationsDieZauberlehrlingeSpeciesCompatMIO}{0.52}
\newcommand{\PValFitnessEvaluationsDieZauberlehrlingeSpeciesCompatMIO}{0.65}
\newcommand{\EffectSizeFitnessEvaluationsDieZauberlehrlingeSpeciesNoveltyMIO}{0.48}
\newcommand{\PValFitnessEvaluationsDieZauberlehrlingeSpeciesNoveltyMIO}{0.60}
\newcommand{\EffectSizeFitnessEvaluationsDodgeballNoveltyCompatMIO}{0.48}
\newcommand{\PValFitnessEvaluationsDodgeballNoveltyCompatMIO}{0.84}
\newcommand{\EffectSizeFitnessEvaluationsDodgeballSpeciesCompatMIO}{0.44}
\newcommand{\PValFitnessEvaluationsDodgeballSpeciesCompatMIO}{0.46}
\newcommand{\EffectSizeFitnessEvaluationsDodgeballSpeciesNoveltyMIO}{0.47}
\newcommand{\PValFitnessEvaluationsDodgeballSpeciesNoveltyMIO}{0.71}
\newcommand{\EffectSizeFitnessEvaluationsDragonsNoveltyCompatMIO}{0.51}
\newcommand{\PValFitnessEvaluationsDragonsNoveltyCompatMIO}{0.89}
\newcommand{\EffectSizeFitnessEvaluationsDragonsSpeciesCompatMIO}{0.61}
\newcommand{\PValFitnessEvaluationsDragonsSpeciesCompatMIO}{0.15}
\newcommand{\EffectSizeFitnessEvaluationsDragonsSpeciesNoveltyMIO}{0.62}
\newcommand{\PValFitnessEvaluationsDragonsSpeciesNoveltyMIO}{0.10}
\newcommand{\EffectSizeFitnessEvaluationsEndlessRunnerNoveltyCompatMIO}{0.50}
\newcommand{\PValFitnessEvaluationsEndlessRunnerNoveltyCompatMIO}{0.99}
\newcommand{\EffectSizeFitnessEvaluationsEndlessRunnerSpeciesCompatMIO}{0.50}
\newcommand{\PValFitnessEvaluationsEndlessRunnerSpeciesCompatMIO}{0.95}
\newcommand{\EffectSizeFitnessEvaluationsEndlessRunnerSpeciesNoveltyMIO}{0.50}
\newcommand{\PValFitnessEvaluationsEndlessRunnerSpeciesNoveltyMIO}{0.95}
\newcommand{\EffectSizeFitnessEvaluationsFallingStarsNoveltyCompatMIO}{0.48}
\newcommand{\PValFitnessEvaluationsFallingStarsNoveltyCompatMIO}{0.82}
\newcommand{\EffectSizeFitnessEvaluationsFallingStarsSpeciesCompatMIO}{0.52}
\newcommand{\PValFitnessEvaluationsFallingStarsSpeciesCompatMIO}{0.78}
\newcommand{\EffectSizeFitnessEvaluationsFallingStarsSpeciesNoveltyMIO}{0.53}
\newcommand{\PValFitnessEvaluationsFallingStarsSpeciesNoveltyMIO}{0.67}
\newcommand{\EffectSizeFitnessEvaluationsFinalFightNoveltyCompatMIO}{\textbf{0.34}}
\newcommand{\PValFitnessEvaluationsFinalFightNoveltyCompatMIO}{\textbf{0.04}}
\newcommand{\EffectSizeFitnessEvaluationsFinalFightSpeciesCompatMIO}{0.53}
\newcommand{\PValFitnessEvaluationsFinalFightSpeciesCompatMIO}{0.67}
\newcommand{\EffectSizeFitnessEvaluationsFinalFightSpeciesNoveltyMIO}{\textbf{0.67}}
\newcommand{\PValFitnessEvaluationsFinalFightSpeciesNoveltyMIO}{\textbf{0.03}}
\newcommand{\EffectSizeFitnessEvaluationsFlappyParrotNoveltyCompatMIO}{0.50}
\newcommand{\PValFitnessEvaluationsFlappyParrotNoveltyCompatMIO}{1.00}
\newcommand{\EffectSizeFitnessEvaluationsFlappyParrotSpeciesCompatMIO}{0.50}
\newcommand{\PValFitnessEvaluationsFlappyParrotSpeciesCompatMIO}{0.99}
\newcommand{\EffectSizeFitnessEvaluationsFlappyParrotSpeciesNoveltyMIO}{0.50}
\newcommand{\PValFitnessEvaluationsFlappyParrotSpeciesNoveltyMIO}{0.98}
\newcommand{\EffectSizeFitnessEvaluationsFroggerNoveltyCompatMIO}{0.53}
\newcommand{\PValFitnessEvaluationsFroggerNoveltyCompatMIO}{0.67}
\newcommand{\EffectSizeFitnessEvaluationsFroggerSpeciesCompatMIO}{0.52}
\newcommand{\PValFitnessEvaluationsFroggerSpeciesCompatMIO}{0.81}
\newcommand{\EffectSizeFitnessEvaluationsFroggerSpeciesNoveltyMIO}{0.49}
\newcommand{\PValFitnessEvaluationsFroggerSpeciesNoveltyMIO}{0.87}
\newcommand{\EffectSizeFitnessEvaluationsFruitCatchingNoveltyCompatMIO}{0.49}
\newcommand{\PValFitnessEvaluationsFruitCatchingNoveltyCompatMIO}{0.91}
\newcommand{\EffectSizeFitnessEvaluationsFruitCatchingSpeciesCompatMIO}{0.43}
\newcommand{\PValFitnessEvaluationsFruitCatchingSpeciesCompatMIO}{0.39}
\newcommand{\EffectSizeFitnessEvaluationsFruitCatchingSpeciesNoveltyMIO}{0.42}
\newcommand{\PValFitnessEvaluationsFruitCatchingSpeciesNoveltyMIO}{0.30}
\newcommand{\EffectSizeFitnessEvaluationsHackAttackNoveltyCompatMIO}{0.53}
\newcommand{\PValFitnessEvaluationsHackAttackNoveltyCompatMIO}{0.68}
\newcommand{\EffectSizeFitnessEvaluationsHackAttackSpeciesCompatMIO}{0.45}
\newcommand{\PValFitnessEvaluationsHackAttackSpeciesCompatMIO}{0.50}
\newcommand{\EffectSizeFitnessEvaluationsHackAttackSpeciesNoveltyMIO}{0.44}
\newcommand{\PValFitnessEvaluationsHackAttackSpeciesNoveltyMIO}{0.46}
\newcommand{\EffectSizeFitnessEvaluationsOceanCleanupNoveltyCompatMIO}{0.48}
\newcommand{\PValFitnessEvaluationsOceanCleanupNoveltyCompatMIO}{0.81}
\newcommand{\EffectSizeFitnessEvaluationsOceanCleanupSpeciesCompatMIO}{0.44}
\newcommand{\PValFitnessEvaluationsOceanCleanupSpeciesCompatMIO}{0.45}
\newcommand{\EffectSizeFitnessEvaluationsOceanCleanupSpeciesNoveltyMIO}{0.46}
\newcommand{\PValFitnessEvaluationsOceanCleanupSpeciesNoveltyMIO}{0.61}
\newcommand{\EffectSizeFitnessEvaluationsPokemonClickerNoveltyCompatMIO}{\textbf{0.19}}
\newcommand{\PValFitnessEvaluationsPokemonClickerNoveltyCompatMIO}{\textbf{< 0.01}}
\newcommand{\EffectSizeFitnessEvaluationsPokemonClickerSpeciesCompatMIO}{\textbf{0.19}}
\newcommand{\PValFitnessEvaluationsPokemonClickerSpeciesCompatMIO}{\textbf{< 0.01}}
\newcommand{\EffectSizeFitnessEvaluationsPokemonClickerSpeciesNoveltyMIO}{0.56}
\newcommand{\PValFitnessEvaluationsPokemonClickerSpeciesNoveltyMIO}{0.45}
\newcommand{\EffectSizeFitnessEvaluationsSnakeNoveltyCompatMIO}{0.46}
\newcommand{\PValFitnessEvaluationsSnakeNoveltyCompatMIO}{0.57}
\newcommand{\EffectSizeFitnessEvaluationsSnakeSpeciesCompatMIO}{\textbf{0.33}}
\newcommand{\PValFitnessEvaluationsSnakeSpeciesCompatMIO}{\textbf{0.02}}
\newcommand{\EffectSizeFitnessEvaluationsSnakeSpeciesNoveltyMIO}{0.45}
\newcommand{\PValFitnessEvaluationsSnakeSpeciesNoveltyMIO}{0.51}
\newcommand{\EffectSizeFitnessEvaluationsSnowballFightNoveltyCompatMIO}{0.58}
\newcommand{\PValFitnessEvaluationsSnowballFightNoveltyCompatMIO}{0.26}
\newcommand{\EffectSizeFitnessEvaluationsSnowballFightSpeciesCompatMIO}{0.40}
\newcommand{\PValFitnessEvaluationsSnowballFightSpeciesCompatMIO}{0.19}
\newcommand{\EffectSizeFitnessEvaluationsSnowballFightSpeciesNoveltyMIO}{\textbf{0.32}}
\newcommand{\PValFitnessEvaluationsSnowballFightSpeciesNoveltyMIO}{\textbf{0.02}}
\newcommand{\EffectSizeFitnessEvaluationsSpaceOdysseyNoveltyCompatMIO}{0.53}
\newcommand{\PValFitnessEvaluationsSpaceOdysseyNoveltyCompatMIO}{0.74}
\newcommand{\EffectSizeFitnessEvaluationsSpaceOdysseySpeciesCompatMIO}{0.49}
\newcommand{\PValFitnessEvaluationsSpaceOdysseySpeciesCompatMIO}{0.86}
\newcommand{\EffectSizeFitnessEvaluationsSpaceOdysseySpeciesNoveltyMIO}{0.45}
\newcommand{\PValFitnessEvaluationsSpaceOdysseySpeciesNoveltyMIO}{0.52}
\newcommand{\EffectSizeFitnessEvaluationsWhackAMoleNoveltyCompatMIO}{0.43}
\newcommand{\PValFitnessEvaluationsWhackAMoleNoveltyCompatMIO}{0.36}
\newcommand{\EffectSizeFitnessEvaluationsWhackAMoleSpeciesCompatMIO}{0.45}
\newcommand{\PValFitnessEvaluationsWhackAMoleSpeciesCompatMIO}{0.50}
\newcommand{\EffectSizeFitnessEvaluationsWhackAMoleSpeciesNoveltyMIO}{0.49}
\newcommand{\PValFitnessEvaluationsWhackAMoleSpeciesNoveltyMIO}{0.88}
\newcommand{\MeanEffectSizeFitnessEvaluationsNoveltyCompatMIO}{0.48}
\newcommand{\MeanEffectSizeFitnessEvaluationsSpeciesCompatMIO}{0.46}
\newcommand{\MeanEffectSizeFitnessEvaluationsSpeciesNoveltyMIO}{0.48}

\newcommand{\WinCountCatchTheDotsCompatMOSA}{11}
\newcommand{\WinCountCatchTheDotsNoveltyMOSA}{7}
\newcommand{\WinCountCatchTheDotsSpeciesMOSA}{8}
\newcommand{\WinCountCityDefenderCompatMOSA}{0}
\newcommand{\WinCountCityDefenderNoveltyMOSA}{0}
\newcommand{\WinCountCityDefenderSpeciesMOSA}{0}
\newcommand{\WinCountCreateYourWorldCompatMOSA}{4}
\newcommand{\WinCountCreateYourWorldNoveltyMOSA}{3}
\newcommand{\WinCountCreateYourWorldSpeciesMOSA}{3}
\newcommand{\WinCountDessertRallyCompatMOSA}{0}
\newcommand{\WinCountDessertRallyNoveltyMOSA}{0}
\newcommand{\WinCountDessertRallySpeciesMOSA}{0}
\newcommand{\WinCountDieZauberlehrlingeCompatMOSA}{19}
\newcommand{\WinCountDieZauberlehrlingeNoveltyMOSA}{21}
\newcommand{\WinCountDieZauberlehrlingeSpeciesMOSA}{19}
\newcommand{\WinCountDodgeballCompatMOSA}{0}
\newcommand{\WinCountDodgeballNoveltyMOSA}{0}
\newcommand{\WinCountDodgeballSpeciesMOSA}{0}
\newcommand{\WinCountDragonsCompatMOSA}{29}
\newcommand{\WinCountDragonsNoveltyMOSA}{30}
\newcommand{\WinCountDragonsSpeciesMOSA}{29}
\newcommand{\WinCountEndlessRunnerCompatMOSA}{30}
\newcommand{\WinCountEndlessRunnerNoveltyMOSA}{30}
\newcommand{\WinCountEndlessRunnerSpeciesMOSA}{30}
\newcommand{\WinCountFallingStarsCompatMOSA}{30}
\newcommand{\WinCountFallingStarsNoveltyMOSA}{30}
\newcommand{\WinCountFallingStarsSpeciesMOSA}{30}
\newcommand{\WinCountFinalFightCompatMOSA}{30}
\newcommand{\WinCountFinalFightNoveltyMOSA}{30}
\newcommand{\WinCountFinalFightSpeciesMOSA}{30}
\newcommand{\WinCountFlappyParrotCompatMOSA}{7}
\newcommand{\WinCountFlappyParrotNoveltyMOSA}{4}
\newcommand{\WinCountFlappyParrotSpeciesMOSA}{6}
\newcommand{\WinCountFroggerCompatMOSA}{4}
\newcommand{\WinCountFroggerNoveltyMOSA}{0}
\newcommand{\WinCountFroggerSpeciesMOSA}{5}
\newcommand{\WinCountFruitCatchingCompatMOSA}{4}
\newcommand{\WinCountFruitCatchingNoveltyMOSA}{5}
\newcommand{\WinCountFruitCatchingSpeciesMOSA}{10}
\newcommand{\WinCountHackAttackCompatMOSA}{11}
\newcommand{\WinCountHackAttackNoveltyMOSA}{6}
\newcommand{\WinCountHackAttackSpeciesMOSA}{10}
\newcommand{\WinCountOceanCleanupCompatMOSA}{19}
\newcommand{\WinCountOceanCleanupNoveltyMOSA}{10}
\newcommand{\WinCountOceanCleanupSpeciesMOSA}{17}
\newcommand{\WinCountPokemonClickerCompatMOSA}{0}
\newcommand{\WinCountPokemonClickerNoveltyMOSA}{0}
\newcommand{\WinCountPokemonClickerSpeciesMOSA}{0}
\newcommand{\WinCountSnakeCompatMOSA}{0}
\newcommand{\WinCountSnakeNoveltyMOSA}{0}
\newcommand{\WinCountSnakeSpeciesMOSA}{0}
\newcommand{\WinCountSnowballFightCompatMOSA}{10}
\newcommand{\WinCountSnowballFightNoveltyMOSA}{12}
\newcommand{\WinCountSnowballFightSpeciesMOSA}{17}
\newcommand{\WinCountSpaceOdysseyCompatMOSA}{0}
\newcommand{\WinCountSpaceOdysseyNoveltyMOSA}{0}
\newcommand{\WinCountSpaceOdysseySpeciesMOSA}{0}
\newcommand{\WinCountWhackAMoleCompatMOSA}{29}
\newcommand{\WinCountWhackAMoleNoveltyMOSA}{27}
\newcommand{\WinCountWhackAMoleSpeciesMOSA}{29}
\newcommand{\MeanWinsCompatMOSA}{11.85}
\newcommand{\MeanWinsNoveltyMOSA}{10.75}
\newcommand{\MeanWinsSpeciesMOSA}{12.15}
\newcommand{\MeanStatementCoverageCatchTheDotsCompatMOSA}{96.75}
\newcommand{\MeanStatementCoverageCityDefenderCompatMOSA}{74.33}
\newcommand{\MeanStatementCoverageCreateYourWorldCompatMOSA}{75.52}
\newcommand{\MeanStatementCoverageDessertRallyCompatMOSA}{93.90}
\newcommand{\MeanStatementCoverageDieZauberlehrlingeCompatMOSA}{85.04}
\newcommand{\MeanStatementCoverageDodgeballCompatMOSA}{96.11}
\newcommand{\MeanStatementCoverageDragonsCompatMOSA}{87.96}
\newcommand{\MeanStatementCoverageEndlessRunnerCompatMOSA}{82.82}
\newcommand{\MeanStatementCoverageFallingStarsCompatMOSA}{99.63}
\newcommand{\MeanStatementCoverageFinalFightCompatMOSA}{99.65}
\newcommand{\MeanStatementCoverageFlappyParrotCompatMOSA}{93.60}
\newcommand{\MeanStatementCoverageFroggerCompatMOSA}{94.49}
\newcommand{\MeanStatementCoverageFruitCatchingCompatMOSA}{87.03}
\newcommand{\MeanStatementCoverageHackAttackCompatMOSA}{95.23}
\newcommand{\MeanStatementCoverageOceanCleanupCompatMOSA}{84.70}
\newcommand{\MeanStatementCoveragePokemonClickerCompatMOSA}{26.21}
\newcommand{\MeanStatementCoverageSnakeCompatMOSA}{94.89}
\newcommand{\MeanStatementCoverageSnowballFightCompatMOSA}{95.73}
\newcommand{\MeanStatementCoverageSpaceOdysseyCompatMOSA}{87.87}
\newcommand{\MeanStatementCoverageWhackAMoleCompatMOSA}{82.66}
\newcommand{\MeanStatementCoverageTotalCompatMOSA}{86.71}
\newcommand{\MeanStatementCoverageCatchTheDotsNoveltyMOSA}{96.83}
\newcommand{\MeanStatementCoverageCityDefenderNoveltyMOSA}{74.33}
\newcommand{\MeanStatementCoverageCreateYourWorldNoveltyMOSA}{75.13}
\newcommand{\MeanStatementCoverageDessertRallyNoveltyMOSA}{93.90}
\newcommand{\MeanStatementCoverageDieZauberlehrlingeNoveltyMOSA}{85.66}
\newcommand{\MeanStatementCoverageDodgeballNoveltyMOSA}{96.07}
\newcommand{\MeanStatementCoverageDragonsNoveltyMOSA}{87.59}
\newcommand{\MeanStatementCoverageEndlessRunnerNoveltyMOSA}{82.88}
\newcommand{\MeanStatementCoverageFallingStarsNoveltyMOSA}{99.34}
\newcommand{\MeanStatementCoverageFinalFightNoveltyMOSA}{99.65}
\newcommand{\MeanStatementCoverageFlappyParrotNoveltyMOSA}{92.97}
\newcommand{\MeanStatementCoverageFroggerNoveltyMOSA}{94.10}
\newcommand{\MeanStatementCoverageFruitCatchingNoveltyMOSA}{88.30}
\newcommand{\MeanStatementCoverageHackAttackNoveltyMOSA}{93.98}
\newcommand{\MeanStatementCoverageOceanCleanupNoveltyMOSA}{82.20}
\newcommand{\MeanStatementCoveragePokemonClickerNoveltyMOSA}{26.21}
\newcommand{\MeanStatementCoverageSnakeNoveltyMOSA}{94.94}
\newcommand{\MeanStatementCoverageSnowballFightNoveltyMOSA}{95.90}
\newcommand{\MeanStatementCoverageSpaceOdysseyNoveltyMOSA}{87.47}
\newcommand{\MeanStatementCoverageWhackAMoleNoveltyMOSA}{83.91}
\newcommand{\MeanStatementCoverageTotalNoveltyMOSA}{86.57}
\newcommand{\MeanStatementCoverageCatchTheDotsSpeciesMOSA}{96.34}
\newcommand{\MeanStatementCoverageCityDefenderSpeciesMOSA}{74.33}
\newcommand{\MeanStatementCoverageCreateYourWorldSpeciesMOSA}{75.39}
\newcommand{\MeanStatementCoverageDessertRallySpeciesMOSA}{93.90}
\newcommand{\MeanStatementCoverageDieZauberlehrlingeSpeciesMOSA}{85.04}
\newcommand{\MeanStatementCoverageDodgeballSpeciesMOSA}{95.94}
\newcommand{\MeanStatementCoverageDragonsSpeciesMOSA}{87.65}
\newcommand{\MeanStatementCoverageEndlessRunnerSpeciesMOSA}{82.65}
\newcommand{\MeanStatementCoverageFallingStarsSpeciesMOSA}{99.49}
\newcommand{\MeanStatementCoverageFinalFightSpeciesMOSA}{99.65}
\newcommand{\MeanStatementCoverageFlappyParrotSpeciesMOSA}{93.51}
\newcommand{\MeanStatementCoverageFroggerSpeciesMOSA}{94.36}
\newcommand{\MeanStatementCoverageFruitCatchingSpeciesMOSA}{92.97}
\newcommand{\MeanStatementCoverageHackAttackSpeciesMOSA}{94.98}
\newcommand{\MeanStatementCoverageOceanCleanupSpeciesMOSA}{84.83}
\newcommand{\MeanStatementCoveragePokemonClickerSpeciesMOSA}{26.21}
\newcommand{\MeanStatementCoverageSnakeSpeciesMOSA}{94.89}
\newcommand{\MeanStatementCoverageSnowballFightSpeciesMOSA}{96.32}
\newcommand{\MeanStatementCoverageSpaceOdysseySpeciesMOSA}{88.30}
\newcommand{\MeanStatementCoverageWhackAMoleSpeciesMOSA}{84.08}
\newcommand{\MeanStatementCoverageTotalSpeciesMOSA}{87.04}
\newcommand{\EffectSizeStatementCoverageCatchTheDotsNoveltyCompatMOSA}{0.50}
\newcommand{\PValStatementCoverageCatchTheDotsNoveltyCompatMOSA}{0.95}
\newcommand{\EffectSizeStatementCoverageCatchTheDotsSpeciesCompatMOSA}{0.55}
\newcommand{\PValStatementCoverageCatchTheDotsSpeciesCompatMOSA}{0.47}
\newcommand{\EffectSizeStatementCoverageCatchTheDotsSpeciesNoveltyMOSA}{0.56}
\newcommand{\PValStatementCoverageCatchTheDotsSpeciesNoveltyMOSA}{0.40}
\newcommand{\EffectSizeStatementCoverageCityDefenderNoveltyCompatMOSA}{0.50}
\newcommand{\PValStatementCoverageCityDefenderNoveltyCompatMOSA}{1.00}
\newcommand{\EffectSizeStatementCoverageCityDefenderSpeciesCompatMOSA}{0.50}
\newcommand{\PValStatementCoverageCityDefenderSpeciesCompatMOSA}{1.00}
\newcommand{\EffectSizeStatementCoverageCityDefenderSpeciesNoveltyMOSA}{0.50}
\newcommand{\PValStatementCoverageCityDefenderSpeciesNoveltyMOSA}{1.00}
\newcommand{\EffectSizeStatementCoverageCreateYourWorldNoveltyCompatMOSA}{0.57}
\newcommand{\PValStatementCoverageCreateYourWorldNoveltyCompatMOSA}{0.22}
\newcommand{\EffectSizeStatementCoverageCreateYourWorldSpeciesCompatMOSA}{0.47}
\newcommand{\PValStatementCoverageCreateYourWorldSpeciesCompatMOSA}{0.68}
\newcommand{\EffectSizeStatementCoverageCreateYourWorldSpeciesNoveltyMOSA}{0.39}
\newcommand{\PValStatementCoverageCreateYourWorldSpeciesNoveltyMOSA}{0.08}
\newcommand{\EffectSizeStatementCoverageDessertRallyNoveltyCompatMOSA}{0.50}
\newcommand{\PValStatementCoverageDessertRallyNoveltyCompatMOSA}{1.00}
\newcommand{\EffectSizeStatementCoverageDessertRallySpeciesCompatMOSA}{0.50}
\newcommand{\PValStatementCoverageDessertRallySpeciesCompatMOSA}{1.00}
\newcommand{\EffectSizeStatementCoverageDessertRallySpeciesNoveltyMOSA}{0.50}
\newcommand{\PValStatementCoverageDessertRallySpeciesNoveltyMOSA}{1.00}
\newcommand{\EffectSizeStatementCoverageDieZauberlehrlingeNoveltyCompatMOSA}{0.49}
\newcommand{\PValStatementCoverageDieZauberlehrlingeNoveltyCompatMOSA}{0.85}
\newcommand{\EffectSizeStatementCoverageDieZauberlehrlingeSpeciesCompatMOSA}{0.50}
\newcommand{\PValStatementCoverageDieZauberlehrlingeSpeciesCompatMOSA}{1.00}
\newcommand{\EffectSizeStatementCoverageDieZauberlehrlingeSpeciesNoveltyMOSA}{0.51}
\newcommand{\PValStatementCoverageDieZauberlehrlingeSpeciesNoveltyMOSA}{0.85}
\newcommand{\EffectSizeStatementCoverageDodgeballNoveltyCompatMOSA}{0.52}
\newcommand{\PValStatementCoverageDodgeballNoveltyCompatMOSA}{0.57}
\newcommand{\EffectSizeStatementCoverageDodgeballSpeciesCompatMOSA}{0.57}
\newcommand{\PValStatementCoverageDodgeballSpeciesCompatMOSA}{0.09}
\newcommand{\EffectSizeStatementCoverageDodgeballSpeciesNoveltyMOSA}{0.55}
\newcommand{\PValStatementCoverageDodgeballSpeciesNoveltyMOSA}{0.24}
\newcommand{\EffectSizeStatementCoverageDragonsNoveltyCompatMOSA}{0.56}
\newcommand{\PValStatementCoverageDragonsNoveltyCompatMOSA}{0.46}
\newcommand{\EffectSizeStatementCoverageDragonsSpeciesCompatMOSA}{0.56}
\newcommand{\PValStatementCoverageDragonsSpeciesCompatMOSA}{0.39}
\newcommand{\EffectSizeStatementCoverageDragonsSpeciesNoveltyMOSA}{0.51}
\newcommand{\PValStatementCoverageDragonsSpeciesNoveltyMOSA}{0.90}
\newcommand{\EffectSizeStatementCoverageEndlessRunnerNoveltyCompatMOSA}{0.50}
\newcommand{\PValStatementCoverageEndlessRunnerNoveltyCompatMOSA}{0.94}
\newcommand{\EffectSizeStatementCoverageEndlessRunnerSpeciesCompatMOSA}{0.51}
\newcommand{\PValStatementCoverageEndlessRunnerSpeciesCompatMOSA}{0.82}
\newcommand{\EffectSizeStatementCoverageEndlessRunnerSpeciesNoveltyMOSA}{0.52}
\newcommand{\PValStatementCoverageEndlessRunnerSpeciesNoveltyMOSA}{0.77}
\newcommand{\EffectSizeStatementCoverageFallingStarsNoveltyCompatMOSA}{\textbf{0.63}}
\newcommand{\PValStatementCoverageFallingStarsNoveltyCompatMOSA}{\textbf{0.04}}
\newcommand{\EffectSizeStatementCoverageFallingStarsSpeciesCompatMOSA}{0.57}
\newcommand{\PValStatementCoverageFallingStarsSpeciesCompatMOSA}{0.30}
\newcommand{\EffectSizeStatementCoverageFallingStarsSpeciesNoveltyMOSA}{0.43}
\newcommand{\PValStatementCoverageFallingStarsSpeciesNoveltyMOSA}{0.31}
\newcommand{\EffectSizeStatementCoverageFinalFightNoveltyCompatMOSA}{0.50}
\newcommand{\PValStatementCoverageFinalFightNoveltyCompatMOSA}{1.00}
\newcommand{\EffectSizeStatementCoverageFinalFightSpeciesCompatMOSA}{0.50}
\newcommand{\PValStatementCoverageFinalFightSpeciesCompatMOSA}{1.00}
\newcommand{\EffectSizeStatementCoverageFinalFightSpeciesNoveltyMOSA}{0.50}
\newcommand{\PValStatementCoverageFinalFightSpeciesNoveltyMOSA}{1.00}
\newcommand{\EffectSizeStatementCoverageFlappyParrotNoveltyCompatMOSA}{0.55}
\newcommand{\PValStatementCoverageFlappyParrotNoveltyCompatMOSA}{0.37}
\newcommand{\EffectSizeStatementCoverageFlappyParrotSpeciesCompatMOSA}{0.51}
\newcommand{\PValStatementCoverageFlappyParrotSpeciesCompatMOSA}{0.86}
\newcommand{\EffectSizeStatementCoverageFlappyParrotSpeciesNoveltyMOSA}{0.47}
\newcommand{\PValStatementCoverageFlappyParrotSpeciesNoveltyMOSA}{0.50}
\newcommand{\EffectSizeStatementCoverageFroggerNoveltyCompatMOSA}{0.56}
\newcommand{\PValStatementCoverageFroggerNoveltyCompatMOSA}{0.16}
\newcommand{\EffectSizeStatementCoverageFroggerSpeciesCompatMOSA}{0.52}
\newcommand{\PValStatementCoverageFroggerSpeciesCompatMOSA}{0.79}
\newcommand{\EffectSizeStatementCoverageFroggerSpeciesNoveltyMOSA}{0.46}
\newcommand{\PValStatementCoverageFroggerSpeciesNoveltyMOSA}{0.39}
\newcommand{\EffectSizeStatementCoverageFruitCatchingNoveltyCompatMOSA}{0.47}
\newcommand{\PValStatementCoverageFruitCatchingNoveltyCompatMOSA}{0.67}
\newcommand{\EffectSizeStatementCoverageFruitCatchingSpeciesCompatMOSA}{\textbf{0.30}}
\newcommand{\PValStatementCoverageFruitCatchingSpeciesCompatMOSA}{\textbf{0.01}}
\newcommand{\EffectSizeStatementCoverageFruitCatchingSpeciesNoveltyMOSA}{\textbf{0.35}}
\newcommand{\PValStatementCoverageFruitCatchingSpeciesNoveltyMOSA}{\textbf{0.04}}
\newcommand{\EffectSizeStatementCoverageHackAttackNoveltyCompatMOSA}{0.58}
\newcommand{\PValStatementCoverageHackAttackNoveltyCompatMOSA}{0.16}
\newcommand{\EffectSizeStatementCoverageHackAttackSpeciesCompatMOSA}{0.52}
\newcommand{\PValStatementCoverageHackAttackSpeciesCompatMOSA}{0.80}
\newcommand{\EffectSizeStatementCoverageHackAttackSpeciesNoveltyMOSA}{0.43}
\newcommand{\PValStatementCoverageHackAttackSpeciesNoveltyMOSA}{0.25}
\newcommand{\EffectSizeStatementCoverageOceanCleanupNoveltyCompatMOSA}{0.62}
\newcommand{\PValStatementCoverageOceanCleanupNoveltyCompatMOSA}{0.10}
\newcommand{\EffectSizeStatementCoverageOceanCleanupSpeciesCompatMOSA}{0.48}
\newcommand{\PValStatementCoverageOceanCleanupSpeciesCompatMOSA}{0.81}
\newcommand{\EffectSizeStatementCoverageOceanCleanupSpeciesNoveltyMOSA}{0.37}
\newcommand{\PValStatementCoverageOceanCleanupSpeciesNoveltyMOSA}{0.08}
\newcommand{\EffectSizeStatementCoveragePokemonClickerNoveltyCompatMOSA}{0.50}
\newcommand{\PValStatementCoveragePokemonClickerNoveltyCompatMOSA}{1.00}
\newcommand{\EffectSizeStatementCoveragePokemonClickerSpeciesCompatMOSA}{0.50}
\newcommand{\PValStatementCoveragePokemonClickerSpeciesCompatMOSA}{1.00}
\newcommand{\EffectSizeStatementCoveragePokemonClickerSpeciesNoveltyMOSA}{0.50}
\newcommand{\PValStatementCoveragePokemonClickerSpeciesNoveltyMOSA}{1.00}
\newcommand{\EffectSizeStatementCoverageSnakeNoveltyCompatMOSA}{0.48}
\newcommand{\PValStatementCoverageSnakeNoveltyCompatMOSA}{0.58}
\newcommand{\EffectSizeStatementCoverageSnakeSpeciesCompatMOSA}{0.50}
\newcommand{\PValStatementCoverageSnakeSpeciesCompatMOSA}{1.00}
\newcommand{\EffectSizeStatementCoverageSnakeSpeciesNoveltyMOSA}{0.52}
\newcommand{\PValStatementCoverageSnakeSpeciesNoveltyMOSA}{0.58}
\newcommand{\EffectSizeStatementCoverageSnowballFightNoveltyCompatMOSA}{0.47}
\newcommand{\PValStatementCoverageSnowballFightNoveltyCompatMOSA}{0.60}
\newcommand{\EffectSizeStatementCoverageSnowballFightSpeciesCompatMOSA}{0.38}
\newcommand{\PValStatementCoverageSnowballFightSpeciesCompatMOSA}{0.07}
\newcommand{\EffectSizeStatementCoverageSnowballFightSpeciesNoveltyMOSA}{0.42}
\newcommand{\PValStatementCoverageSnowballFightSpeciesNoveltyMOSA}{0.20}
\newcommand{\EffectSizeStatementCoverageSpaceOdysseyNoveltyCompatMOSA}{0.53}
\newcommand{\PValStatementCoverageSpaceOdysseyNoveltyCompatMOSA}{0.40}
\newcommand{\EffectSizeStatementCoverageSpaceOdysseySpeciesCompatMOSA}{0.48}
\newcommand{\PValStatementCoverageSpaceOdysseySpeciesCompatMOSA}{0.69}
\newcommand{\EffectSizeStatementCoverageSpaceOdysseySpeciesNoveltyMOSA}{0.45}
\newcommand{\PValStatementCoverageSpaceOdysseySpeciesNoveltyMOSA}{0.23}
\newcommand{\EffectSizeStatementCoverageWhackAMoleNoveltyCompatMOSA}{0.38}
\newcommand{\PValStatementCoverageWhackAMoleNoveltyCompatMOSA}{0.11}
\newcommand{\EffectSizeStatementCoverageWhackAMoleSpeciesCompatMOSA}{0.39}
\newcommand{\PValStatementCoverageWhackAMoleSpeciesCompatMOSA}{0.13}
\newcommand{\EffectSizeStatementCoverageWhackAMoleSpeciesNoveltyMOSA}{0.53}
\newcommand{\PValStatementCoverageWhackAMoleSpeciesNoveltyMOSA}{0.71}
\newcommand{\MeanEffectSizeStatementCoverageNoveltyCompatMOSA}{0.52}
\newcommand{\MeanEffectSizeStatementCoverageSpeciesCompatMOSA}{0.49}
\newcommand{\MeanEffectSizeStatementCoverageSpeciesNoveltyMOSA}{0.47}
\newcommand{\MeanBranchCoverageCatchTheDotsCompatMOSA}{87.71}
\newcommand{\MeanBranchCoverageCityDefenderCompatMOSA}{67.66}
\newcommand{\MeanBranchCoverageCreateYourWorldCompatMOSA}{71.79}
\newcommand{\MeanBranchCoverageDessertRallyCompatMOSA}{96.15}
\newcommand{\MeanBranchCoverageDieZauberlehrlingeCompatMOSA}{70.75}
\newcommand{\MeanBranchCoverageDodgeballCompatMOSA}{92.95}
\newcommand{\MeanBranchCoverageDragonsCompatMOSA}{82.57}
\newcommand{\MeanBranchCoverageEndlessRunnerCompatMOSA}{70.33}
\newcommand{\MeanBranchCoverageFallingStarsCompatMOSA}{94.70}
\newcommand{\MeanBranchCoverageFinalFightCompatMOSA}{99.23}
\newcommand{\MeanBranchCoverageFlappyParrotCompatMOSA}{85.45}
\newcommand{\MeanBranchCoverageFroggerCompatMOSA}{90.33}
\newcommand{\MeanBranchCoverageFruitCatchingCompatMOSA}{80.00}
\newcommand{\MeanBranchCoverageHackAttackCompatMOSA}{98.33}
\newcommand{\MeanBranchCoverageOceanCleanupCompatMOSA}{74.89}
\newcommand{\MeanBranchCoveragePokemonClickerCompatMOSA}{17.46}
\newcommand{\MeanBranchCoverageSnakeCompatMOSA}{78.57}
\newcommand{\MeanBranchCoverageSnowballFightCompatMOSA}{85.51}
\newcommand{\MeanBranchCoverageSpaceOdysseyCompatMOSA}{91.93}
\newcommand{\MeanBranchCoverageWhackAMoleCompatMOSA}{82.73}
\newcommand{\MeanBranchCoverageTotalCompatMOSA}{80.95}
\newcommand{\MeanBranchCoverageCatchTheDotsNoveltyMOSA}{87.81}
\newcommand{\MeanBranchCoverageCityDefenderNoveltyMOSA}{67.66}
\newcommand{\MeanBranchCoverageCreateYourWorldNoveltyMOSA}{71.17}
\newcommand{\MeanBranchCoverageDessertRallyNoveltyMOSA}{96.15}
\newcommand{\MeanBranchCoverageDieZauberlehrlingeNoveltyMOSA}{71.29}
\newcommand{\MeanBranchCoverageDodgeballNoveltyMOSA}{92.87}
\newcommand{\MeanBranchCoverageDragonsNoveltyMOSA}{82.04}
\newcommand{\MeanBranchCoverageEndlessRunnerNoveltyMOSA}{70.27}
\newcommand{\MeanBranchCoverageFallingStarsNoveltyMOSA}{94.09}
\newcommand{\MeanBranchCoverageFinalFightNoveltyMOSA}{99.23}
\newcommand{\MeanBranchCoverageFlappyParrotNoveltyMOSA}{84.24}
\newcommand{\MeanBranchCoverageFroggerNoveltyMOSA}{90.00}
\newcommand{\MeanBranchCoverageFruitCatchingNoveltyMOSA}{80.48}
\newcommand{\MeanBranchCoverageHackAttackNoveltyMOSA}{97.89}
\newcommand{\MeanBranchCoverageOceanCleanupNoveltyMOSA}{72.15}
\newcommand{\MeanBranchCoveragePokemonClickerNoveltyMOSA}{17.46}
\newcommand{\MeanBranchCoverageSnakeNoveltyMOSA}{78.81}
\newcommand{\MeanBranchCoverageSnowballFightNoveltyMOSA}{86.09}
\newcommand{\MeanBranchCoverageSpaceOdysseyNoveltyMOSA}{91.58}
\newcommand{\MeanBranchCoverageWhackAMoleNoveltyMOSA}{83.09}
\newcommand{\MeanBranchCoverageTotalNoveltyMOSA}{80.72}
\newcommand{\MeanBranchCoverageCatchTheDotsSpeciesMOSA}{87.33}
\newcommand{\MeanBranchCoverageCityDefenderSpeciesMOSA}{67.66}
\newcommand{\MeanBranchCoverageCreateYourWorldSpeciesMOSA}{71.65}
\newcommand{\MeanBranchCoverageDessertRallySpeciesMOSA}{96.15}
\newcommand{\MeanBranchCoverageDieZauberlehrlingeSpeciesMOSA}{70.75}
\newcommand{\MeanBranchCoverageDodgeballSpeciesMOSA}{92.64}
\newcommand{\MeanBranchCoverageDragonsSpeciesMOSA}{82.18}
\newcommand{\MeanBranchCoverageEndlessRunnerSpeciesMOSA}{70.13}
\newcommand{\MeanBranchCoverageFallingStarsSpeciesMOSA}{94.39}
\newcommand{\MeanBranchCoverageFinalFightSpeciesMOSA}{99.23}
\newcommand{\MeanBranchCoverageFlappyParrotSpeciesMOSA}{85.45}
\newcommand{\MeanBranchCoverageFroggerSpeciesMOSA}{90.50}
\newcommand{\MeanBranchCoverageFruitCatchingSpeciesMOSA}{85.24}
\newcommand{\MeanBranchCoverageHackAttackSpeciesMOSA}{98.25}
\newcommand{\MeanBranchCoverageOceanCleanupSpeciesMOSA}{75.05}
\newcommand{\MeanBranchCoveragePokemonClickerSpeciesMOSA}{17.47}
\newcommand{\MeanBranchCoverageSnakeSpeciesMOSA}{78.57}
\newcommand{\MeanBranchCoverageSnowballFightSpeciesMOSA}{87.54}
\newcommand{\MeanBranchCoverageSpaceOdysseySpeciesMOSA}{92.22}
\newcommand{\MeanBranchCoverageWhackAMoleSpeciesMOSA}{84.21}
\newcommand{\MeanBranchCoverageTotalSpeciesMOSA}{81.33}
\newcommand{\EffectSizeBranchCoverageCatchTheDotsNoveltyCompatMOSA}{0.49}
\newcommand{\PValBranchCoverageCatchTheDotsNoveltyCompatMOSA}{0.93}
\newcommand{\EffectSizeBranchCoverageCatchTheDotsSpeciesCompatMOSA}{0.55}
\newcommand{\PValBranchCoverageCatchTheDotsSpeciesCompatMOSA}{0.52}
\newcommand{\EffectSizeBranchCoverageCatchTheDotsSpeciesNoveltyMOSA}{0.56}
\newcommand{\PValBranchCoverageCatchTheDotsSpeciesNoveltyMOSA}{0.39}
\newcommand{\EffectSizeBranchCoverageCityDefenderNoveltyCompatMOSA}{0.50}
\newcommand{\PValBranchCoverageCityDefenderNoveltyCompatMOSA}{1.00}
\newcommand{\EffectSizeBranchCoverageCityDefenderSpeciesCompatMOSA}{0.50}
\newcommand{\PValBranchCoverageCityDefenderSpeciesCompatMOSA}{1.00}
\newcommand{\EffectSizeBranchCoverageCityDefenderSpeciesNoveltyMOSA}{0.50}
\newcommand{\PValBranchCoverageCityDefenderSpeciesNoveltyMOSA}{1.00}
\newcommand{\EffectSizeBranchCoverageCreateYourWorldNoveltyCompatMOSA}{0.57}
\newcommand{\PValBranchCoverageCreateYourWorldNoveltyCompatMOSA}{0.22}
\newcommand{\EffectSizeBranchCoverageCreateYourWorldSpeciesCompatMOSA}{0.47}
\newcommand{\PValBranchCoverageCreateYourWorldSpeciesCompatMOSA}{0.68}
\newcommand{\EffectSizeBranchCoverageCreateYourWorldSpeciesNoveltyMOSA}{0.39}
\newcommand{\PValBranchCoverageCreateYourWorldSpeciesNoveltyMOSA}{0.08}
\newcommand{\EffectSizeBranchCoverageDessertRallyNoveltyCompatMOSA}{0.50}
\newcommand{\PValBranchCoverageDessertRallyNoveltyCompatMOSA}{1.00}
\newcommand{\EffectSizeBranchCoverageDessertRallySpeciesCompatMOSA}{0.50}
\newcommand{\PValBranchCoverageDessertRallySpeciesCompatMOSA}{1.00}
\newcommand{\EffectSizeBranchCoverageDessertRallySpeciesNoveltyMOSA}{0.50}
\newcommand{\PValBranchCoverageDessertRallySpeciesNoveltyMOSA}{1.00}
\newcommand{\EffectSizeBranchCoverageDieZauberlehrlingeNoveltyCompatMOSA}{0.48}
\newcommand{\PValBranchCoverageDieZauberlehrlingeNoveltyCompatMOSA}{0.76}
\newcommand{\EffectSizeBranchCoverageDieZauberlehrlingeSpeciesCompatMOSA}{0.50}
\newcommand{\PValBranchCoverageDieZauberlehrlingeSpeciesCompatMOSA}{1.00}
\newcommand{\EffectSizeBranchCoverageDieZauberlehrlingeSpeciesNoveltyMOSA}{0.52}
\newcommand{\PValBranchCoverageDieZauberlehrlingeSpeciesNoveltyMOSA}{0.76}
\newcommand{\EffectSizeBranchCoverageDodgeballNoveltyCompatMOSA}{0.52}
\newcommand{\PValBranchCoverageDodgeballNoveltyCompatMOSA}{0.57}
\newcommand{\EffectSizeBranchCoverageDodgeballSpeciesCompatMOSA}{0.57}
\newcommand{\PValBranchCoverageDodgeballSpeciesCompatMOSA}{0.09}
\newcommand{\EffectSizeBranchCoverageDodgeballSpeciesNoveltyMOSA}{0.55}
\newcommand{\PValBranchCoverageDodgeballSpeciesNoveltyMOSA}{0.24}
\newcommand{\EffectSizeBranchCoverageDragonsNoveltyCompatMOSA}{0.54}
\newcommand{\PValBranchCoverageDragonsNoveltyCompatMOSA}{0.59}
\newcommand{\EffectSizeBranchCoverageDragonsSpeciesCompatMOSA}{0.55}
\newcommand{\PValBranchCoverageDragonsSpeciesCompatMOSA}{0.51}
\newcommand{\EffectSizeBranchCoverageDragonsSpeciesNoveltyMOSA}{0.51}
\newcommand{\PValBranchCoverageDragonsSpeciesNoveltyMOSA}{0.93}
\newcommand{\EffectSizeBranchCoverageEndlessRunnerNoveltyCompatMOSA}{0.52}
\newcommand{\PValBranchCoverageEndlessRunnerNoveltyCompatMOSA}{0.78}
\newcommand{\EffectSizeBranchCoverageEndlessRunnerSpeciesCompatMOSA}{0.51}
\newcommand{\PValBranchCoverageEndlessRunnerSpeciesCompatMOSA}{0.83}
\newcommand{\EffectSizeBranchCoverageEndlessRunnerSpeciesNoveltyMOSA}{0.49}
\newcommand{\PValBranchCoverageEndlessRunnerSpeciesNoveltyMOSA}{0.93}
\newcommand{\EffectSizeBranchCoverageFallingStarsNoveltyCompatMOSA}{\textbf{0.63}}
\newcommand{\PValBranchCoverageFallingStarsNoveltyCompatMOSA}{\textbf{0.04}}
\newcommand{\EffectSizeBranchCoverageFallingStarsSpeciesCompatMOSA}{0.57}
\newcommand{\PValBranchCoverageFallingStarsSpeciesCompatMOSA}{0.30}
\newcommand{\EffectSizeBranchCoverageFallingStarsSpeciesNoveltyMOSA}{0.43}
\newcommand{\PValBranchCoverageFallingStarsSpeciesNoveltyMOSA}{0.31}
\newcommand{\EffectSizeBranchCoverageFinalFightNoveltyCompatMOSA}{0.50}
\newcommand{\PValBranchCoverageFinalFightNoveltyCompatMOSA}{1.00}
\newcommand{\EffectSizeBranchCoverageFinalFightSpeciesCompatMOSA}{0.50}
\newcommand{\PValBranchCoverageFinalFightSpeciesCompatMOSA}{1.00}
\newcommand{\EffectSizeBranchCoverageFinalFightSpeciesNoveltyMOSA}{0.50}
\newcommand{\PValBranchCoverageFinalFightSpeciesNoveltyMOSA}{1.00}
\newcommand{\EffectSizeBranchCoverageFlappyParrotNoveltyCompatMOSA}{0.55}
\newcommand{\PValBranchCoverageFlappyParrotNoveltyCompatMOSA}{0.37}
\newcommand{\EffectSizeBranchCoverageFlappyParrotSpeciesCompatMOSA}{0.51}
\newcommand{\PValBranchCoverageFlappyParrotSpeciesCompatMOSA}{0.86}
\newcommand{\EffectSizeBranchCoverageFlappyParrotSpeciesNoveltyMOSA}{0.47}
\newcommand{\PValBranchCoverageFlappyParrotSpeciesNoveltyMOSA}{0.50}
\newcommand{\EffectSizeBranchCoverageFroggerNoveltyCompatMOSA}{\textbf{0.57}}
\newcommand{\PValBranchCoverageFroggerNoveltyCompatMOSA}{\textbf{0.04}}
\newcommand{\EffectSizeBranchCoverageFroggerSpeciesCompatMOSA}{0.48}
\newcommand{\PValBranchCoverageFroggerSpeciesCompatMOSA}{0.69}
\newcommand{\EffectSizeBranchCoverageFroggerSpeciesNoveltyMOSA}{\textbf{0.42}}
\newcommand{\PValBranchCoverageFroggerSpeciesNoveltyMOSA}{\textbf{0.02}}
\newcommand{\EffectSizeBranchCoverageFruitCatchingNoveltyCompatMOSA}{0.48}
\newcommand{\PValBranchCoverageFruitCatchingNoveltyCompatMOSA}{0.81}
\newcommand{\EffectSizeBranchCoverageFruitCatchingSpeciesCompatMOSA}{\textbf{0.34}}
\newcommand{\PValBranchCoverageFruitCatchingSpeciesCompatMOSA}{\textbf{0.03}}
\newcommand{\EffectSizeBranchCoverageFruitCatchingSpeciesNoveltyMOSA}{0.36}
\newcommand{\PValBranchCoverageFruitCatchingSpeciesNoveltyMOSA}{0.06}
\newcommand{\EffectSizeBranchCoverageHackAttackNoveltyCompatMOSA}{0.58}
\newcommand{\PValBranchCoverageHackAttackNoveltyCompatMOSA}{0.16}
\newcommand{\EffectSizeBranchCoverageHackAttackSpeciesCompatMOSA}{0.52}
\newcommand{\PValBranchCoverageHackAttackSpeciesCompatMOSA}{0.80}
\newcommand{\EffectSizeBranchCoverageHackAttackSpeciesNoveltyMOSA}{0.43}
\newcommand{\PValBranchCoverageHackAttackSpeciesNoveltyMOSA}{0.25}
\newcommand{\EffectSizeBranchCoverageOceanCleanupNoveltyCompatMOSA}{0.61}
\newcommand{\PValBranchCoverageOceanCleanupNoveltyCompatMOSA}{0.14}
\newcommand{\EffectSizeBranchCoverageOceanCleanupSpeciesCompatMOSA}{0.48}
\newcommand{\PValBranchCoverageOceanCleanupSpeciesCompatMOSA}{0.82}
\newcommand{\EffectSizeBranchCoverageOceanCleanupSpeciesNoveltyMOSA}{0.38}
\newcommand{\PValBranchCoverageOceanCleanupSpeciesNoveltyMOSA}{0.10}
\newcommand{\EffectSizeBranchCoveragePokemonClickerNoveltyCompatMOSA}{0.50}
\newcommand{\PValBranchCoveragePokemonClickerNoveltyCompatMOSA}{1.00}
\newcommand{\EffectSizeBranchCoveragePokemonClickerSpeciesCompatMOSA}{0.48}
\newcommand{\PValBranchCoveragePokemonClickerSpeciesCompatMOSA}{0.33}
\newcommand{\EffectSizeBranchCoveragePokemonClickerSpeciesNoveltyMOSA}{0.48}
\newcommand{\PValBranchCoveragePokemonClickerSpeciesNoveltyMOSA}{0.33}
\newcommand{\EffectSizeBranchCoverageSnakeNoveltyCompatMOSA}{0.48}
\newcommand{\PValBranchCoverageSnakeNoveltyCompatMOSA}{0.33}
\newcommand{\EffectSizeBranchCoverageSnakeSpeciesCompatMOSA}{0.50}
\newcommand{\PValBranchCoverageSnakeSpeciesCompatMOSA}{1.00}
\newcommand{\EffectSizeBranchCoverageSnakeSpeciesNoveltyMOSA}{0.52}
\newcommand{\PValBranchCoverageSnakeSpeciesNoveltyMOSA}{0.33}
\newcommand{\EffectSizeBranchCoverageSnowballFightNoveltyCompatMOSA}{0.47}
\newcommand{\PValBranchCoverageSnowballFightNoveltyCompatMOSA}{0.60}
\newcommand{\EffectSizeBranchCoverageSnowballFightSpeciesCompatMOSA}{0.38}
\newcommand{\PValBranchCoverageSnowballFightSpeciesCompatMOSA}{0.07}
\newcommand{\EffectSizeBranchCoverageSnowballFightSpeciesNoveltyMOSA}{0.42}
\newcommand{\PValBranchCoverageSnowballFightSpeciesNoveltyMOSA}{0.20}
\newcommand{\EffectSizeBranchCoverageSpaceOdysseyNoveltyCompatMOSA}{0.53}
\newcommand{\PValBranchCoverageSpaceOdysseyNoveltyCompatMOSA}{0.40}
\newcommand{\EffectSizeBranchCoverageSpaceOdysseySpeciesCompatMOSA}{0.48}
\newcommand{\PValBranchCoverageSpaceOdysseySpeciesCompatMOSA}{0.69}
\newcommand{\EffectSizeBranchCoverageSpaceOdysseySpeciesNoveltyMOSA}{0.45}
\newcommand{\PValBranchCoverageSpaceOdysseySpeciesNoveltyMOSA}{0.23}
\newcommand{\EffectSizeBranchCoverageWhackAMoleNoveltyCompatMOSA}{0.48}
\newcommand{\PValBranchCoverageWhackAMoleNoveltyCompatMOSA}{0.82}
\newcommand{\EffectSizeBranchCoverageWhackAMoleSpeciesCompatMOSA}{0.39}
\newcommand{\PValBranchCoverageWhackAMoleSpeciesCompatMOSA}{0.14}
\newcommand{\EffectSizeBranchCoverageWhackAMoleSpeciesNoveltyMOSA}{0.41}
\newcommand{\PValBranchCoverageWhackAMoleSpeciesNoveltyMOSA}{0.21}
\newcommand{\MeanEffectSizeBranchCoverageNoveltyCompatMOSA}{0.53}
\newcommand{\MeanEffectSizeBranchCoverageSpeciesCompatMOSA}{0.49}
\newcommand{\MeanEffectSizeBranchCoverageSpeciesNoveltyMOSA}{0.46}
\newcommand{\MeanFitnessEvaluationsCatchTheDotsCompatMOSA}{586.67}
\newcommand{\MeanFitnessEvaluationsCityDefenderCompatMOSA}{112.57}
\newcommand{\MeanFitnessEvaluationsCreateYourWorldCompatMOSA}{649.37}
\newcommand{\MeanFitnessEvaluationsDessertRallyCompatMOSA}{825.67}
\newcommand{\MeanFitnessEvaluationsDieZauberlehrlingeCompatMOSA}{82.80}
\newcommand{\MeanFitnessEvaluationsDodgeballCompatMOSA}{573.50}
\newcommand{\MeanFitnessEvaluationsDragonsCompatMOSA}{83.80}
\newcommand{\MeanFitnessEvaluationsEndlessRunnerCompatMOSA}{84.23}
\newcommand{\MeanFitnessEvaluationsFallingStarsCompatMOSA}{620.07}
\newcommand{\MeanFitnessEvaluationsFinalFightCompatMOSA}{581.40}
\newcommand{\MeanFitnessEvaluationsFlappyParrotCompatMOSA}{234.10}
\newcommand{\MeanFitnessEvaluationsFroggerCompatMOSA}{520.87}
\newcommand{\MeanFitnessEvaluationsFruitCatchingCompatMOSA}{1749.77}
\newcommand{\MeanFitnessEvaluationsHackAttackCompatMOSA}{468.97}
\newcommand{\MeanFitnessEvaluationsOceanCleanupCompatMOSA}{2380.70}
\newcommand{\MeanFitnessEvaluationsPokemonClickerCompatMOSA}{555.97}
\newcommand{\MeanFitnessEvaluationsSnakeCompatMOSA}{7206.13}
\newcommand{\MeanFitnessEvaluationsSnowballFightCompatMOSA}{649.27}
\newcommand{\MeanFitnessEvaluationsSpaceOdysseyCompatMOSA}{231.33}
\newcommand{\MeanFitnessEvaluationsWhackAMoleCompatMOSA}{1058.23}
\newcommand{\MeanFitnessEvaluationsTotalCompatMOSA}{962.77}
\newcommand{\MeanFitnessEvaluationsCatchTheDotsNoveltyMOSA}{607.63}
\newcommand{\MeanFitnessEvaluationsCityDefenderNoveltyMOSA}{110.43}
\newcommand{\MeanFitnessEvaluationsCreateYourWorldNoveltyMOSA}{649.43}
\newcommand{\MeanFitnessEvaluationsDessertRallyNoveltyMOSA}{872.37}
\newcommand{\MeanFitnessEvaluationsDieZauberlehrlingeNoveltyMOSA}{96.63}
\newcommand{\MeanFitnessEvaluationsDodgeballNoveltyMOSA}{612.27}
\newcommand{\MeanFitnessEvaluationsDragonsNoveltyMOSA}{81.40}
\newcommand{\MeanFitnessEvaluationsEndlessRunnerNoveltyMOSA}{79.93}
\newcommand{\MeanFitnessEvaluationsFallingStarsNoveltyMOSA}{657.47}
\newcommand{\MeanFitnessEvaluationsFinalFightNoveltyMOSA}{710.17}
\newcommand{\MeanFitnessEvaluationsFlappyParrotNoveltyMOSA}{276.90}
\newcommand{\MeanFitnessEvaluationsFroggerNoveltyMOSA}{581.90}
\newcommand{\MeanFitnessEvaluationsFruitCatchingNoveltyMOSA}{2138.10}
\newcommand{\MeanFitnessEvaluationsHackAttackNoveltyMOSA}{590.03}
\newcommand{\MeanFitnessEvaluationsOceanCleanupNoveltyMOSA}{6829.60}
\newcommand{\MeanFitnessEvaluationsPokemonClickerNoveltyMOSA}{655.77}
\newcommand{\MeanFitnessEvaluationsSnakeNoveltyMOSA}{13827.40}
\newcommand{\MeanFitnessEvaluationsSnowballFightNoveltyMOSA}{668.83}
\newcommand{\MeanFitnessEvaluationsSpaceOdysseyNoveltyMOSA}{296.20}
\newcommand{\MeanFitnessEvaluationsWhackAMoleNoveltyMOSA}{1135}
\newcommand{\MeanFitnessEvaluationsTotalNoveltyMOSA}{1573.87}
\newcommand{\MeanFitnessEvaluationsCatchTheDotsSpeciesMOSA}{585.30}
\newcommand{\MeanFitnessEvaluationsCityDefenderSpeciesMOSA}{112.07}
\newcommand{\MeanFitnessEvaluationsCreateYourWorldSpeciesMOSA}{650.33}
\newcommand{\MeanFitnessEvaluationsDessertRallySpeciesMOSA}{877.47}
\newcommand{\MeanFitnessEvaluationsDieZauberlehrlingeSpeciesMOSA}{83.53}
\newcommand{\MeanFitnessEvaluationsDodgeballSpeciesMOSA}{536.60}
\newcommand{\MeanFitnessEvaluationsDragonsSpeciesMOSA}{81.80}
\newcommand{\MeanFitnessEvaluationsEndlessRunnerSpeciesMOSA}{84.93}
\newcommand{\MeanFitnessEvaluationsFallingStarsSpeciesMOSA}{633.80}
\newcommand{\MeanFitnessEvaluationsFinalFightSpeciesMOSA}{597.80}
\newcommand{\MeanFitnessEvaluationsFlappyParrotSpeciesMOSA}{239}
\newcommand{\MeanFitnessEvaluationsFroggerSpeciesMOSA}{556.57}
\newcommand{\MeanFitnessEvaluationsFruitCatchingSpeciesMOSA}{1649.53}
\newcommand{\MeanFitnessEvaluationsHackAttackSpeciesMOSA}{479.77}
\newcommand{\MeanFitnessEvaluationsOceanCleanupSpeciesMOSA}{2302.83}
\newcommand{\MeanFitnessEvaluationsPokemonClickerSpeciesMOSA}{624.27}
\newcommand{\MeanFitnessEvaluationsSnakeSpeciesMOSA}{7258.80}
\newcommand{\MeanFitnessEvaluationsSnowballFightSpeciesMOSA}{662.17}
\newcommand{\MeanFitnessEvaluationsSpaceOdysseySpeciesMOSA}{240.77}
\newcommand{\MeanFitnessEvaluationsWhackAMoleSpeciesMOSA}{1153.63}
\newcommand{\MeanFitnessEvaluationsTotalSpeciesMOSA}{970.55}
\newcommand{\EffectSizeFitnessEvaluationsCatchTheDotsNoveltyCompatMOSA}{0.46}
\newcommand{\PValFitnessEvaluationsCatchTheDotsNoveltyCompatMOSA}{0.57}
\newcommand{\EffectSizeFitnessEvaluationsCatchTheDotsSpeciesCompatMOSA}{0.53}
\newcommand{\PValFitnessEvaluationsCatchTheDotsSpeciesCompatMOSA}{0.73}
\newcommand{\EffectSizeFitnessEvaluationsCatchTheDotsSpeciesNoveltyMOSA}{0.58}
\newcommand{\PValFitnessEvaluationsCatchTheDotsSpeciesNoveltyMOSA}{0.29}
\newcommand{\EffectSizeFitnessEvaluationsCityDefenderNoveltyCompatMOSA}{0.52}
\newcommand{\PValFitnessEvaluationsCityDefenderNoveltyCompatMOSA}{0.78}
\newcommand{\EffectSizeFitnessEvaluationsCityDefenderSpeciesCompatMOSA}{0.47}
\newcommand{\PValFitnessEvaluationsCityDefenderSpeciesCompatMOSA}{0.69}
\newcommand{\EffectSizeFitnessEvaluationsCityDefenderSpeciesNoveltyMOSA}{0.44}
\newcommand{\PValFitnessEvaluationsCityDefenderSpeciesNoveltyMOSA}{0.46}
\newcommand{\EffectSizeFitnessEvaluationsCreateYourWorldNoveltyCompatMOSA}{0.55}
\newcommand{\PValFitnessEvaluationsCreateYourWorldNoveltyCompatMOSA}{0.47}
\newcommand{\EffectSizeFitnessEvaluationsCreateYourWorldSpeciesCompatMOSA}{0.43}
\newcommand{\PValFitnessEvaluationsCreateYourWorldSpeciesCompatMOSA}{0.37}
\newcommand{\EffectSizeFitnessEvaluationsCreateYourWorldSpeciesNoveltyMOSA}{0.42}
\newcommand{\PValFitnessEvaluationsCreateYourWorldSpeciesNoveltyMOSA}{0.30}
\newcommand{\EffectSizeFitnessEvaluationsDessertRallyNoveltyCompatMOSA}{\textbf{0.31}}
\newcommand{\PValFitnessEvaluationsDessertRallyNoveltyCompatMOSA}{\textbf{0.01}}
\newcommand{\EffectSizeFitnessEvaluationsDessertRallySpeciesCompatMOSA}{\textbf{0.28}}
\newcommand{\PValFitnessEvaluationsDessertRallySpeciesCompatMOSA}{\textbf{< 0.01}}
\newcommand{\EffectSizeFitnessEvaluationsDessertRallySpeciesNoveltyMOSA}{0.50}
\newcommand{\PValFitnessEvaluationsDessertRallySpeciesNoveltyMOSA}{0.98}
\newcommand{\EffectSizeFitnessEvaluationsDieZauberlehrlingeNoveltyCompatMOSA}{0.48}
\newcommand{\PValFitnessEvaluationsDieZauberlehrlingeNoveltyCompatMOSA}{0.64}
\newcommand{\EffectSizeFitnessEvaluationsDieZauberlehrlingeSpeciesCompatMOSA}{0.50}
\newcommand{\PValFitnessEvaluationsDieZauberlehrlingeSpeciesCompatMOSA}{1.00}
\newcommand{\EffectSizeFitnessEvaluationsDieZauberlehrlingeSpeciesNoveltyMOSA}{0.52}
\newcommand{\PValFitnessEvaluationsDieZauberlehrlingeSpeciesNoveltyMOSA}{0.66}
\newcommand{\EffectSizeFitnessEvaluationsDodgeballNoveltyCompatMOSA}{0.48}
\newcommand{\PValFitnessEvaluationsDodgeballNoveltyCompatMOSA}{0.80}
\newcommand{\EffectSizeFitnessEvaluationsDodgeballSpeciesCompatMOSA}{0.59}
\newcommand{\PValFitnessEvaluationsDodgeballSpeciesCompatMOSA}{0.23}
\newcommand{\EffectSizeFitnessEvaluationsDodgeballSpeciesNoveltyMOSA}{0.64}
\newcommand{\PValFitnessEvaluationsDodgeballSpeciesNoveltyMOSA}{0.07}
\newcommand{\EffectSizeFitnessEvaluationsDragonsNoveltyCompatMOSA}{0.59}
\newcommand{\PValFitnessEvaluationsDragonsNoveltyCompatMOSA}{0.22}
\newcommand{\EffectSizeFitnessEvaluationsDragonsSpeciesCompatMOSA}{0.54}
\newcommand{\PValFitnessEvaluationsDragonsSpeciesCompatMOSA}{0.59}
\newcommand{\EffectSizeFitnessEvaluationsDragonsSpeciesNoveltyMOSA}{0.45}
\newcommand{\PValFitnessEvaluationsDragonsSpeciesNoveltyMOSA}{0.48}
\newcommand{\EffectSizeFitnessEvaluationsEndlessRunnerNoveltyCompatMOSA}{0.50}
\newcommand{\PValFitnessEvaluationsEndlessRunnerNoveltyCompatMOSA}{0.95}
\newcommand{\EffectSizeFitnessEvaluationsEndlessRunnerSpeciesCompatMOSA}{0.50}
\newcommand{\PValFitnessEvaluationsEndlessRunnerSpeciesCompatMOSA}{1.00}
\newcommand{\EffectSizeFitnessEvaluationsEndlessRunnerSpeciesNoveltyMOSA}{0.50}
\newcommand{\PValFitnessEvaluationsEndlessRunnerSpeciesNoveltyMOSA}{0.95}
\newcommand{\EffectSizeFitnessEvaluationsFallingStarsNoveltyCompatMOSA}{0.36}
\newcommand{\PValFitnessEvaluationsFallingStarsNoveltyCompatMOSA}{0.06}
\newcommand{\EffectSizeFitnessEvaluationsFallingStarsSpeciesCompatMOSA}{0.40}
\newcommand{\PValFitnessEvaluationsFallingStarsSpeciesCompatMOSA}{0.20}
\newcommand{\EffectSizeFitnessEvaluationsFallingStarsSpeciesNoveltyMOSA}{0.52}
\newcommand{\PValFitnessEvaluationsFallingStarsSpeciesNoveltyMOSA}{0.79}
\newcommand{\EffectSizeFitnessEvaluationsFinalFightNoveltyCompatMOSA}{\textbf{0.26}}
\newcommand{\PValFitnessEvaluationsFinalFightNoveltyCompatMOSA}{\textbf{< 0.01}}
\newcommand{\EffectSizeFitnessEvaluationsFinalFightSpeciesCompatMOSA}{0.45}
\newcommand{\PValFitnessEvaluationsFinalFightSpeciesCompatMOSA}{0.52}
\newcommand{\EffectSizeFitnessEvaluationsFinalFightSpeciesNoveltyMOSA}{\textbf{0.71}}
\newcommand{\PValFitnessEvaluationsFinalFightSpeciesNoveltyMOSA}{\textbf{0.01}}
\newcommand{\EffectSizeFitnessEvaluationsFlappyParrotNoveltyCompatMOSA}{0.37}
\newcommand{\PValFitnessEvaluationsFlappyParrotNoveltyCompatMOSA}{0.08}
\newcommand{\EffectSizeFitnessEvaluationsFlappyParrotSpeciesCompatMOSA}{0.49}
\newcommand{\PValFitnessEvaluationsFlappyParrotSpeciesCompatMOSA}{0.90}
\newcommand{\EffectSizeFitnessEvaluationsFlappyParrotSpeciesNoveltyMOSA}{0.62}
\newcommand{\PValFitnessEvaluationsFlappyParrotSpeciesNoveltyMOSA}{0.12}
\newcommand{\EffectSizeFitnessEvaluationsFroggerNoveltyCompatMOSA}{\textbf{0.32}}
\newcommand{\PValFitnessEvaluationsFroggerNoveltyCompatMOSA}{\textbf{0.01}}
\newcommand{\EffectSizeFitnessEvaluationsFroggerSpeciesCompatMOSA}{0.40}
\newcommand{\PValFitnessEvaluationsFroggerSpeciesCompatMOSA}{0.17}
\newcommand{\EffectSizeFitnessEvaluationsFroggerSpeciesNoveltyMOSA}{0.60}
\newcommand{\PValFitnessEvaluationsFroggerSpeciesNoveltyMOSA}{0.21}
\newcommand{\EffectSizeFitnessEvaluationsFruitCatchingNoveltyCompatMOSA}{\textbf{0.32}}
\newcommand{\PValFitnessEvaluationsFruitCatchingNoveltyCompatMOSA}{\textbf{0.02}}
\newcommand{\EffectSizeFitnessEvaluationsFruitCatchingSpeciesCompatMOSA}{0.56}
\newcommand{\PValFitnessEvaluationsFruitCatchingSpeciesCompatMOSA}{0.44}
\newcommand{\EffectSizeFitnessEvaluationsFruitCatchingSpeciesNoveltyMOSA}{\textbf{0.72}}
\newcommand{\PValFitnessEvaluationsFruitCatchingSpeciesNoveltyMOSA}{\textbf{< 0.01}}
\newcommand{\EffectSizeFitnessEvaluationsHackAttackNoveltyCompatMOSA}{\textbf{0.27}}
\newcommand{\PValFitnessEvaluationsHackAttackNoveltyCompatMOSA}{\textbf{< 0.01}}
\newcommand{\EffectSizeFitnessEvaluationsHackAttackSpeciesCompatMOSA}{0.47}
\newcommand{\PValFitnessEvaluationsHackAttackSpeciesCompatMOSA}{0.70}
\newcommand{\EffectSizeFitnessEvaluationsHackAttackSpeciesNoveltyMOSA}{\textbf{0.75}}
\newcommand{\PValFitnessEvaluationsHackAttackSpeciesNoveltyMOSA}{\textbf{< 0.01}}
\newcommand{\EffectSizeFitnessEvaluationsOceanCleanupNoveltyCompatMOSA}{0.46}
\newcommand{\PValFitnessEvaluationsOceanCleanupNoveltyCompatMOSA}{0.59}
\newcommand{\EffectSizeFitnessEvaluationsOceanCleanupSpeciesCompatMOSA}{0.49}
\newcommand{\PValFitnessEvaluationsOceanCleanupSpeciesCompatMOSA}{0.94}
\newcommand{\EffectSizeFitnessEvaluationsOceanCleanupSpeciesNoveltyMOSA}{0.55}
\newcommand{\PValFitnessEvaluationsOceanCleanupSpeciesNoveltyMOSA}{0.54}
\newcommand{\EffectSizeFitnessEvaluationsPokemonClickerNoveltyCompatMOSA}{\textbf{0.10}}
\newcommand{\PValFitnessEvaluationsPokemonClickerNoveltyCompatMOSA}{\textbf{< 0.01}}
\newcommand{\EffectSizeFitnessEvaluationsPokemonClickerSpeciesCompatMOSA}{\textbf{0.17}}
\newcommand{\PValFitnessEvaluationsPokemonClickerSpeciesCompatMOSA}{\textbf{< 0.01}}
\newcommand{\EffectSizeFitnessEvaluationsPokemonClickerSpeciesNoveltyMOSA}{0.61}
\newcommand{\PValFitnessEvaluationsPokemonClickerSpeciesNoveltyMOSA}{0.14}
\newcommand{\EffectSizeFitnessEvaluationsSnakeNoveltyCompatMOSA}{\textbf{0.09}}
\newcommand{\PValFitnessEvaluationsSnakeNoveltyCompatMOSA}{\textbf{< 0.01}}
\newcommand{\EffectSizeFitnessEvaluationsSnakeSpeciesCompatMOSA}{0.50}
\newcommand{\PValFitnessEvaluationsSnakeSpeciesCompatMOSA}{0.97}
\newcommand{\EffectSizeFitnessEvaluationsSnakeSpeciesNoveltyMOSA}{\textbf{0.92}}
\newcommand{\PValFitnessEvaluationsSnakeSpeciesNoveltyMOSA}{\textbf{< 0.01}}
\newcommand{\EffectSizeFitnessEvaluationsSnowballFightNoveltyCompatMOSA}{0.35}
\newcommand{\PValFitnessEvaluationsSnowballFightNoveltyCompatMOSA}{0.05}
\newcommand{\EffectSizeFitnessEvaluationsSnowballFightSpeciesCompatMOSA}{0.38}
\newcommand{\PValFitnessEvaluationsSnowballFightSpeciesCompatMOSA}{0.10}
\newcommand{\EffectSizeFitnessEvaluationsSnowballFightSpeciesNoveltyMOSA}{0.53}
\newcommand{\PValFitnessEvaluationsSnowballFightSpeciesNoveltyMOSA}{0.74}
\newcommand{\EffectSizeFitnessEvaluationsSpaceOdysseyNoveltyCompatMOSA}{0.36}
\newcommand{\PValFitnessEvaluationsSpaceOdysseyNoveltyCompatMOSA}{0.06}
\newcommand{\EffectSizeFitnessEvaluationsSpaceOdysseySpeciesCompatMOSA}{0.45}
\newcommand{\PValFitnessEvaluationsSpaceOdysseySpeciesCompatMOSA}{0.53}
\newcommand{\EffectSizeFitnessEvaluationsSpaceOdysseySpeciesNoveltyMOSA}{0.62}
\newcommand{\PValFitnessEvaluationsSpaceOdysseySpeciesNoveltyMOSA}{0.12}
\newcommand{\EffectSizeFitnessEvaluationsWhackAMoleNoveltyCompatMOSA}{\textbf{0.33}}
\newcommand{\PValFitnessEvaluationsWhackAMoleNoveltyCompatMOSA}{\textbf{0.02}}
\newcommand{\EffectSizeFitnessEvaluationsWhackAMoleSpeciesCompatMOSA}{0.39}
\newcommand{\PValFitnessEvaluationsWhackAMoleSpeciesCompatMOSA}{0.14}
\newcommand{\EffectSizeFitnessEvaluationsWhackAMoleSpeciesNoveltyMOSA}{0.56}
\newcommand{\PValFitnessEvaluationsWhackAMoleSpeciesNoveltyMOSA}{0.41}
\newcommand{\MeanEffectSizeFitnessEvaluationsNoveltyCompatMOSA}{0.37}
\newcommand{\MeanEffectSizeFitnessEvaluationsSpeciesCompatMOSA}{0.45}
\newcommand{\MeanEffectSizeFitnessEvaluationsSpeciesNoveltyMOSA}{0.59}

\newcommand{\WinCountCatchTheDotsMIO}{7}
\newcommand{\WinCountCatchTheDotsMOSA}{8}
\newcommand{\WinCountCatchTheDotsNEWSD}{10}
\newcommand{\WinCountCatchTheDotsNeatest}{20}
\newcommand{\WinCountCityDefenderMIO}{0}
\newcommand{\WinCountCityDefenderMOSA}{0}
\newcommand{\WinCountCityDefenderNEWSD}{0}
\newcommand{\WinCountCityDefenderNeatest}{0}
\newcommand{\WinCountCreateYourWorldMIO}{3}
\newcommand{\WinCountCreateYourWorldMOSA}{3}
\newcommand{\WinCountCreateYourWorldNEWSD}{3}
\newcommand{\WinCountCreateYourWorldNeatest}{3}
\newcommand{\WinCountDessertRallyMIO}{0}
\newcommand{\WinCountDessertRallyMOSA}{0}
\newcommand{\WinCountDessertRallyNEWSD}{0}
\newcommand{\WinCountDessertRallyNeatest}{0}
\newcommand{\WinCountDieZauberlehrlingeMIO}{17}
\newcommand{\WinCountDieZauberlehrlingeMOSA}{19}
\newcommand{\WinCountDieZauberlehrlingeNEWSD}{20}
\newcommand{\WinCountDieZauberlehrlingeNeatest}{12}
\newcommand{\WinCountDodgeballMIO}{0}
\newcommand{\WinCountDodgeballMOSA}{0}
\newcommand{\WinCountDodgeballNEWSD}{0}
\newcommand{\WinCountDodgeballNeatest}{0}
\newcommand{\WinCountDragonsMIO}{28}
\newcommand{\WinCountDragonsMOSA}{29}
\newcommand{\WinCountDragonsNEWSD}{29}
\newcommand{\WinCountDragonsNeatest}{28}
\newcommand{\WinCountEndlessRunnerMIO}{30}
\newcommand{\WinCountEndlessRunnerMOSA}{30}
\newcommand{\WinCountEndlessRunnerNEWSD}{30}
\newcommand{\WinCountEndlessRunnerNeatest}{30}
\newcommand{\WinCountFallingStarsMIO}{30}
\newcommand{\WinCountFallingStarsMOSA}{30}
\newcommand{\WinCountFallingStarsNEWSD}{30}
\newcommand{\WinCountFallingStarsNeatest}{16}
\newcommand{\WinCountFinalFightMIO}{30}
\newcommand{\WinCountFinalFightMOSA}{30}
\newcommand{\WinCountFinalFightNEWSD}{30}
\newcommand{\WinCountFinalFightNeatest}{30}
\newcommand{\WinCountFlappyParrotMIO}{4}
\newcommand{\WinCountFlappyParrotMOSA}{6}
\newcommand{\WinCountFlappyParrotNEWSD}{1}
\newcommand{\WinCountFlappyParrotNeatest}{6}
\newcommand{\WinCountFroggerMIO}{0}
\newcommand{\WinCountFroggerMOSA}{5}
\newcommand{\WinCountFroggerNEWSD}{2}
\newcommand{\WinCountFroggerNeatest}{2}
\newcommand{\WinCountFruitCatchingMIO}{2}
\newcommand{\WinCountFruitCatchingMOSA}{10}
\newcommand{\WinCountFruitCatchingNEWSD}{1}
\newcommand{\WinCountFruitCatchingNeatest}{11}
\newcommand{\WinCountHackAttackMIO}{1}
\newcommand{\WinCountHackAttackMOSA}{10}
\newcommand{\WinCountHackAttackNEWSD}{2}
\newcommand{\WinCountHackAttackNeatest}{16}
\newcommand{\WinCountOceanCleanupMIO}{12}
\newcommand{\WinCountOceanCleanupMOSA}{17}
\newcommand{\WinCountOceanCleanupNEWSD}{19}
\newcommand{\WinCountOceanCleanupNeatest}{17}
\newcommand{\WinCountPokemonClickerMIO}{0}
\newcommand{\WinCountPokemonClickerMOSA}{0}
\newcommand{\WinCountPokemonClickerNEWSD}{0}
\newcommand{\WinCountPokemonClickerNeatest}{0}
\newcommand{\WinCountSnakeMIO}{0}
\newcommand{\WinCountSnakeMOSA}{0}
\newcommand{\WinCountSnakeNEWSD}{0}
\newcommand{\WinCountSnakeNeatest}{0}
\newcommand{\WinCountSnowballFightMIO}{14}
\newcommand{\WinCountSnowballFightMOSA}{17}
\newcommand{\WinCountSnowballFightNEWSD}{18}
\newcommand{\WinCountSnowballFightNeatest}{8}
\newcommand{\WinCountSpaceOdysseyMIO}{0}
\newcommand{\WinCountSpaceOdysseyMOSA}{0}
\newcommand{\WinCountSpaceOdysseyNEWSD}{0}
\newcommand{\WinCountSpaceOdysseyNeatest}{0}
\newcommand{\WinCountWhackAMoleMIO}{27}
\newcommand{\WinCountWhackAMoleMOSA}{29}
\newcommand{\WinCountWhackAMoleNEWSD}{28}
\newcommand{\WinCountWhackAMoleNeatest}{9}
\newcommand{\MeanWinsMIO}{10.25}
\newcommand{\MeanWinsMOSA}{12.15}
\newcommand{\MeanWinsNEWSD}{11.15}
\newcommand{\MeanWinsNeatest}{10.40}
\newcommand{\SameStatementCoverageMIO}{10}
\newcommand{\SpeedUpStatementMIO}{91.60}
\newcommand{\SameStatementCoverageMOSA}{7}
\newcommand{\SpeedUpStatementMOSA}{94.12}
\newcommand{\SameStatementCoverageNEWSD}{7}
\newcommand{\SpeedUpStatementNEWSD}{94.12}
\newcommand{\SameBranchCoverageMIO}{12}
\newcommand{\SpeedUpBranchMIO}{89.92}
\newcommand{\SameBranchCoverageMOSA}{8}
\newcommand{\SpeedUpBranchMOSA}{93.28}
\newcommand{\SameBranchCoverageNEWSD}{8}
\newcommand{\SpeedUpBranchNEWSD}{93.28}
\newcommand{\MeanStatementCoverageCatchTheDotsMIO}{96.99}
\newcommand{\MeanStatementCoverageCityDefenderMIO}{74.33}
\newcommand{\MeanStatementCoverageCreateYourWorldMIO}{75.23}
\newcommand{\MeanStatementCoverageDessertRallyMIO}{93.90}
\newcommand{\MeanStatementCoverageDieZauberlehrlingeMIO}{84.22}
\newcommand{\MeanStatementCoverageDodgeballMIO}{95.98}
\newcommand{\MeanStatementCoverageDragonsMIO}{87.46}
\newcommand{\MeanStatementCoverageEndlessRunnerMIO}{84.43}
\newcommand{\MeanStatementCoverageFallingStarsMIO}{99.30}
\newcommand{\MeanStatementCoverageFinalFightMIO}{99.65}
\newcommand{\MeanStatementCoverageFlappyParrotMIO}{92.97}
\newcommand{\MeanStatementCoverageFroggerMIO}{94.10}
\newcommand{\MeanStatementCoverageFruitCatchingMIO}{82.06}
\newcommand{\MeanStatementCoverageHackAttackMIO}{92.72}
\newcommand{\MeanStatementCoverageOceanCleanupMIO}{84.32}
\newcommand{\MeanStatementCoveragePokemonClickerMIO}{26.21}
\newcommand{\MeanStatementCoverageSnakeMIO}{95.00}
\newcommand{\MeanStatementCoverageSnowballFightMIO}{96.07}
\newcommand{\MeanStatementCoverageSpaceOdysseyMIO}{87.27}
\newcommand{\MeanStatementCoverageWhackAMoleMIO}{84.61}
\newcommand{\MeanStatementCoverageTotalMIO}{86.34}
\newcommand{\MeanStatementCoverageCatchTheDotsMOSA}{96.34}
\newcommand{\MeanStatementCoverageCityDefenderMOSA}{74.33}
\newcommand{\MeanStatementCoverageCreateYourWorldMOSA}{75.39}
\newcommand{\MeanStatementCoverageDessertRallyMOSA}{93.90}
\newcommand{\MeanStatementCoverageDieZauberlehrlingeMOSA}{85.04}
\newcommand{\MeanStatementCoverageDodgeballMOSA}{95.94}
\newcommand{\MeanStatementCoverageDragonsMOSA}{87.65}
\newcommand{\MeanStatementCoverageEndlessRunnerMOSA}{82.65}
\newcommand{\MeanStatementCoverageFallingStarsMOSA}{99.49}
\newcommand{\MeanStatementCoverageFinalFightMOSA}{99.65}
\newcommand{\MeanStatementCoverageFlappyParrotMOSA}{93.51}
\newcommand{\MeanStatementCoverageFroggerMOSA}{94.36}
\newcommand{\MeanStatementCoverageFruitCatchingMOSA}{92.97}
\newcommand{\MeanStatementCoverageHackAttackMOSA}{94.98}
\newcommand{\MeanStatementCoverageOceanCleanupMOSA}{84.83}
\newcommand{\MeanStatementCoveragePokemonClickerMOSA}{26.21}
\newcommand{\MeanStatementCoverageSnakeMOSA}{94.89}
\newcommand{\MeanStatementCoverageSnowballFightMOSA}{96.32}
\newcommand{\MeanStatementCoverageSpaceOdysseyMOSA}{88.30}
\newcommand{\MeanStatementCoverageWhackAMoleMOSA}{84.08}
\newcommand{\MeanStatementCoverageTotalMOSA}{87.04}
\newcommand{\MeanStatementCoverageCatchTheDotsNEWSD}{96.99}
\newcommand{\MeanStatementCoverageCityDefenderNEWSD}{74.23}
\newcommand{\MeanStatementCoverageCreateYourWorldNEWSD}{75.35}
\newcommand{\MeanStatementCoverageDessertRallyNEWSD}{93.90}
\newcommand{\MeanStatementCoverageDieZauberlehrlingeNEWSD}{85.12}
\newcommand{\MeanStatementCoverageDodgeballNEWSD}{96.11}
\newcommand{\MeanStatementCoverageDragonsNEWSD}{88.14}
\newcommand{\MeanStatementCoverageEndlessRunnerNEWSD}{82.98}
\newcommand{\MeanStatementCoverageFallingStarsNEWSD}{99.56}
\newcommand{\MeanStatementCoverageFinalFightNEWSD}{99.65}
\newcommand{\MeanStatementCoverageFlappyParrotNEWSD}{92.16}
\newcommand{\MeanStatementCoverageFroggerNEWSD}{94.10}
\newcommand{\MeanStatementCoverageFruitCatchingNEWSD}{86.36}
\newcommand{\MeanStatementCoverageHackAttackNEWSD}{92.97}
\newcommand{\MeanStatementCoverageOceanCleanupNEWSD}{85.66}
\newcommand{\MeanStatementCoveragePokemonClickerNEWSD}{26.21}
\newcommand{\MeanStatementCoverageSnakeNEWSD}{94.89}
\newcommand{\MeanStatementCoverageSnowballFightNEWSD}{96.41}
\newcommand{\MeanStatementCoverageSpaceOdysseyNEWSD}{87.67}
\newcommand{\MeanStatementCoverageWhackAMoleNEWSD}{86.03}
\newcommand{\MeanStatementCoverageTotalNEWSD}{86.72}
\newcommand{\MeanStatementCoverageCatchTheDotsNeatest}{98.54}
\newcommand{\MeanStatementCoverageCityDefenderNeatest}{60.52}
\newcommand{\MeanStatementCoverageCreateYourWorldNeatest}{62.26}
\newcommand{\MeanStatementCoverageDessertRallyNeatest}{93.55}
\newcommand{\MeanStatementCoverageDieZauberlehrlingeNeatest}{68.84}
\newcommand{\MeanStatementCoverageDodgeballNeatest}{90.30}
\newcommand{\MeanStatementCoverageDragonsNeatest}{88.58}
\newcommand{\MeanStatementCoverageEndlessRunnerNeatest}{84.22}
\newcommand{\MeanStatementCoverageFallingStarsNeatest}{57.36}
\newcommand{\MeanStatementCoverageFinalFightNeatest}{98.97}
\newcommand{\MeanStatementCoverageFlappyParrotNeatest}{93.33}
\newcommand{\MeanStatementCoverageFroggerNeatest}{92.69}
\newcommand{\MeanStatementCoverageFruitCatchingNeatest}{91.03}
\newcommand{\MeanStatementCoverageHackAttackNeatest}{96.49}
\newcommand{\MeanStatementCoverageOceanCleanupNeatest}{84.49}
\newcommand{\MeanStatementCoveragePokemonClickerNeatest}{26.21}
\newcommand{\MeanStatementCoverageSnakeNeatest}{95.00}
\newcommand{\MeanStatementCoverageSnowballFightNeatest}{95.56}
\newcommand{\MeanStatementCoverageSpaceOdysseyNeatest}{89.37}
\newcommand{\MeanStatementCoverageWhackAMoleNeatest}{74.40}
\newcommand{\MeanStatementCoverageTotalNeatest}{82.09}
\newcommand{\EffectSizeStatementCoverageCatchTheDotsMOSAMIO}{0.61}
\newcommand{\PValStatementCoverageCatchTheDotsMOSAMIO}{0.12}
\newcommand{\EffectSizeStatementCoverageCatchTheDotsNEWSDMIO}{0.49}
\newcommand{\PValStatementCoverageCatchTheDotsNEWSDMIO}{0.88}
\newcommand{\EffectSizeStatementCoverageCatchTheDotsNeatestMIO}{\textbf{0.27}}
\newcommand{\PValStatementCoverageCatchTheDotsNeatestMIO}{\textbf{< 0.01}}
\newcommand{\EffectSizeStatementCoverageCatchTheDotsNEWSDMOSA}{0.41}
\newcommand{\PValStatementCoverageCatchTheDotsNEWSDMOSA}{0.23}
\newcommand{\EffectSizeStatementCoverageCatchTheDotsNeatestMOSA}{\textbf{0.22}}
\newcommand{\PValStatementCoverageCatchTheDotsNeatestMOSA}{\textbf{< 0.01}}
\newcommand{\EffectSizeStatementCoverageCatchTheDotsNeatestNEWSD}{\textbf{0.32}}
\newcommand{\PValStatementCoverageCatchTheDotsNeatestNEWSD}{\textbf{0.01}}
\newcommand{\EffectSizeStatementCoverageCityDefenderMOSAMIO}{0.50}
\newcommand{\PValStatementCoverageCityDefenderMOSAMIO}{1.00}
\newcommand{\EffectSizeStatementCoverageCityDefenderNEWSDMIO}{0.52}
\newcommand{\PValStatementCoverageCityDefenderNEWSDMIO}{0.33}
\newcommand{\EffectSizeStatementCoverageCityDefenderNeatestMIO}{\textbf{0.65}}
\newcommand{\PValStatementCoverageCityDefenderNeatestMIO}{\textbf{< 0.01}}
\newcommand{\EffectSizeStatementCoverageCityDefenderNEWSDMOSA}{0.52}
\newcommand{\PValStatementCoverageCityDefenderNEWSDMOSA}{0.33}
\newcommand{\EffectSizeStatementCoverageCityDefenderNeatestMOSA}{\textbf{0.65}}
\newcommand{\PValStatementCoverageCityDefenderNeatestMOSA}{\textbf{< 0.01}}
\newcommand{\EffectSizeStatementCoverageCityDefenderNeatestNEWSD}{\textbf{0.63}}
\newcommand{\PValStatementCoverageCityDefenderNeatestNEWSD}{\textbf{< 0.01}}
\newcommand{\EffectSizeStatementCoverageCreateYourWorldMOSAMIO}{0.42}
\newcommand{\PValStatementCoverageCreateYourWorldMOSAMIO}{0.20}
\newcommand{\EffectSizeStatementCoverageCreateYourWorldNEWSDMIO}{0.42}
\newcommand{\PValStatementCoverageCreateYourWorldNEWSDMIO}{0.23}
\newcommand{\EffectSizeStatementCoverageCreateYourWorldNeatestMIO}{\textbf{0.88}}
\newcommand{\PValStatementCoverageCreateYourWorldNeatestMIO}{\textbf{< 0.01}}
\newcommand{\EffectSizeStatementCoverageCreateYourWorldNEWSDMOSA}{0.52}
\newcommand{\PValStatementCoverageCreateYourWorldNEWSDMOSA}{0.82}
\newcommand{\EffectSizeStatementCoverageCreateYourWorldNeatestMOSA}{\textbf{0.88}}
\newcommand{\PValStatementCoverageCreateYourWorldNeatestMOSA}{\textbf{< 0.01}}
\newcommand{\EffectSizeStatementCoverageCreateYourWorldNeatestNEWSD}{\textbf{0.89}}
\newcommand{\PValStatementCoverageCreateYourWorldNeatestNEWSD}{\textbf{< 0.01}}
\newcommand{\EffectSizeStatementCoverageDessertRallyMOSAMIO}{0.50}
\newcommand{\PValStatementCoverageDessertRallyMOSAMIO}{1.00}
\newcommand{\EffectSizeStatementCoverageDessertRallyNEWSDMIO}{0.50}
\newcommand{\PValStatementCoverageDessertRallyNEWSDMIO}{1.00}
\newcommand{\EffectSizeStatementCoverageDessertRallyNeatestMIO}{\textbf{0.62}}
\newcommand{\PValStatementCoverageDessertRallyNeatestMIO}{\textbf{0.01}}
\newcommand{\EffectSizeStatementCoverageDessertRallyNEWSDMOSA}{0.50}
\newcommand{\PValStatementCoverageDessertRallyNEWSDMOSA}{1.00}
\newcommand{\EffectSizeStatementCoverageDessertRallyNeatestMOSA}{\textbf{0.62}}
\newcommand{\PValStatementCoverageDessertRallyNeatestMOSA}{\textbf{0.01}}
\newcommand{\EffectSizeStatementCoverageDessertRallyNeatestNEWSD}{\textbf{0.62}}
\newcommand{\PValStatementCoverageDessertRallyNeatestNEWSD}{\textbf{0.01}}
\newcommand{\EffectSizeStatementCoverageDieZauberlehrlingeMOSAMIO}{0.44}
\newcommand{\PValStatementCoverageDieZauberlehrlingeMOSAMIO}{0.42}
\newcommand{\EffectSizeStatementCoverageDieZauberlehrlingeNEWSDMIO}{0.46}
\newcommand{\PValStatementCoverageDieZauberlehrlingeNEWSDMIO}{0.53}
\newcommand{\EffectSizeStatementCoverageDieZauberlehrlingeNeatestMIO}{\textbf{0.66}}
\newcommand{\PValStatementCoverageDieZauberlehrlingeNeatestMIO}{\textbf{0.03}}
\newcommand{\EffectSizeStatementCoverageDieZauberlehrlingeNEWSDMOSA}{0.52}
\newcommand{\PValStatementCoverageDieZauberlehrlingeNEWSDMOSA}{0.79}
\newcommand{\EffectSizeStatementCoverageDieZauberlehrlingeNeatestMOSA}{\textbf{0.69}}
\newcommand{\PValStatementCoverageDieZauberlehrlingeNeatestMOSA}{\textbf{0.01}}
\newcommand{\EffectSizeStatementCoverageDieZauberlehrlingeNeatestNEWSD}{\textbf{0.69}}
\newcommand{\PValStatementCoverageDieZauberlehrlingeNeatestNEWSD}{\textbf{0.01}}
\newcommand{\EffectSizeStatementCoverageDodgeballMOSAMIO}{0.52}
\newcommand{\PValStatementCoverageDodgeballMOSAMIO}{0.73}
\newcommand{\EffectSizeStatementCoverageDodgeballNEWSDMIO}{0.45}
\newcommand{\PValStatementCoverageDodgeballNEWSDMIO}{0.17}
\newcommand{\EffectSizeStatementCoverageDodgeballNeatestMIO}{\textbf{0.67}}
\newcommand{\PValStatementCoverageDodgeballNeatestMIO}{\textbf{< 0.01}}
\newcommand{\EffectSizeStatementCoverageDodgeballNEWSDMOSA}{0.43}
\newcommand{\PValStatementCoverageDodgeballNEWSDMOSA}{0.09}
\newcommand{\EffectSizeStatementCoverageDodgeballNeatestMOSA}{\textbf{0.66}}
\newcommand{\PValStatementCoverageDodgeballNeatestMOSA}{\textbf{0.01}}
\newcommand{\EffectSizeStatementCoverageDodgeballNeatestNEWSD}{\textbf{0.70}}
\newcommand{\PValStatementCoverageDodgeballNeatestNEWSD}{\textbf{< 0.01}}
\newcommand{\EffectSizeStatementCoverageDragonsMOSAMIO}{0.48}
\newcommand{\PValStatementCoverageDragonsMOSAMIO}{0.74}
\newcommand{\EffectSizeStatementCoverageDragonsNEWSDMIO}{0.39}
\newcommand{\PValStatementCoverageDragonsNEWSDMIO}{0.14}
\newcommand{\EffectSizeStatementCoverageDragonsNeatestMIO}{\textbf{0.29}}
\newcommand{\PValStatementCoverageDragonsNeatestMIO}{\textbf{< 0.01}}
\newcommand{\EffectSizeStatementCoverageDragonsNEWSDMOSA}{0.41}
\newcommand{\PValStatementCoverageDragonsNEWSDMOSA}{0.25}
\newcommand{\EffectSizeStatementCoverageDragonsNeatestMOSA}{\textbf{0.34}}
\newcommand{\PValStatementCoverageDragonsNeatestMOSA}{\textbf{0.03}}
\newcommand{\EffectSizeStatementCoverageDragonsNeatestNEWSD}{0.38}
\newcommand{\PValStatementCoverageDragonsNeatestNEWSD}{0.12}
\newcommand{\EffectSizeStatementCoverageEndlessRunnerMOSAMIO}{0.62}
\newcommand{\PValStatementCoverageEndlessRunnerMOSAMIO}{0.07}
\newcommand{\EffectSizeStatementCoverageEndlessRunnerNEWSDMIO}{0.60}
\newcommand{\PValStatementCoverageEndlessRunnerNEWSDMIO}{0.16}
\newcommand{\EffectSizeStatementCoverageEndlessRunnerNeatestMIO}{0.53}
\newcommand{\PValStatementCoverageEndlessRunnerNeatestMIO}{0.70}
\newcommand{\EffectSizeStatementCoverageEndlessRunnerNEWSDMOSA}{0.47}
\newcommand{\PValStatementCoverageEndlessRunnerNEWSDMOSA}{0.65}
\newcommand{\EffectSizeStatementCoverageEndlessRunnerNeatestMOSA}{0.42}
\newcommand{\PValStatementCoverageEndlessRunnerNeatestMOSA}{0.25}
\newcommand{\EffectSizeStatementCoverageEndlessRunnerNeatestNEWSD}{0.45}
\newcommand{\PValStatementCoverageEndlessRunnerNeatestNEWSD}{0.44}
\newcommand{\EffectSizeStatementCoverageFallingStarsMOSAMIO}{0.42}
\newcommand{\PValStatementCoverageFallingStarsMOSAMIO}{0.20}
\newcommand{\EffectSizeStatementCoverageFallingStarsNEWSDMIO}{0.38}
\newcommand{\PValStatementCoverageFallingStarsNEWSDMIO}{0.07}
\newcommand{\EffectSizeStatementCoverageFallingStarsNeatestMIO}{\textbf{0.79}}
\newcommand{\PValStatementCoverageFallingStarsNeatestMIO}{\textbf{< 0.01}}
\newcommand{\EffectSizeStatementCoverageFallingStarsNEWSDMOSA}{0.47}
\newcommand{\PValStatementCoverageFallingStarsNEWSDMOSA}{0.61}
\newcommand{\EffectSizeStatementCoverageFallingStarsNeatestMOSA}{\textbf{0.82}}
\newcommand{\PValStatementCoverageFallingStarsNeatestMOSA}{\textbf{< 0.01}}
\newcommand{\EffectSizeStatementCoverageFallingStarsNeatestNEWSD}{\textbf{0.84}}
\newcommand{\PValStatementCoverageFallingStarsNeatestNEWSD}{\textbf{< 0.01}}
\newcommand{\EffectSizeStatementCoverageFinalFightMOSAMIO}{0.50}
\newcommand{\PValStatementCoverageFinalFightMOSAMIO}{1.00}
\newcommand{\EffectSizeStatementCoverageFinalFightNEWSDMIO}{0.50}
\newcommand{\PValStatementCoverageFinalFightNEWSDMIO}{1.00}
\newcommand{\EffectSizeStatementCoverageFinalFightNeatestMIO}{\textbf{0.72}}
\newcommand{\PValStatementCoverageFinalFightNeatestMIO}{\textbf{< 0.01}}
\newcommand{\EffectSizeStatementCoverageFinalFightNEWSDMOSA}{0.50}
\newcommand{\PValStatementCoverageFinalFightNEWSDMOSA}{1.00}
\newcommand{\EffectSizeStatementCoverageFinalFightNeatestMOSA}{\textbf{0.72}}
\newcommand{\PValStatementCoverageFinalFightNeatestMOSA}{\textbf{< 0.01}}
\newcommand{\EffectSizeStatementCoverageFinalFightNeatestNEWSD}{\textbf{0.72}}
\newcommand{\PValStatementCoverageFinalFightNeatestNEWSD}{\textbf{< 0.01}}
\newcommand{\EffectSizeStatementCoverageFlappyParrotMOSAMIO}{0.47}
\newcommand{\PValStatementCoverageFlappyParrotMOSAMIO}{0.50}
\newcommand{\EffectSizeStatementCoverageFlappyParrotNEWSDMIO}{0.55}
\newcommand{\PValStatementCoverageFlappyParrotNEWSDMIO}{0.17}
\newcommand{\EffectSizeStatementCoverageFlappyParrotNeatestMIO}{0.47}
\newcommand{\PValStatementCoverageFlappyParrotNeatestMIO}{0.56}
\newcommand{\EffectSizeStatementCoverageFlappyParrotNEWSDMOSA}{\textbf{0.58}}
\newcommand{\PValStatementCoverageFlappyParrotNEWSDMOSA}{\textbf{0.05}}
\newcommand{\EffectSizeStatementCoverageFlappyParrotNeatestMOSA}{0.51}
\newcommand{\PValStatementCoverageFlappyParrotNeatestMOSA}{0.91}
\newcommand{\EffectSizeStatementCoverageFlappyParrotNeatestNEWSD}{0.42}
\newcommand{\PValStatementCoverageFlappyParrotNeatestNEWSD}{0.05}
\newcommand{\EffectSizeStatementCoverageFroggerMOSAMIO}{0.46}
\newcommand{\PValStatementCoverageFroggerMOSAMIO}{0.39}
\newcommand{\EffectSizeStatementCoverageFroggerNEWSDMIO}{0.52}
\newcommand{\PValStatementCoverageFroggerNEWSDMIO}{0.69}
\newcommand{\EffectSizeStatementCoverageFroggerNeatestMIO}{\textbf{0.74}}
\newcommand{\PValStatementCoverageFroggerNeatestMIO}{\textbf{< 0.01}}
\newcommand{\EffectSizeStatementCoverageFroggerNEWSDMOSA}{0.56}
\newcommand{\PValStatementCoverageFroggerNEWSDMOSA}{0.34}
\newcommand{\EffectSizeStatementCoverageFroggerNeatestMOSA}{\textbf{0.74}}
\newcommand{\PValStatementCoverageFroggerNeatestMOSA}{\textbf{< 0.01}}
\newcommand{\EffectSizeStatementCoverageFroggerNeatestNEWSD}{\textbf{0.71}}
\newcommand{\PValStatementCoverageFroggerNeatestNEWSD}{\textbf{< 0.01}}
\newcommand{\EffectSizeStatementCoverageFruitCatchingMOSAMIO}{\textbf{0.20}}
\newcommand{\PValStatementCoverageFruitCatchingMOSAMIO}{\textbf{< 0.01}}
\newcommand{\EffectSizeStatementCoverageFruitCatchingNEWSDMIO}{0.37}
\newcommand{\PValStatementCoverageFruitCatchingNEWSDMIO}{0.07}
\newcommand{\EffectSizeStatementCoverageFruitCatchingNeatestMIO}{\textbf{0.25}}
\newcommand{\PValStatementCoverageFruitCatchingNeatestMIO}{\textbf{< 0.01}}
\newcommand{\EffectSizeStatementCoverageFruitCatchingNEWSDMOSA}{\textbf{0.73}}
\newcommand{\PValStatementCoverageFruitCatchingNEWSDMOSA}{\textbf{< 0.01}}
\newcommand{\EffectSizeStatementCoverageFruitCatchingNeatestMOSA}{0.52}
\newcommand{\PValStatementCoverageFruitCatchingNeatestMOSA}{0.81}
\newcommand{\EffectSizeStatementCoverageFruitCatchingNeatestNEWSD}{\textbf{0.31}}
\newcommand{\PValStatementCoverageFruitCatchingNeatestNEWSD}{\textbf{0.01}}
\newcommand{\EffectSizeStatementCoverageHackAttackMOSAMIO}{\textbf{0.35}}
\newcommand{\PValStatementCoverageHackAttackMOSAMIO}{\textbf{< 0.01}}
\newcommand{\EffectSizeStatementCoverageHackAttackNEWSDMIO}{0.48}
\newcommand{\PValStatementCoverageHackAttackNEWSDMIO}{0.57}
\newcommand{\EffectSizeStatementCoverageHackAttackNeatestMIO}{\textbf{0.25}}
\newcommand{\PValStatementCoverageHackAttackNeatestMIO}{\textbf{< 0.01}}
\newcommand{\EffectSizeStatementCoverageHackAttackNEWSDMOSA}{\textbf{0.63}}
\newcommand{\PValStatementCoverageHackAttackNEWSDMOSA}{\textbf{0.01}}
\newcommand{\EffectSizeStatementCoverageHackAttackNeatestMOSA}{0.40}
\newcommand{\PValStatementCoverageHackAttackNeatestMOSA}{0.12}
\newcommand{\EffectSizeStatementCoverageHackAttackNeatestNEWSD}{\textbf{0.27}}
\newcommand{\PValStatementCoverageHackAttackNeatestNEWSD}{\textbf{< 0.01}}
\newcommand{\EffectSizeStatementCoverageOceanCleanupMOSAMIO}{0.51}
\newcommand{\PValStatementCoverageOceanCleanupMOSAMIO}{0.93}
\newcommand{\EffectSizeStatementCoverageOceanCleanupNEWSDMIO}{0.40}
\newcommand{\PValStatementCoverageOceanCleanupNEWSDMIO}{0.20}
\newcommand{\EffectSizeStatementCoverageOceanCleanupNeatestMIO}{0.44}
\newcommand{\PValStatementCoverageOceanCleanupNeatestMIO}{0.43}
\newcommand{\EffectSizeStatementCoverageOceanCleanupNEWSDMOSA}{0.39}
\newcommand{\PValStatementCoverageOceanCleanupNEWSDMOSA}{0.15}
\newcommand{\EffectSizeStatementCoverageOceanCleanupNeatestMOSA}{0.44}
\newcommand{\PValStatementCoverageOceanCleanupNeatestMOSA}{0.41}
\newcommand{\EffectSizeStatementCoverageOceanCleanupNeatestNEWSD}{0.51}
\newcommand{\PValStatementCoverageOceanCleanupNeatestNEWSD}{0.89}
\newcommand{\EffectSizeStatementCoveragePokemonClickerMOSAMIO}{0.50}
\newcommand{\PValStatementCoveragePokemonClickerMOSAMIO}{1.00}
\newcommand{\EffectSizeStatementCoveragePokemonClickerNEWSDMIO}{0.50}
\newcommand{\PValStatementCoveragePokemonClickerNEWSDMIO}{1.00}
\newcommand{\EffectSizeStatementCoveragePokemonClickerNeatestMIO}{0.50}
\newcommand{\PValStatementCoveragePokemonClickerNeatestMIO}{1.00}
\newcommand{\EffectSizeStatementCoveragePokemonClickerNEWSDMOSA}{0.50}
\newcommand{\PValStatementCoveragePokemonClickerNEWSDMOSA}{1.00}
\newcommand{\EffectSizeStatementCoveragePokemonClickerNeatestMOSA}{0.50}
\newcommand{\PValStatementCoveragePokemonClickerNeatestMOSA}{1.00}
\newcommand{\EffectSizeStatementCoveragePokemonClickerNeatestNEWSD}{0.50}
\newcommand{\PValStatementCoveragePokemonClickerNeatestNEWSD}{1.00}
\newcommand{\EffectSizeStatementCoverageSnakeMOSAMIO}{0.52}
\newcommand{\PValStatementCoverageSnakeMOSAMIO}{0.33}
\newcommand{\EffectSizeStatementCoverageSnakeNEWSDMIO}{0.52}
\newcommand{\PValStatementCoverageSnakeNEWSDMIO}{0.33}
\newcommand{\EffectSizeStatementCoverageSnakeNeatestMIO}{0.50}
\newcommand{\PValStatementCoverageSnakeNeatestMIO}{1.00}
\newcommand{\EffectSizeStatementCoverageSnakeNEWSDMOSA}{0.50}
\newcommand{\PValStatementCoverageSnakeNEWSDMOSA}{1.00}
\newcommand{\EffectSizeStatementCoverageSnakeNeatestMOSA}{0.48}
\newcommand{\PValStatementCoverageSnakeNeatestMOSA}{0.33}
\newcommand{\EffectSizeStatementCoverageSnakeNeatestNEWSD}{0.48}
\newcommand{\PValStatementCoverageSnakeNeatestNEWSD}{0.33}
\newcommand{\EffectSizeStatementCoverageSnowballFightMOSAMIO}{0.45}
\newcommand{\PValStatementCoverageSnowballFightMOSAMIO}{0.45}
\newcommand{\EffectSizeStatementCoverageSnowballFightNEWSDMIO}{0.43}
\newcommand{\PValStatementCoverageSnowballFightNEWSDMIO}{0.31}
\newcommand{\EffectSizeStatementCoverageSnowballFightNeatestMIO}{0.60}
\newcommand{\PValStatementCoverageSnowballFightNeatestMIO}{0.11}
\newcommand{\EffectSizeStatementCoverageSnowballFightNEWSDMOSA}{0.48}
\newcommand{\PValStatementCoverageSnowballFightNEWSDMOSA}{0.80}
\newcommand{\EffectSizeStatementCoverageSnowballFightNeatestMOSA}{\textbf{0.65}}
\newcommand{\PValStatementCoverageSnowballFightNeatestMOSA}{\textbf{0.02}}
\newcommand{\EffectSizeStatementCoverageSnowballFightNeatestNEWSD}{\textbf{0.67}}
\newcommand{\PValStatementCoverageSnowballFightNeatestNEWSD}{\textbf{0.01}}
\newcommand{\EffectSizeStatementCoverageSpaceOdysseyMOSAMIO}{0.43}
\newcommand{\PValStatementCoverageSpaceOdysseyMOSAMIO}{0.09}
\newcommand{\EffectSizeStatementCoverageSpaceOdysseyNEWSDMIO}{0.47}
\newcommand{\PValStatementCoverageSpaceOdysseyNEWSDMIO}{0.31}
\newcommand{\EffectSizeStatementCoverageSpaceOdysseyNeatestMIO}{\textbf{0.38}}
\newcommand{\PValStatementCoverageSpaceOdysseyNeatestMIO}{\textbf{0.01}}
\newcommand{\EffectSizeStatementCoverageSpaceOdysseyNEWSDMOSA}{0.54}
\newcommand{\PValStatementCoverageSpaceOdysseyNEWSDMOSA}{0.44}
\newcommand{\EffectSizeStatementCoverageSpaceOdysseyNeatestMOSA}{0.45}
\newcommand{\PValStatementCoverageSpaceOdysseyNeatestMOSA}{0.32}
\newcommand{\EffectSizeStatementCoverageSpaceOdysseyNeatestNEWSD}{0.41}
\newcommand{\PValStatementCoverageSpaceOdysseyNeatestNEWSD}{0.08}
\newcommand{\EffectSizeStatementCoverageWhackAMoleMOSAMIO}{0.57}
\newcommand{\PValStatementCoverageWhackAMoleMOSAMIO}{0.39}
\newcommand{\EffectSizeStatementCoverageWhackAMoleNEWSDMIO}{\textbf{0.28}}
\newcommand{\PValStatementCoverageWhackAMoleNEWSDMIO}{\textbf{< 0.01}}
\newcommand{\EffectSizeStatementCoverageWhackAMoleNeatestMIO}{\textbf{0.97}}
\newcommand{\PValStatementCoverageWhackAMoleNeatestMIO}{\textbf{< 0.01}}
\newcommand{\EffectSizeStatementCoverageWhackAMoleNEWSDMOSA}{\textbf{0.21}}
\newcommand{\PValStatementCoverageWhackAMoleNEWSDMOSA}{\textbf{< 0.01}}
\newcommand{\EffectSizeStatementCoverageWhackAMoleNeatestMOSA}{\textbf{0.96}}
\newcommand{\PValStatementCoverageWhackAMoleNeatestMOSA}{\textbf{< 0.01}}
\newcommand{\EffectSizeStatementCoverageWhackAMoleNeatestNEWSD}{\textbf{0.99}}
\newcommand{\PValStatementCoverageWhackAMoleNeatestNEWSD}{\textbf{< 0.01}}
\newcommand{\MeanEffectSizeStatementCoverageMOSAMIO}{0.47}
\newcommand{\MeanEffectSizeStatementCoverageNEWSDMIO}{0.46}
\newcommand{\MeanEffectSizeStatementCoverageNeatestMIO}{0.56}
\newcommand{\MeanEffectSizeStatementCoverageNEWSDMOSA}{0.49}
\newcommand{\MeanEffectSizeStatementCoverageNeatestMOSA}{0.58}
\newcommand{\MeanEffectSizeStatementCoverageNeatestNEWSD}{0.58}
\newcommand{\MeanBranchCoverageCatchTheDotsMIO}{87.90}
\newcommand{\MeanBranchCoverageCityDefenderMIO}{67.66}
\newcommand{\MeanBranchCoverageCreateYourWorldMIO}{71.36}
\newcommand{\MeanBranchCoverageDessertRallyMIO}{96.15}
\newcommand{\MeanBranchCoverageDieZauberlehrlingeMIO}{70.54}
\newcommand{\MeanBranchCoverageDodgeballMIO}{92.71}
\newcommand{\MeanBranchCoverageDragonsMIO}{81.79}
\newcommand{\MeanBranchCoverageEndlessRunnerMIO}{72.20}
\newcommand{\MeanBranchCoverageFallingStarsMIO}{94.02}
\newcommand{\MeanBranchCoverageFinalFightMIO}{99.23}
\newcommand{\MeanBranchCoverageFlappyParrotMIO}{84.24}
\newcommand{\MeanBranchCoverageFroggerMIO}{90.00}
\newcommand{\MeanBranchCoverageFruitCatchingMIO}{75.36}
\newcommand{\MeanBranchCoverageHackAttackMIO}{97.46}
\newcommand{\MeanBranchCoverageOceanCleanupMIO}{74.41}
\newcommand{\MeanBranchCoveragePokemonClickerMIO}{17.46}
\newcommand{\MeanBranchCoverageSnakeMIO}{78.57}
\newcommand{\MeanBranchCoverageSnowballFightMIO}{86.52}
\newcommand{\MeanBranchCoverageSpaceOdysseyMIO}{91.40}
\newcommand{\MeanBranchCoverageWhackAMoleMIO}{83.47}
\newcommand{\MeanBranchCoverageTotalMIO}{80.62}
\newcommand{\MeanBranchCoverageCatchTheDotsMOSA}{87.33}
\newcommand{\MeanBranchCoverageCityDefenderMOSA}{67.66}
\newcommand{\MeanBranchCoverageCreateYourWorldMOSA}{71.65}
\newcommand{\MeanBranchCoverageDessertRallyMOSA}{96.15}
\newcommand{\MeanBranchCoverageDieZauberlehrlingeMOSA}{70.75}
\newcommand{\MeanBranchCoverageDodgeballMOSA}{92.64}
\newcommand{\MeanBranchCoverageDragonsMOSA}{82.18}
\newcommand{\MeanBranchCoverageEndlessRunnerMOSA}{70.13}
\newcommand{\MeanBranchCoverageFallingStarsMOSA}{94.39}
\newcommand{\MeanBranchCoverageFinalFightMOSA}{99.23}
\newcommand{\MeanBranchCoverageFlappyParrotMOSA}{85.45}
\newcommand{\MeanBranchCoverageFroggerMOSA}{90.50}
\newcommand{\MeanBranchCoverageFruitCatchingMOSA}{85.24}
\newcommand{\MeanBranchCoverageHackAttackMOSA}{98.25}
\newcommand{\MeanBranchCoverageOceanCleanupMOSA}{75.05}
\newcommand{\MeanBranchCoveragePokemonClickerMOSA}{17.47}
\newcommand{\MeanBranchCoverageSnakeMOSA}{78.57}
\newcommand{\MeanBranchCoverageSnowballFightMOSA}{87.54}
\newcommand{\MeanBranchCoverageSpaceOdysseyMOSA}{92.22}
\newcommand{\MeanBranchCoverageWhackAMoleMOSA}{84.21}
\newcommand{\MeanBranchCoverageTotalMOSA}{81.33}
\newcommand{\MeanBranchCoverageCatchTheDotsNEWSD}{88.00}
\newcommand{\MeanBranchCoverageCityDefenderNEWSD}{67.57}
\newcommand{\MeanBranchCoverageCreateYourWorldNEWSD}{71.58}
\newcommand{\MeanBranchCoverageDessertRallyNEWSD}{96.15}
\newcommand{\MeanBranchCoverageDieZauberlehrlingeNEWSD}{70.65}
\newcommand{\MeanBranchCoverageDodgeballNEWSD}{92.95}
\newcommand{\MeanBranchCoverageDragonsNEWSD}{83.01}
\newcommand{\MeanBranchCoverageEndlessRunnerNEWSD}{70.60}
\newcommand{\MeanBranchCoverageFallingStarsNEWSD}{94.55}
\newcommand{\MeanBranchCoverageFinalFightNEWSD}{99.23}
\newcommand{\MeanBranchCoverageFlappyParrotNEWSD}{82.42}
\newcommand{\MeanBranchCoverageFroggerNEWSD}{90.17}
\newcommand{\MeanBranchCoverageFruitCatchingNEWSD}{77.86}
\newcommand{\MeanBranchCoverageHackAttackNEWSD}{97.54}
\newcommand{\MeanBranchCoverageOceanCleanupNEWSD}{75.91}
\newcommand{\MeanBranchCoveragePokemonClickerNEWSD}{17.46}
\newcommand{\MeanBranchCoverageSnakeNEWSD}{78.57}
\newcommand{\MeanBranchCoverageSnowballFightNEWSD}{87.83}
\newcommand{\MeanBranchCoverageSpaceOdysseyNEWSD}{91.75}
\newcommand{\MeanBranchCoverageWhackAMoleNEWSD}{85.43}
\newcommand{\MeanBranchCoverageTotalNEWSD}{80.96}
\newcommand{\MeanBranchCoverageCatchTheDotsNeatest}{89.71}
\newcommand{\MeanBranchCoverageCityDefenderNeatest}{54.95}
\newcommand{\MeanBranchCoverageCreateYourWorldNeatest}{56.74}
\newcommand{\MeanBranchCoverageDessertRallyNeatest}{95.64}
\newcommand{\MeanBranchCoverageDieZauberlehrlingeNeatest}{47.31}
\newcommand{\MeanBranchCoverageDodgeballNeatest}{87.83}
\newcommand{\MeanBranchCoverageDragonsNeatest}{84.17}
\newcommand{\MeanBranchCoverageEndlessRunnerNeatest}{71.80}
\newcommand{\MeanBranchCoverageFallingStarsNeatest}{52.73}
\newcommand{\MeanBranchCoverageFinalFightNeatest}{98.23}
\newcommand{\MeanBranchCoverageFlappyParrotNeatest}{84.85}
\newcommand{\MeanBranchCoverageFroggerNeatest}{88.67}
\newcommand{\MeanBranchCoverageFruitCatchingNeatest}{85.12}
\newcommand{\MeanBranchCoverageHackAttackNeatest}{98.77}
\newcommand{\MeanBranchCoverageOceanCleanupNeatest}{74.25}
\newcommand{\MeanBranchCoveragePokemonClickerNeatest}{17.46}
\newcommand{\MeanBranchCoverageSnakeNeatest}{78.57}
\newcommand{\MeanBranchCoverageSnowballFightNeatest}{84.93}
\newcommand{\MeanBranchCoverageSpaceOdysseyNeatest}{92.98}
\newcommand{\MeanBranchCoverageWhackAMoleNeatest}{72.79}
\newcommand{\MeanBranchCoverageTotalNeatest}{75.88}
\newcommand{\EffectSizeBranchCoverageCatchTheDotsMOSAMIO}{0.59}
\newcommand{\PValBranchCoverageCatchTheDotsMOSAMIO}{0.18}
\newcommand{\EffectSizeBranchCoverageCatchTheDotsNEWSDMIO}{0.49}
\newcommand{\PValBranchCoverageCatchTheDotsNEWSDMIO}{0.87}
\newcommand{\EffectSizeBranchCoverageCatchTheDotsNeatestMIO}{\textbf{0.27}}
\newcommand{\PValBranchCoverageCatchTheDotsNeatestMIO}{\textbf{< 0.01}}
\newcommand{\EffectSizeBranchCoverageCatchTheDotsNEWSDMOSA}{0.43}
\newcommand{\PValBranchCoverageCatchTheDotsNEWSDMOSA}{0.31}
\newcommand{\EffectSizeBranchCoverageCatchTheDotsNeatestMOSA}{\textbf{0.25}}
\newcommand{\PValBranchCoverageCatchTheDotsNeatestMOSA}{\textbf{< 0.01}}
\newcommand{\EffectSizeBranchCoverageCatchTheDotsNeatestNEWSD}{\textbf{0.33}}
\newcommand{\PValBranchCoverageCatchTheDotsNeatestNEWSD}{\textbf{0.02}}
\newcommand{\EffectSizeBranchCoverageCityDefenderMOSAMIO}{0.50}
\newcommand{\PValBranchCoverageCityDefenderMOSAMIO}{1.00}
\newcommand{\EffectSizeBranchCoverageCityDefenderNEWSDMIO}{0.52}
\newcommand{\PValBranchCoverageCityDefenderNEWSDMIO}{0.33}
\newcommand{\EffectSizeBranchCoverageCityDefenderNeatestMIO}{\textbf{0.65}}
\newcommand{\PValBranchCoverageCityDefenderNeatestMIO}{\textbf{< 0.01}}
\newcommand{\EffectSizeBranchCoverageCityDefenderNEWSDMOSA}{0.52}
\newcommand{\PValBranchCoverageCityDefenderNEWSDMOSA}{0.33}
\newcommand{\EffectSizeBranchCoverageCityDefenderNeatestMOSA}{\textbf{0.65}}
\newcommand{\PValBranchCoverageCityDefenderNeatestMOSA}{\textbf{< 0.01}}
\newcommand{\EffectSizeBranchCoverageCityDefenderNeatestNEWSD}{\textbf{0.63}}
\newcommand{\PValBranchCoverageCityDefenderNeatestNEWSD}{\textbf{< 0.01}}
\newcommand{\EffectSizeBranchCoverageCreateYourWorldMOSAMIO}{0.42}
\newcommand{\PValBranchCoverageCreateYourWorldMOSAMIO}{0.20}
\newcommand{\EffectSizeBranchCoverageCreateYourWorldNEWSDMIO}{0.42}
\newcommand{\PValBranchCoverageCreateYourWorldNEWSDMIO}{0.23}
\newcommand{\EffectSizeBranchCoverageCreateYourWorldNeatestMIO}{\textbf{0.88}}
\newcommand{\PValBranchCoverageCreateYourWorldNeatestMIO}{\textbf{< 0.01}}
\newcommand{\EffectSizeBranchCoverageCreateYourWorldNEWSDMOSA}{0.52}
\newcommand{\PValBranchCoverageCreateYourWorldNEWSDMOSA}{0.82}
\newcommand{\EffectSizeBranchCoverageCreateYourWorldNeatestMOSA}{\textbf{0.88}}
\newcommand{\PValBranchCoverageCreateYourWorldNeatestMOSA}{\textbf{< 0.01}}
\newcommand{\EffectSizeBranchCoverageCreateYourWorldNeatestNEWSD}{\textbf{0.89}}
\newcommand{\PValBranchCoverageCreateYourWorldNeatestNEWSD}{\textbf{< 0.01}}
\newcommand{\EffectSizeBranchCoverageDessertRallyMOSAMIO}{0.50}
\newcommand{\PValBranchCoverageDessertRallyMOSAMIO}{1.00}
\newcommand{\EffectSizeBranchCoverageDessertRallyNEWSDMIO}{0.50}
\newcommand{\PValBranchCoverageDessertRallyNEWSDMIO}{1.00}
\newcommand{\EffectSizeBranchCoverageDessertRallyNeatestMIO}{\textbf{0.62}}
\newcommand{\PValBranchCoverageDessertRallyNeatestMIO}{\textbf{0.01}}
\newcommand{\EffectSizeBranchCoverageDessertRallyNEWSDMOSA}{0.50}
\newcommand{\PValBranchCoverageDessertRallyNEWSDMOSA}{1.00}
\newcommand{\EffectSizeBranchCoverageDessertRallyNeatestMOSA}{\textbf{0.62}}
\newcommand{\PValBranchCoverageDessertRallyNeatestMOSA}{\textbf{0.01}}
\newcommand{\EffectSizeBranchCoverageDessertRallyNeatestNEWSD}{\textbf{0.62}}
\newcommand{\PValBranchCoverageDessertRallyNeatestNEWSD}{\textbf{0.01}}
\newcommand{\EffectSizeBranchCoverageDieZauberlehrlingeMOSAMIO}{0.49}
\newcommand{\PValBranchCoverageDieZauberlehrlingeMOSAMIO}{0.94}
\newcommand{\EffectSizeBranchCoverageDieZauberlehrlingeNEWSDMIO}{0.49}
\newcommand{\PValBranchCoverageDieZauberlehrlingeNEWSDMIO}{0.84}
\newcommand{\EffectSizeBranchCoverageDieZauberlehrlingeNeatestMIO}{\textbf{0.70}}
\newcommand{\PValBranchCoverageDieZauberlehrlingeNeatestMIO}{\textbf{0.01}}
\newcommand{\EffectSizeBranchCoverageDieZauberlehrlingeNEWSDMOSA}{0.49}
\newcommand{\PValBranchCoverageDieZauberlehrlingeNEWSDMOSA}{0.91}
\newcommand{\EffectSizeBranchCoverageDieZauberlehrlingeNeatestMOSA}{\textbf{0.70}}
\newcommand{\PValBranchCoverageDieZauberlehrlingeNeatestMOSA}{\textbf{0.01}}
\newcommand{\EffectSizeBranchCoverageDieZauberlehrlingeNeatestNEWSD}{\textbf{0.71}}
\newcommand{\PValBranchCoverageDieZauberlehrlingeNeatestNEWSD}{\textbf{< 0.01}}
\newcommand{\EffectSizeBranchCoverageDodgeballMOSAMIO}{0.52}
\newcommand{\PValBranchCoverageDodgeballMOSAMIO}{0.73}
\newcommand{\EffectSizeBranchCoverageDodgeballNEWSDMIO}{0.45}
\newcommand{\PValBranchCoverageDodgeballNEWSDMIO}{0.17}
\newcommand{\EffectSizeBranchCoverageDodgeballNeatestMIO}{\textbf{0.67}}
\newcommand{\PValBranchCoverageDodgeballNeatestMIO}{\textbf{< 0.01}}
\newcommand{\EffectSizeBranchCoverageDodgeballNEWSDMOSA}{0.43}
\newcommand{\PValBranchCoverageDodgeballNEWSDMOSA}{0.09}
\newcommand{\EffectSizeBranchCoverageDodgeballNeatestMOSA}{\textbf{0.66}}
\newcommand{\PValBranchCoverageDodgeballNeatestMOSA}{\textbf{0.01}}
\newcommand{\EffectSizeBranchCoverageDodgeballNeatestNEWSD}{\textbf{0.70}}
\newcommand{\PValBranchCoverageDodgeballNeatestNEWSD}{\textbf{< 0.01}}
\newcommand{\EffectSizeBranchCoverageDragonsMOSAMIO}{0.47}
\newcommand{\PValBranchCoverageDragonsMOSAMIO}{0.71}
\newcommand{\EffectSizeBranchCoverageDragonsNEWSDMIO}{0.38}
\newcommand{\PValBranchCoverageDragonsNEWSDMIO}{0.12}
\newcommand{\EffectSizeBranchCoverageDragonsNeatestMIO}{\textbf{0.27}}
\newcommand{\PValBranchCoverageDragonsNeatestMIO}{\textbf{< 0.01}}
\newcommand{\EffectSizeBranchCoverageDragonsNEWSDMOSA}{0.42}
\newcommand{\PValBranchCoverageDragonsNEWSDMOSA}{0.28}
\newcommand{\EffectSizeBranchCoverageDragonsNeatestMOSA}{\textbf{0.32}}
\newcommand{\PValBranchCoverageDragonsNeatestMOSA}{\textbf{0.02}}
\newcommand{\EffectSizeBranchCoverageDragonsNeatestNEWSD}{0.38}
\newcommand{\PValBranchCoverageDragonsNeatestNEWSD}{0.11}
\newcommand{\EffectSizeBranchCoverageEndlessRunnerMOSAMIO}{0.62}
\newcommand{\PValBranchCoverageEndlessRunnerMOSAMIO}{0.07}
\newcommand{\EffectSizeBranchCoverageEndlessRunnerNEWSDMIO}{0.59}
\newcommand{\PValBranchCoverageEndlessRunnerNEWSDMIO}{0.19}
\newcommand{\EffectSizeBranchCoverageEndlessRunnerNeatestMIO}{0.54}
\newcommand{\PValBranchCoverageEndlessRunnerNeatestMIO}{0.54}
\newcommand{\EffectSizeBranchCoverageEndlessRunnerNEWSDMOSA}{0.47}
\newcommand{\PValBranchCoverageEndlessRunnerNEWSDMOSA}{0.63}
\newcommand{\EffectSizeBranchCoverageEndlessRunnerNeatestMOSA}{0.44}
\newcommand{\PValBranchCoverageEndlessRunnerNeatestMOSA}{0.41}
\newcommand{\EffectSizeBranchCoverageEndlessRunnerNeatestNEWSD}{0.47}
\newcommand{\PValBranchCoverageEndlessRunnerNeatestNEWSD}{0.68}
\newcommand{\EffectSizeBranchCoverageFallingStarsMOSAMIO}{0.42}
\newcommand{\PValBranchCoverageFallingStarsMOSAMIO}{0.20}
\newcommand{\EffectSizeBranchCoverageFallingStarsNEWSDMIO}{0.38}
\newcommand{\PValBranchCoverageFallingStarsNEWSDMIO}{0.07}
\newcommand{\EffectSizeBranchCoverageFallingStarsNeatestMIO}{\textbf{0.79}}
\newcommand{\PValBranchCoverageFallingStarsNeatestMIO}{\textbf{< 0.01}}
\newcommand{\EffectSizeBranchCoverageFallingStarsNEWSDMOSA}{0.47}
\newcommand{\PValBranchCoverageFallingStarsNEWSDMOSA}{0.61}
\newcommand{\EffectSizeBranchCoverageFallingStarsNeatestMOSA}{\textbf{0.82}}
\newcommand{\PValBranchCoverageFallingStarsNeatestMOSA}{\textbf{< 0.01}}
\newcommand{\EffectSizeBranchCoverageFallingStarsNeatestNEWSD}{\textbf{0.84}}
\newcommand{\PValBranchCoverageFallingStarsNeatestNEWSD}{\textbf{< 0.01}}
\newcommand{\EffectSizeBranchCoverageFinalFightMOSAMIO}{0.50}
\newcommand{\PValBranchCoverageFinalFightMOSAMIO}{1.00}
\newcommand{\EffectSizeBranchCoverageFinalFightNEWSDMIO}{0.50}
\newcommand{\PValBranchCoverageFinalFightNEWSDMIO}{1.00}
\newcommand{\EffectSizeBranchCoverageFinalFightNeatestMIO}{\textbf{0.72}}
\newcommand{\PValBranchCoverageFinalFightNeatestMIO}{\textbf{< 0.01}}
\newcommand{\EffectSizeBranchCoverageFinalFightNEWSDMOSA}{0.50}
\newcommand{\PValBranchCoverageFinalFightNEWSDMOSA}{1.00}
\newcommand{\EffectSizeBranchCoverageFinalFightNeatestMOSA}{\textbf{0.72}}
\newcommand{\PValBranchCoverageFinalFightNeatestMOSA}{\textbf{< 0.01}}
\newcommand{\EffectSizeBranchCoverageFinalFightNeatestNEWSD}{\textbf{0.72}}
\newcommand{\PValBranchCoverageFinalFightNeatestNEWSD}{\textbf{< 0.01}}
\newcommand{\EffectSizeBranchCoverageFlappyParrotMOSAMIO}{0.47}
\newcommand{\PValBranchCoverageFlappyParrotMOSAMIO}{0.50}
\newcommand{\EffectSizeBranchCoverageFlappyParrotNEWSDMIO}{0.55}
\newcommand{\PValBranchCoverageFlappyParrotNEWSDMIO}{0.17}
\newcommand{\EffectSizeBranchCoverageFlappyParrotNeatestMIO}{0.47}
\newcommand{\PValBranchCoverageFlappyParrotNeatestMIO}{0.56}
\newcommand{\EffectSizeBranchCoverageFlappyParrotNEWSDMOSA}{\textbf{0.58}}
\newcommand{\PValBranchCoverageFlappyParrotNEWSDMOSA}{\textbf{0.05}}
\newcommand{\EffectSizeBranchCoverageFlappyParrotNeatestMOSA}{0.51}
\newcommand{\PValBranchCoverageFlappyParrotNeatestMOSA}{0.91}
\newcommand{\EffectSizeBranchCoverageFlappyParrotNeatestNEWSD}{0.42}
\newcommand{\PValBranchCoverageFlappyParrotNeatestNEWSD}{0.05}
\newcommand{\EffectSizeBranchCoverageFroggerMOSAMIO}{\textbf{0.42}}
\newcommand{\PValBranchCoverageFroggerMOSAMIO}{\textbf{0.02}}
\newcommand{\EffectSizeBranchCoverageFroggerNEWSDMIO}{0.47}
\newcommand{\PValBranchCoverageFroggerNEWSDMIO}{0.16}
\newcommand{\EffectSizeBranchCoverageFroggerNeatestMIO}{\textbf{0.65}}
\newcommand{\PValBranchCoverageFroggerNeatestMIO}{\textbf{0.01}}
\newcommand{\EffectSizeBranchCoverageFroggerNEWSDMOSA}{0.55}
\newcommand{\PValBranchCoverageFroggerNEWSDMOSA}{0.23}
\newcommand{\EffectSizeBranchCoverageFroggerNeatestMOSA}{\textbf{0.70}}
\newcommand{\PValBranchCoverageFroggerNeatestMOSA}{\textbf{< 0.01}}
\newcommand{\EffectSizeBranchCoverageFroggerNeatestNEWSD}{\textbf{0.67}}
\newcommand{\PValBranchCoverageFroggerNeatestNEWSD}{\textbf{< 0.01}}
\newcommand{\EffectSizeBranchCoverageFruitCatchingMOSAMIO}{\textbf{0.21}}
\newcommand{\PValBranchCoverageFruitCatchingMOSAMIO}{\textbf{< 0.01}}
\newcommand{\EffectSizeBranchCoverageFruitCatchingNEWSDMIO}{\textbf{0.36}}
\newcommand{\PValBranchCoverageFruitCatchingNEWSDMIO}{\textbf{0.05}}
\newcommand{\EffectSizeBranchCoverageFruitCatchingNeatestMIO}{\textbf{0.24}}
\newcommand{\PValBranchCoverageFruitCatchingNeatestMIO}{\textbf{< 0.01}}
\newcommand{\EffectSizeBranchCoverageFruitCatchingNEWSDMOSA}{\textbf{0.71}}
\newcommand{\PValBranchCoverageFruitCatchingNEWSDMOSA}{\textbf{< 0.01}}
\newcommand{\EffectSizeBranchCoverageFruitCatchingNeatestMOSA}{0.49}
\newcommand{\PValBranchCoverageFruitCatchingNeatestMOSA}{0.94}
\newcommand{\EffectSizeBranchCoverageFruitCatchingNeatestNEWSD}{\textbf{0.32}}
\newcommand{\PValBranchCoverageFruitCatchingNeatestNEWSD}{\textbf{0.02}}
\newcommand{\EffectSizeBranchCoverageHackAttackMOSAMIO}{\textbf{0.35}}
\newcommand{\PValBranchCoverageHackAttackMOSAMIO}{\textbf{< 0.01}}
\newcommand{\EffectSizeBranchCoverageHackAttackNEWSDMIO}{0.48}
\newcommand{\PValBranchCoverageHackAttackNEWSDMIO}{0.57}
\newcommand{\EffectSizeBranchCoverageHackAttackNeatestMIO}{\textbf{0.25}}
\newcommand{\PValBranchCoverageHackAttackNeatestMIO}{\textbf{< 0.01}}
\newcommand{\EffectSizeBranchCoverageHackAttackNEWSDMOSA}{\textbf{0.63}}
\newcommand{\PValBranchCoverageHackAttackNEWSDMOSA}{\textbf{0.01}}
\newcommand{\EffectSizeBranchCoverageHackAttackNeatestMOSA}{0.40}
\newcommand{\PValBranchCoverageHackAttackNeatestMOSA}{0.12}
\newcommand{\EffectSizeBranchCoverageHackAttackNeatestNEWSD}{\textbf{0.27}}
\newcommand{\PValBranchCoverageHackAttackNeatestNEWSD}{\textbf{< 0.01}}
\newcommand{\EffectSizeBranchCoverageOceanCleanupMOSAMIO}{0.52}
\newcommand{\PValBranchCoverageOceanCleanupMOSAMIO}{0.84}
\newcommand{\EffectSizeBranchCoverageOceanCleanupNEWSDMIO}{0.38}
\newcommand{\PValBranchCoverageOceanCleanupNEWSDMIO}{0.12}
\newcommand{\EffectSizeBranchCoverageOceanCleanupNeatestMIO}{0.44}
\newcommand{\PValBranchCoverageOceanCleanupNeatestMIO}{0.44}
\newcommand{\EffectSizeBranchCoverageOceanCleanupNEWSDMOSA}{0.39}
\newcommand{\PValBranchCoverageOceanCleanupNEWSDMOSA}{0.14}
\newcommand{\EffectSizeBranchCoverageOceanCleanupNeatestMOSA}{0.45}
\newcommand{\PValBranchCoverageOceanCleanupNeatestMOSA}{0.47}
\newcommand{\EffectSizeBranchCoverageOceanCleanupNeatestNEWSD}{0.54}
\newcommand{\PValBranchCoverageOceanCleanupNeatestNEWSD}{0.57}
\newcommand{\EffectSizeBranchCoveragePokemonClickerMOSAMIO}{0.48}
\newcommand{\PValBranchCoveragePokemonClickerMOSAMIO}{0.33}
\newcommand{\EffectSizeBranchCoveragePokemonClickerNEWSDMIO}{0.50}
\newcommand{\PValBranchCoveragePokemonClickerNEWSDMIO}{1.00}
\newcommand{\EffectSizeBranchCoveragePokemonClickerNeatestMIO}{0.50}
\newcommand{\PValBranchCoveragePokemonClickerNeatestMIO}{1.00}
\newcommand{\EffectSizeBranchCoveragePokemonClickerNEWSDMOSA}{0.52}
\newcommand{\PValBranchCoveragePokemonClickerNEWSDMOSA}{0.33}
\newcommand{\EffectSizeBranchCoveragePokemonClickerNeatestMOSA}{0.52}
\newcommand{\PValBranchCoveragePokemonClickerNeatestMOSA}{0.33}
\newcommand{\EffectSizeBranchCoveragePokemonClickerNeatestNEWSD}{0.50}
\newcommand{\PValBranchCoveragePokemonClickerNeatestNEWSD}{1.00}
\newcommand{\EffectSizeBranchCoverageSnakeMOSAMIO}{0.50}
\newcommand{\PValBranchCoverageSnakeMOSAMIO}{1.00}
\newcommand{\EffectSizeBranchCoverageSnakeNEWSDMIO}{0.50}
\newcommand{\PValBranchCoverageSnakeNEWSDMIO}{1.00}
\newcommand{\EffectSizeBranchCoverageSnakeNeatestMIO}{0.50}
\newcommand{\PValBranchCoverageSnakeNeatestMIO}{1.00}
\newcommand{\EffectSizeBranchCoverageSnakeNEWSDMOSA}{0.50}
\newcommand{\PValBranchCoverageSnakeNEWSDMOSA}{1.00}
\newcommand{\EffectSizeBranchCoverageSnakeNeatestMOSA}{0.50}
\newcommand{\PValBranchCoverageSnakeNeatestMOSA}{1.00}
\newcommand{\EffectSizeBranchCoverageSnakeNeatestNEWSD}{0.50}
\newcommand{\PValBranchCoverageSnakeNeatestNEWSD}{1.00}
\newcommand{\EffectSizeBranchCoverageSnowballFightMOSAMIO}{0.44}
\newcommand{\PValBranchCoverageSnowballFightMOSAMIO}{0.37}
\newcommand{\EffectSizeBranchCoverageSnowballFightNEWSDMIO}{0.42}
\newcommand{\PValBranchCoverageSnowballFightNEWSDMIO}{0.25}
\newcommand{\EffectSizeBranchCoverageSnowballFightNeatestMIO}{0.60}
\newcommand{\PValBranchCoverageSnowballFightNeatestMIO}{0.13}
\newcommand{\EffectSizeBranchCoverageSnowballFightNEWSDMOSA}{0.48}
\newcommand{\PValBranchCoverageSnowballFightNEWSDMOSA}{0.80}
\newcommand{\EffectSizeBranchCoverageSnowballFightNeatestMOSA}{\textbf{0.65}}
\newcommand{\PValBranchCoverageSnowballFightNeatestMOSA}{\textbf{0.02}}
\newcommand{\EffectSizeBranchCoverageSnowballFightNeatestNEWSD}{\textbf{0.67}}
\newcommand{\PValBranchCoverageSnowballFightNeatestNEWSD}{\textbf{0.01}}
\newcommand{\EffectSizeBranchCoverageSpaceOdysseyMOSAMIO}{0.43}
\newcommand{\PValBranchCoverageSpaceOdysseyMOSAMIO}{0.09}
\newcommand{\EffectSizeBranchCoverageSpaceOdysseyNEWSDMIO}{0.47}
\newcommand{\PValBranchCoverageSpaceOdysseyNEWSDMIO}{0.31}
\newcommand{\EffectSizeBranchCoverageSpaceOdysseyNeatestMIO}{\textbf{0.38}}
\newcommand{\PValBranchCoverageSpaceOdysseyNeatestMIO}{\textbf{0.01}}
\newcommand{\EffectSizeBranchCoverageSpaceOdysseyNEWSDMOSA}{0.54}
\newcommand{\PValBranchCoverageSpaceOdysseyNEWSDMOSA}{0.44}
\newcommand{\EffectSizeBranchCoverageSpaceOdysseyNeatestMOSA}{0.45}
\newcommand{\PValBranchCoverageSpaceOdysseyNeatestMOSA}{0.32}
\newcommand{\EffectSizeBranchCoverageSpaceOdysseyNeatestNEWSD}{0.41}
\newcommand{\PValBranchCoverageSpaceOdysseyNeatestNEWSD}{0.08}
\newcommand{\EffectSizeBranchCoverageWhackAMoleMOSAMIO}{0.39}
\newcommand{\PValBranchCoverageWhackAMoleMOSAMIO}{0.16}
\newcommand{\EffectSizeBranchCoverageWhackAMoleNEWSDMIO}{\textbf{0.26}}
\newcommand{\PValBranchCoverageWhackAMoleNEWSDMIO}{\textbf{< 0.01}}
\newcommand{\EffectSizeBranchCoverageWhackAMoleNeatestMIO}{\textbf{0.95}}
\newcommand{\PValBranchCoverageWhackAMoleNeatestMIO}{\textbf{< 0.01}}
\newcommand{\EffectSizeBranchCoverageWhackAMoleNEWSDMOSA}{0.36}
\newcommand{\PValBranchCoverageWhackAMoleNEWSDMOSA}{0.06}
\newcommand{\EffectSizeBranchCoverageWhackAMoleNeatestMOSA}{\textbf{0.96}}
\newcommand{\PValBranchCoverageWhackAMoleNeatestMOSA}{\textbf{< 0.01}}
\newcommand{\EffectSizeBranchCoverageWhackAMoleNeatestNEWSD}{\textbf{0.98}}
\newcommand{\PValBranchCoverageWhackAMoleNeatestNEWSD}{\textbf{< 0.01}}
\newcommand{\MeanEffectSizeBranchCoverageMOSAMIO}{0.46}
\newcommand{\MeanEffectSizeBranchCoverageNEWSDMIO}{0.46}
\newcommand{\MeanEffectSizeBranchCoverageNeatestMIO}{0.55}
\newcommand{\MeanEffectSizeBranchCoverageNEWSDMOSA}{0.50}
\newcommand{\MeanEffectSizeBranchCoverageNeatestMOSA}{0.58}
\newcommand{\MeanEffectSizeBranchCoverageNeatestNEWSD}{0.58}
\newcommand{\MeanFitnessEvaluationsCatchTheDotsMIO}{669.67}
\newcommand{\MeanFitnessEvaluationsCityDefenderMIO}{114.73}
\newcommand{\MeanFitnessEvaluationsCreateYourWorldMIO}{620.93}
\newcommand{\MeanFitnessEvaluationsDessertRallyMIO}{824.90}
\newcommand{\MeanFitnessEvaluationsDieZauberlehrlingeMIO}{77.27}
\newcommand{\MeanFitnessEvaluationsDodgeballMIO}{591.40}
\newcommand{\MeanFitnessEvaluationsDragonsMIO}{77.40}
\newcommand{\MeanFitnessEvaluationsEndlessRunnerMIO}{87.03}
\newcommand{\MeanFitnessEvaluationsFallingStarsMIO}{629.60}
\newcommand{\MeanFitnessEvaluationsFinalFightMIO}{867.87}
\newcommand{\MeanFitnessEvaluationsFlappyParrotMIO}{346.37}
\newcommand{\MeanFitnessEvaluationsFroggerMIO}{587.40}
\newcommand{\MeanFitnessEvaluationsFruitCatchingMIO}{3872.90}
\newcommand{\MeanFitnessEvaluationsHackAttackMIO}{632.73}
\newcommand{\MeanFitnessEvaluationsOceanCleanupMIO}{3812.63}
\newcommand{\MeanFitnessEvaluationsPokemonClickerMIO}{633.10}
\newcommand{\MeanFitnessEvaluationsSnakeMIO}{23749.33}
\newcommand{\MeanFitnessEvaluationsSnowballFightMIO}{654.63}
\newcommand{\MeanFitnessEvaluationsSpaceOdysseyMIO}{269.90}
\newcommand{\MeanFitnessEvaluationsWhackAMoleMIO}{1765.03}
\newcommand{\MeanFitnessEvaluationsTotalMIO}{2044.24}
\newcommand{\MeanFitnessEvaluationsCatchTheDotsMOSA}{585.30}
\newcommand{\MeanFitnessEvaluationsCityDefenderMOSA}{112.07}
\newcommand{\MeanFitnessEvaluationsCreateYourWorldMOSA}{650.33}
\newcommand{\MeanFitnessEvaluationsDessertRallyMOSA}{877.47}
\newcommand{\MeanFitnessEvaluationsDieZauberlehrlingeMOSA}{83.53}
\newcommand{\MeanFitnessEvaluationsDodgeballMOSA}{536.60}
\newcommand{\MeanFitnessEvaluationsDragonsMOSA}{81.80}
\newcommand{\MeanFitnessEvaluationsEndlessRunnerMOSA}{84.93}
\newcommand{\MeanFitnessEvaluationsFallingStarsMOSA}{633.80}
\newcommand{\MeanFitnessEvaluationsFinalFightMOSA}{597.80}
\newcommand{\MeanFitnessEvaluationsFlappyParrotMOSA}{239}
\newcommand{\MeanFitnessEvaluationsFroggerMOSA}{556.57}
\newcommand{\MeanFitnessEvaluationsFruitCatchingMOSA}{1649.53}
\newcommand{\MeanFitnessEvaluationsHackAttackMOSA}{479.77}
\newcommand{\MeanFitnessEvaluationsOceanCleanupMOSA}{2302.83}
\newcommand{\MeanFitnessEvaluationsPokemonClickerMOSA}{624.27}
\newcommand{\MeanFitnessEvaluationsSnakeMOSA}{7258.80}
\newcommand{\MeanFitnessEvaluationsSnowballFightMOSA}{662.17}
\newcommand{\MeanFitnessEvaluationsSpaceOdysseyMOSA}{240.77}
\newcommand{\MeanFitnessEvaluationsWhackAMoleMOSA}{1153.63}
\newcommand{\MeanFitnessEvaluationsTotalMOSA}{970.55}
\newcommand{\MeanFitnessEvaluationsCatchTheDotsNEWSD}{697.63}
\newcommand{\MeanFitnessEvaluationsCityDefenderNEWSD}{114.77}
\newcommand{\MeanFitnessEvaluationsCreateYourWorldNEWSD}{649.53}
\newcommand{\MeanFitnessEvaluationsDessertRallyNEWSD}{860.53}
\newcommand{\MeanFitnessEvaluationsDieZauberlehrlingeNEWSD}{84.17}
\newcommand{\MeanFitnessEvaluationsDodgeballNEWSD}{630.47}
\newcommand{\MeanFitnessEvaluationsDragonsNEWSD}{82.73}
\newcommand{\MeanFitnessEvaluationsEndlessRunnerNEWSD}{91.50}
\newcommand{\MeanFitnessEvaluationsFallingStarsNEWSD}{657.10}
\newcommand{\MeanFitnessEvaluationsFinalFightNEWSD}{822.57}
\newcommand{\MeanFitnessEvaluationsFlappyParrotNEWSD}{372.57}
\newcommand{\MeanFitnessEvaluationsFroggerNEWSD}{593.90}
\newcommand{\MeanFitnessEvaluationsFruitCatchingNEWSD}{2138.97}
\newcommand{\MeanFitnessEvaluationsHackAttackNEWSD}{654.17}
\newcommand{\MeanFitnessEvaluationsOceanCleanupNEWSD}{4753.80}
\newcommand{\MeanFitnessEvaluationsPokemonClickerNEWSD}{627.37}
\newcommand{\MeanFitnessEvaluationsSnakeNEWSD}{6124.07}
\newcommand{\MeanFitnessEvaluationsSnowballFightNEWSD}{669.43}
\newcommand{\MeanFitnessEvaluationsSpaceOdysseyNEWSD}{311.30}
\newcommand{\MeanFitnessEvaluationsWhackAMoleNEWSD}{1277.93}
\newcommand{\MeanFitnessEvaluationsTotalNEWSD}{1110.73}
\newcommand{\MeanFitnessEvaluationsCatchTheDotsNeatest}{1373.10}
\newcommand{\MeanFitnessEvaluationsCityDefenderNeatest}{456.43}
\newcommand{\MeanFitnessEvaluationsCreateYourWorldNeatest}{697.03}
\newcommand{\MeanFitnessEvaluationsDessertRallyNeatest}{925.17}
\newcommand{\MeanFitnessEvaluationsDieZauberlehrlingeNeatest}{395.87}
\newcommand{\MeanFitnessEvaluationsDodgeballNeatest}{577.93}
\newcommand{\MeanFitnessEvaluationsDragonsNeatest}{159.93}
\newcommand{\MeanFitnessEvaluationsEndlessRunnerNeatest}{357.67}
\newcommand{\MeanFitnessEvaluationsFallingStarsNeatest}{629.80}
\newcommand{\MeanFitnessEvaluationsFinalFightNeatest}{694.30}
\newcommand{\MeanFitnessEvaluationsFlappyParrotNeatest}{324.50}
\newcommand{\MeanFitnessEvaluationsFroggerNeatest}{608.47}
\newcommand{\MeanFitnessEvaluationsFruitCatchingNeatest}{3350.73}
\newcommand{\MeanFitnessEvaluationsHackAttackNeatest}{501.50}
\newcommand{\MeanFitnessEvaluationsOceanCleanupNeatest}{9678.37}
\newcommand{\MeanFitnessEvaluationsPokemonClickerNeatest}{598.20}
\newcommand{\MeanFitnessEvaluationsSnakeNeatest}{13676.70}
\newcommand{\MeanFitnessEvaluationsSnowballFightNeatest}{670.10}
\newcommand{\MeanFitnessEvaluationsSpaceOdysseyNeatest}{258.67}
\newcommand{\MeanFitnessEvaluationsWhackAMoleNeatest}{3791.30}
\newcommand{\MeanFitnessEvaluationsTotalNeatest}{1986.29}
\newcommand{\EffectSizeFitnessEvaluationsCatchTheDotsMOSAMIO}{\textbf{0.78}}
\newcommand{\PValFitnessEvaluationsCatchTheDotsMOSAMIO}{\textbf{< 0.01}}
\newcommand{\EffectSizeFitnessEvaluationsCatchTheDotsNEWSDMIO}{0.60}
\newcommand{\PValFitnessEvaluationsCatchTheDotsNEWSDMIO}{0.19}
\newcommand{\EffectSizeFitnessEvaluationsCatchTheDotsNeatestMIO}{\textbf{0.04}}
\newcommand{\PValFitnessEvaluationsCatchTheDotsNeatestMIO}{\textbf{< 0.01}}
\newcommand{\EffectSizeFitnessEvaluationsCatchTheDotsNEWSDMOSA}{\textbf{0.31}}
\newcommand{\PValFitnessEvaluationsCatchTheDotsNEWSDMOSA}{\textbf{0.01}}
\newcommand{\EffectSizeFitnessEvaluationsCatchTheDotsNeatestMOSA}{\textbf{0.03}}
\newcommand{\PValFitnessEvaluationsCatchTheDotsNeatestMOSA}{\textbf{< 0.01}}
\newcommand{\EffectSizeFitnessEvaluationsCatchTheDotsNeatestNEWSD}{\textbf{0.08}}
\newcommand{\PValFitnessEvaluationsCatchTheDotsNeatestNEWSD}{\textbf{< 0.01}}
\newcommand{\EffectSizeFitnessEvaluationsCityDefenderMOSAMIO}{0.43}
\newcommand{\PValFitnessEvaluationsCityDefenderMOSAMIO}{0.33}
\newcommand{\EffectSizeFitnessEvaluationsCityDefenderNEWSDMIO}{0.47}
\newcommand{\PValFitnessEvaluationsCityDefenderNEWSDMIO}{0.67}
\newcommand{\EffectSizeFitnessEvaluationsCityDefenderNeatestMIO}{\textbf{0.06}}
\newcommand{\PValFitnessEvaluationsCityDefenderNeatestMIO}{\textbf{< 0.01}}
\newcommand{\EffectSizeFitnessEvaluationsCityDefenderNEWSDMOSA}{0.55}
\newcommand{\PValFitnessEvaluationsCityDefenderNEWSDMOSA}{0.53}
\newcommand{\EffectSizeFitnessEvaluationsCityDefenderNeatestMOSA}{\textbf{0.08}}
\newcommand{\PValFitnessEvaluationsCityDefenderNeatestMOSA}{\textbf{< 0.01}}
\newcommand{\EffectSizeFitnessEvaluationsCityDefenderNeatestNEWSD}{\textbf{0.07}}
\newcommand{\PValFitnessEvaluationsCityDefenderNeatestNEWSD}{\textbf{< 0.01}}
\newcommand{\EffectSizeFitnessEvaluationsCreateYourWorldMOSAMIO}{\textbf{0.05}}
\newcommand{\PValFitnessEvaluationsCreateYourWorldMOSAMIO}{\textbf{< 0.01}}
\newcommand{\EffectSizeFitnessEvaluationsCreateYourWorldNEWSDMIO}{\textbf{0.04}}
\newcommand{\PValFitnessEvaluationsCreateYourWorldNEWSDMIO}{\textbf{< 0.01}}
\newcommand{\EffectSizeFitnessEvaluationsCreateYourWorldNeatestMIO}{\textbf{0.00}}
\newcommand{\PValFitnessEvaluationsCreateYourWorldNeatestMIO}{\textbf{< 0.01}}
\newcommand{\EffectSizeFitnessEvaluationsCreateYourWorldNEWSDMOSA}{0.51}
\newcommand{\PValFitnessEvaluationsCreateYourWorldNEWSDMOSA}{0.89}
\newcommand{\EffectSizeFitnessEvaluationsCreateYourWorldNeatestMOSA}{\textbf{0.01}}
\newcommand{\PValFitnessEvaluationsCreateYourWorldNeatestMOSA}{\textbf{< 0.01}}
\newcommand{\EffectSizeFitnessEvaluationsCreateYourWorldNeatestNEWSD}{\textbf{0.01}}
\newcommand{\PValFitnessEvaluationsCreateYourWorldNeatestNEWSD}{\textbf{< 0.01}}
\newcommand{\EffectSizeFitnessEvaluationsDessertRallyMOSAMIO}{\textbf{0.23}}
\newcommand{\PValFitnessEvaluationsDessertRallyMOSAMIO}{\textbf{< 0.01}}
\newcommand{\EffectSizeFitnessEvaluationsDessertRallyNEWSDMIO}{\textbf{0.31}}
\newcommand{\PValFitnessEvaluationsDessertRallyNEWSDMIO}{\textbf{0.01}}
\newcommand{\EffectSizeFitnessEvaluationsDessertRallyNeatestMIO}{\textbf{0.03}}
\newcommand{\PValFitnessEvaluationsDessertRallyNeatestMIO}{\textbf{< 0.01}}
\newcommand{\EffectSizeFitnessEvaluationsDessertRallyNEWSDMOSA}{0.60}
\newcommand{\PValFitnessEvaluationsDessertRallyNEWSDMOSA}{0.19}
\newcommand{\EffectSizeFitnessEvaluationsDessertRallyNeatestMOSA}{\textbf{0.31}}
\newcommand{\PValFitnessEvaluationsDessertRallyNeatestMOSA}{\textbf{0.01}}
\newcommand{\EffectSizeFitnessEvaluationsDessertRallyNeatestNEWSD}{\textbf{0.20}}
\newcommand{\PValFitnessEvaluationsDessertRallyNeatestNEWSD}{\textbf{< 0.01}}
\newcommand{\EffectSizeFitnessEvaluationsDieZauberlehrlingeMOSAMIO}{0.47}
\newcommand{\PValFitnessEvaluationsDieZauberlehrlingeMOSAMIO}{0.32}
\newcommand{\EffectSizeFitnessEvaluationsDieZauberlehrlingeNEWSDMIO}{0.48}
\newcommand{\PValFitnessEvaluationsDieZauberlehrlingeNEWSDMIO}{0.57}
\newcommand{\EffectSizeFitnessEvaluationsDieZauberlehrlingeNeatestMIO}{\textbf{0.16}}
\newcommand{\PValFitnessEvaluationsDieZauberlehrlingeNeatestMIO}{\textbf{< 0.01}}
\newcommand{\EffectSizeFitnessEvaluationsDieZauberlehrlingeNEWSDMOSA}{0.52}
\newcommand{\PValFitnessEvaluationsDieZauberlehrlingeNEWSDMOSA}{0.69}
\newcommand{\EffectSizeFitnessEvaluationsDieZauberlehrlingeNeatestMOSA}{\textbf{0.18}}
\newcommand{\PValFitnessEvaluationsDieZauberlehrlingeNeatestMOSA}{\textbf{< 0.01}}
\newcommand{\EffectSizeFitnessEvaluationsDieZauberlehrlingeNeatestNEWSD}{\textbf{0.17}}
\newcommand{\PValFitnessEvaluationsDieZauberlehrlingeNeatestNEWSD}{\textbf{< 0.01}}
\newcommand{\EffectSizeFitnessEvaluationsDodgeballMOSAMIO}{0.55}
\newcommand{\PValFitnessEvaluationsDodgeballMOSAMIO}{0.48}
\newcommand{\EffectSizeFitnessEvaluationsDodgeballNEWSDMIO}{\textbf{0.28}}
\newcommand{\PValFitnessEvaluationsDodgeballNEWSDMIO}{\textbf{< 0.01}}
\newcommand{\EffectSizeFitnessEvaluationsDodgeballNeatestMIO}{0.48}
\newcommand{\PValFitnessEvaluationsDodgeballNeatestMIO}{0.76}
\newcommand{\EffectSizeFitnessEvaluationsDodgeballNEWSDMOSA}{\textbf{0.31}}
\newcommand{\PValFitnessEvaluationsDodgeballNEWSDMOSA}{\textbf{0.01}}
\newcommand{\EffectSizeFitnessEvaluationsDodgeballNeatestMOSA}{0.40}
\newcommand{\PValFitnessEvaluationsDodgeballNeatestMOSA}{0.18}
\newcommand{\EffectSizeFitnessEvaluationsDodgeballNeatestNEWSD}{0.59}
\newcommand{\PValFitnessEvaluationsDodgeballNeatestNEWSD}{0.22}
\newcommand{\EffectSizeFitnessEvaluationsDragonsMOSAMIO}{0.37}
\newcommand{\PValFitnessEvaluationsDragonsMOSAMIO}{0.09}
\newcommand{\EffectSizeFitnessEvaluationsDragonsNEWSDMIO}{\textbf{0.30}}
\newcommand{\PValFitnessEvaluationsDragonsNEWSDMIO}{\textbf{0.01}}
\newcommand{\EffectSizeFitnessEvaluationsDragonsNeatestMIO}{\textbf{0.00}}
\newcommand{\PValFitnessEvaluationsDragonsNeatestMIO}{\textbf{< 0.01}}
\newcommand{\EffectSizeFitnessEvaluationsDragonsNEWSDMOSA}{0.45}
\newcommand{\PValFitnessEvaluationsDragonsNEWSDMOSA}{0.51}
\newcommand{\EffectSizeFitnessEvaluationsDragonsNeatestMOSA}{\textbf{0.02}}
\newcommand{\PValFitnessEvaluationsDragonsNeatestMOSA}{\textbf{< 0.01}}
\newcommand{\EffectSizeFitnessEvaluationsDragonsNeatestNEWSD}{\textbf{0.01}}
\newcommand{\PValFitnessEvaluationsDragonsNeatestNEWSD}{\textbf{< 0.01}}
\newcommand{\EffectSizeFitnessEvaluationsEndlessRunnerMOSAMIO}{0.50}
\newcommand{\PValFitnessEvaluationsEndlessRunnerMOSAMIO}{1.00}
\newcommand{\EffectSizeFitnessEvaluationsEndlessRunnerNEWSDMIO}{0.50}
\newcommand{\PValFitnessEvaluationsEndlessRunnerNEWSDMIO}{0.97}
\newcommand{\EffectSizeFitnessEvaluationsEndlessRunnerNeatestMIO}{\textbf{0.04}}
\newcommand{\PValFitnessEvaluationsEndlessRunnerNeatestMIO}{\textbf{< 0.01}}
\newcommand{\EffectSizeFitnessEvaluationsEndlessRunnerNEWSDMOSA}{0.50}
\newcommand{\PValFitnessEvaluationsEndlessRunnerNEWSDMOSA}{0.97}
\newcommand{\EffectSizeFitnessEvaluationsEndlessRunnerNeatestMOSA}{\textbf{0.02}}
\newcommand{\PValFitnessEvaluationsEndlessRunnerNeatestMOSA}{\textbf{< 0.01}}
\newcommand{\EffectSizeFitnessEvaluationsEndlessRunnerNeatestNEWSD}{\textbf{0.04}}
\newcommand{\PValFitnessEvaluationsEndlessRunnerNeatestNEWSD}{\textbf{< 0.01}}
\newcommand{\EffectSizeFitnessEvaluationsFallingStarsMOSAMIO}{\textbf{0.24}}
\newcommand{\PValFitnessEvaluationsFallingStarsMOSAMIO}{\textbf{< 0.01}}
\newcommand{\EffectSizeFitnessEvaluationsFallingStarsNEWSDMIO}{\textbf{0.13}}
\newcommand{\PValFitnessEvaluationsFallingStarsNEWSDMIO}{\textbf{< 0.01}}
\newcommand{\EffectSizeFitnessEvaluationsFallingStarsNeatestMIO}{0.37}
\newcommand{\PValFitnessEvaluationsFallingStarsNeatestMIO}{0.08}
\newcommand{\EffectSizeFitnessEvaluationsFallingStarsNEWSDMOSA}{0.55}
\newcommand{\PValFitnessEvaluationsFallingStarsNEWSDMOSA}{0.52}
\newcommand{\EffectSizeFitnessEvaluationsFallingStarsNeatestMOSA}{0.36}
\newcommand{\PValFitnessEvaluationsFallingStarsNeatestMOSA}{0.06}
\newcommand{\EffectSizeFitnessEvaluationsFallingStarsNeatestNEWSD}{0.40}
\newcommand{\PValFitnessEvaluationsFallingStarsNeatestNEWSD}{0.18}
\newcommand{\EffectSizeFitnessEvaluationsFinalFightMOSAMIO}{\textbf{0.91}}
\newcommand{\PValFitnessEvaluationsFinalFightMOSAMIO}{\textbf{< 0.01}}
\newcommand{\EffectSizeFitnessEvaluationsFinalFightNEWSDMIO}{0.60}
\newcommand{\PValFitnessEvaluationsFinalFightNEWSDMIO}{0.19}
\newcommand{\EffectSizeFitnessEvaluationsFinalFightNeatestMIO}{\textbf{0.86}}
\newcommand{\PValFitnessEvaluationsFinalFightNeatestMIO}{\textbf{< 0.01}}
\newcommand{\EffectSizeFitnessEvaluationsFinalFightNEWSDMOSA}{\textbf{0.16}}
\newcommand{\PValFitnessEvaluationsFinalFightNEWSDMOSA}{\textbf{< 0.01}}
\newcommand{\EffectSizeFitnessEvaluationsFinalFightNeatestMOSA}{\textbf{0.30}}
\newcommand{\PValFitnessEvaluationsFinalFightNeatestMOSA}{\textbf{0.01}}
\newcommand{\EffectSizeFitnessEvaluationsFinalFightNeatestNEWSD}{\textbf{0.73}}
\newcommand{\PValFitnessEvaluationsFinalFightNeatestNEWSD}{\textbf{< 0.01}}
\newcommand{\EffectSizeFitnessEvaluationsFlappyParrotMOSAMIO}{\textbf{0.71}}
\newcommand{\PValFitnessEvaluationsFlappyParrotMOSAMIO}{\textbf{< 0.01}}
\newcommand{\EffectSizeFitnessEvaluationsFlappyParrotNEWSDMIO}{0.44}
\newcommand{\PValFitnessEvaluationsFlappyParrotNEWSDMIO}{0.47}
\newcommand{\EffectSizeFitnessEvaluationsFlappyParrotNeatestMIO}{0.54}
\newcommand{\PValFitnessEvaluationsFlappyParrotNeatestMIO}{0.61}
\newcommand{\EffectSizeFitnessEvaluationsFlappyParrotNEWSDMOSA}{\textbf{0.24}}
\newcommand{\PValFitnessEvaluationsFlappyParrotNEWSDMOSA}{\textbf{< 0.01}}
\newcommand{\EffectSizeFitnessEvaluationsFlappyParrotNeatestMOSA}{\textbf{0.22}}
\newcommand{\PValFitnessEvaluationsFlappyParrotNeatestMOSA}{\textbf{< 0.01}}
\newcommand{\EffectSizeFitnessEvaluationsFlappyParrotNeatestNEWSD}{0.60}
\newcommand{\PValFitnessEvaluationsFlappyParrotNeatestNEWSD}{0.18}
\newcommand{\EffectSizeFitnessEvaluationsFroggerMOSAMIO}{0.61}
\newcommand{\PValFitnessEvaluationsFroggerMOSAMIO}{0.13}
\newcommand{\EffectSizeFitnessEvaluationsFroggerNEWSDMIO}{0.44}
\newcommand{\PValFitnessEvaluationsFroggerNEWSDMIO}{0.42}
\newcommand{\EffectSizeFitnessEvaluationsFroggerNeatestMIO}{0.37}
\newcommand{\PValFitnessEvaluationsFroggerNeatestMIO}{0.08}
\newcommand{\EffectSizeFitnessEvaluationsFroggerNEWSDMOSA}{\textbf{0.34}}
\newcommand{\PValFitnessEvaluationsFroggerNEWSDMOSA}{\textbf{0.04}}
\newcommand{\EffectSizeFitnessEvaluationsFroggerNeatestMOSA}{\textbf{0.30}}
\newcommand{\PValFitnessEvaluationsFroggerNeatestMOSA}{\textbf{0.01}}
\newcommand{\EffectSizeFitnessEvaluationsFroggerNeatestNEWSD}{0.41}
\newcommand{\PValFitnessEvaluationsFroggerNeatestNEWSD}{0.24}
\newcommand{\EffectSizeFitnessEvaluationsFruitCatchingMOSAMIO}{\textbf{0.96}}
\newcommand{\PValFitnessEvaluationsFruitCatchingMOSAMIO}{\textbf{< 0.01}}
\newcommand{\EffectSizeFitnessEvaluationsFruitCatchingNEWSDMIO}{\textbf{0.89}}
\newcommand{\PValFitnessEvaluationsFruitCatchingNEWSDMIO}{\textbf{< 0.01}}
\newcommand{\EffectSizeFitnessEvaluationsFruitCatchingNeatestMIO}{0.59}
\newcommand{\PValFitnessEvaluationsFruitCatchingNeatestMIO}{0.23}
\newcommand{\EffectSizeFitnessEvaluationsFruitCatchingNEWSDMOSA}{\textbf{0.29}}
\newcommand{\PValFitnessEvaluationsFruitCatchingNEWSDMOSA}{\textbf{< 0.01}}
\newcommand{\EffectSizeFitnessEvaluationsFruitCatchingNeatestMOSA}{\textbf{0.06}}
\newcommand{\PValFitnessEvaluationsFruitCatchingNeatestMOSA}{\textbf{< 0.01}}
\newcommand{\EffectSizeFitnessEvaluationsFruitCatchingNeatestNEWSD}{\textbf{0.16}}
\newcommand{\PValFitnessEvaluationsFruitCatchingNeatestNEWSD}{\textbf{< 0.01}}
\newcommand{\EffectSizeFitnessEvaluationsHackAttackMOSAMIO}{\textbf{0.82}}
\newcommand{\PValFitnessEvaluationsHackAttackMOSAMIO}{\textbf{< 0.01}}
\newcommand{\EffectSizeFitnessEvaluationsHackAttackNEWSDMIO}{0.35}
\newcommand{\PValFitnessEvaluationsHackAttackNEWSDMIO}{0.05}
\newcommand{\EffectSizeFitnessEvaluationsHackAttackNeatestMIO}{\textbf{0.77}}
\newcommand{\PValFitnessEvaluationsHackAttackNeatestMIO}{\textbf{< 0.01}}
\newcommand{\EffectSizeFitnessEvaluationsHackAttackNEWSDMOSA}{\textbf{0.12}}
\newcommand{\PValFitnessEvaluationsHackAttackNEWSDMOSA}{\textbf{< 0.01}}
\newcommand{\EffectSizeFitnessEvaluationsHackAttackNeatestMOSA}{0.42}
\newcommand{\PValFitnessEvaluationsHackAttackNeatestMOSA}{0.30}
\newcommand{\EffectSizeFitnessEvaluationsHackAttackNeatestNEWSD}{\textbf{0.81}}
\newcommand{\PValFitnessEvaluationsHackAttackNeatestNEWSD}{\textbf{< 0.01}}
\newcommand{\EffectSizeFitnessEvaluationsOceanCleanupMOSAMIO}{\textbf{0.67}}
\newcommand{\PValFitnessEvaluationsOceanCleanupMOSAMIO}{\textbf{0.03}}
\newcommand{\EffectSizeFitnessEvaluationsOceanCleanupNEWSDMIO}{\textbf{0.65}}
\newcommand{\PValFitnessEvaluationsOceanCleanupNEWSDMIO}{\textbf{0.05}}
\newcommand{\EffectSizeFitnessEvaluationsOceanCleanupNeatestMIO}{\textbf{0.18}}
\newcommand{\PValFitnessEvaluationsOceanCleanupNeatestMIO}{\textbf{< 0.01}}
\newcommand{\EffectSizeFitnessEvaluationsOceanCleanupNEWSDMOSA}{0.53}
\newcommand{\PValFitnessEvaluationsOceanCleanupNEWSDMOSA}{0.70}
\newcommand{\EffectSizeFitnessEvaluationsOceanCleanupNeatestMOSA}{\textbf{0.05}}
\newcommand{\PValFitnessEvaluationsOceanCleanupNeatestMOSA}{\textbf{< 0.01}}
\newcommand{\EffectSizeFitnessEvaluationsOceanCleanupNeatestNEWSD}{\textbf{0.09}}
\newcommand{\PValFitnessEvaluationsOceanCleanupNeatestNEWSD}{\textbf{< 0.01}}
\newcommand{\EffectSizeFitnessEvaluationsPokemonClickerMOSAMIO}{0.45}
\newcommand{\PValFitnessEvaluationsPokemonClickerMOSAMIO}{0.47}
\newcommand{\EffectSizeFitnessEvaluationsPokemonClickerNEWSDMIO}{\textbf{0.81}}
\newcommand{\PValFitnessEvaluationsPokemonClickerNEWSDMIO}{\textbf{< 0.01}}
\newcommand{\EffectSizeFitnessEvaluationsPokemonClickerNeatestMIO}{\textbf{0.13}}
\newcommand{\PValFitnessEvaluationsPokemonClickerNeatestMIO}{\textbf{< 0.01}}
\newcommand{\EffectSizeFitnessEvaluationsPokemonClickerNEWSDMOSA}{\textbf{0.76}}
\newcommand{\PValFitnessEvaluationsPokemonClickerNEWSDMOSA}{\textbf{< 0.01}}
\newcommand{\EffectSizeFitnessEvaluationsPokemonClickerNeatestMOSA}{\textbf{0.28}}
\newcommand{\PValFitnessEvaluationsPokemonClickerNeatestMOSA}{\textbf{< 0.01}}
\newcommand{\EffectSizeFitnessEvaluationsPokemonClickerNeatestNEWSD}{\textbf{0.17}}
\newcommand{\PValFitnessEvaluationsPokemonClickerNeatestNEWSD}{\textbf{< 0.01}}
\newcommand{\EffectSizeFitnessEvaluationsSnakeMOSAMIO}{\textbf{0.99}}
\newcommand{\PValFitnessEvaluationsSnakeMOSAMIO}{\textbf{< 0.01}}
\newcommand{\EffectSizeFitnessEvaluationsSnakeNEWSDMIO}{\textbf{1.00}}
\newcommand{\PValFitnessEvaluationsSnakeNEWSDMIO}{\textbf{< 0.01}}
\newcommand{\EffectSizeFitnessEvaluationsSnakeNeatestMIO}{\textbf{0.90}}
\newcommand{\PValFitnessEvaluationsSnakeNeatestMIO}{\textbf{< 0.01}}
\newcommand{\EffectSizeFitnessEvaluationsSnakeNEWSDMOSA}{\textbf{0.72}}
\newcommand{\PValFitnessEvaluationsSnakeNEWSDMOSA}{\textbf{< 0.01}}
\newcommand{\EffectSizeFitnessEvaluationsSnakeNeatestMOSA}{\textbf{0.01}}
\newcommand{\PValFitnessEvaluationsSnakeNeatestMOSA}{\textbf{< 0.01}}
\newcommand{\EffectSizeFitnessEvaluationsSnakeNeatestNEWSD}{\textbf{0.00}}
\newcommand{\PValFitnessEvaluationsSnakeNeatestNEWSD}{\textbf{< 0.01}}
\newcommand{\EffectSizeFitnessEvaluationsSnowballFightMOSAMIO}{\textbf{0.29}}
\newcommand{\PValFitnessEvaluationsSnowballFightMOSAMIO}{\textbf{< 0.01}}
\newcommand{\EffectSizeFitnessEvaluationsSnowballFightNEWSDMIO}{\textbf{0.26}}
\newcommand{\PValFitnessEvaluationsSnowballFightNEWSDMIO}{\textbf{< 0.01}}
\newcommand{\EffectSizeFitnessEvaluationsSnowballFightNeatestMIO}{\textbf{0.26}}
\newcommand{\PValFitnessEvaluationsSnowballFightNeatestMIO}{\textbf{< 0.01}}
\newcommand{\EffectSizeFitnessEvaluationsSnowballFightNEWSDMOSA}{0.57}
\newcommand{\PValFitnessEvaluationsSnowballFightNEWSDMOSA}{0.32}
\newcommand{\EffectSizeFitnessEvaluationsSnowballFightNeatestMOSA}{0.46}
\newcommand{\PValFitnessEvaluationsSnowballFightNeatestMOSA}{0.63}
\newcommand{\EffectSizeFitnessEvaluationsSnowballFightNeatestNEWSD}{0.38}
\newcommand{\PValFitnessEvaluationsSnowballFightNeatestNEWSD}{0.11}
\newcommand{\EffectSizeFitnessEvaluationsSpaceOdysseyMOSAMIO}{0.52}
\newcommand{\PValFitnessEvaluationsSpaceOdysseyMOSAMIO}{0.74}
\newcommand{\EffectSizeFitnessEvaluationsSpaceOdysseyNEWSDMIO}{0.42}
\newcommand{\PValFitnessEvaluationsSpaceOdysseyNEWSDMIO}{0.28}
\newcommand{\EffectSizeFitnessEvaluationsSpaceOdysseyNeatestMIO}{0.53}
\newcommand{\PValFitnessEvaluationsSpaceOdysseyNeatestMIO}{0.73}
\newcommand{\EffectSizeFitnessEvaluationsSpaceOdysseyNEWSDMOSA}{0.39}
\newcommand{\PValFitnessEvaluationsSpaceOdysseyNEWSDMOSA}{0.13}
\newcommand{\EffectSizeFitnessEvaluationsSpaceOdysseyNeatestMOSA}{0.49}
\newcommand{\PValFitnessEvaluationsSpaceOdysseyNeatestMOSA}{0.85}
\newcommand{\EffectSizeFitnessEvaluationsSpaceOdysseyNeatestNEWSD}{0.60}
\newcommand{\PValFitnessEvaluationsSpaceOdysseyNeatestNEWSD}{0.16}
\newcommand{\EffectSizeFitnessEvaluationsWhackAMoleMOSAMIO}{\textbf{0.86}}
\newcommand{\PValFitnessEvaluationsWhackAMoleMOSAMIO}{\textbf{< 0.01}}
\newcommand{\EffectSizeFitnessEvaluationsWhackAMoleNEWSDMIO}{\textbf{0.69}}
\newcommand{\PValFitnessEvaluationsWhackAMoleNEWSDMIO}{\textbf{0.01}}
\newcommand{\EffectSizeFitnessEvaluationsWhackAMoleNeatestMIO}{\textbf{0.13}}
\newcommand{\PValFitnessEvaluationsWhackAMoleNeatestMIO}{\textbf{< 0.01}}
\newcommand{\EffectSizeFitnessEvaluationsWhackAMoleNEWSDMOSA}{\textbf{0.20}}
\newcommand{\PValFitnessEvaluationsWhackAMoleNEWSDMOSA}{\textbf{< 0.01}}
\newcommand{\EffectSizeFitnessEvaluationsWhackAMoleNeatestMOSA}{\textbf{0.03}}
\newcommand{\PValFitnessEvaluationsWhackAMoleNeatestMOSA}{\textbf{< 0.01}}
\newcommand{\EffectSizeFitnessEvaluationsWhackAMoleNeatestNEWSD}{\textbf{0.04}}
\newcommand{\PValFitnessEvaluationsWhackAMoleNeatestNEWSD}{\textbf{< 0.01}}
\newcommand{\MeanEffectSizeFitnessEvaluationsMOSAMIO}{0.57}
\newcommand{\MeanEffectSizeFitnessEvaluationsNEWSDMIO}{0.48}
\newcommand{\MeanEffectSizeFitnessEvaluationsNeatestMIO}{0.32}
\newcommand{\MeanEffectSizeFitnessEvaluationsNEWSDMOSA}{0.43}
\newcommand{\MeanEffectSizeFitnessEvaluationsNeatestMOSA}{0.20}
\newcommand{\MeanEffectSizeFitnessEvaluationsNeatestNEWSD}{0.28}

%% file: sections/1-Introduction.tex
\section{Introduction}

A common approach in automated test generation is to create tests for
individual coverage goals of the program under test, aiming to cover
as many goals as possible overall. If the program under test is a
game, this is challenging as many coverage goals will require
meaningful gameplay to reach advanced program states. 
Even if a test
generator would manage to find a sequence of user inputs that leads to
such an advanced program state, the resulting sequence would likely be
unsuitable as a test case due to games being inherently randomised, such
that the same sequence of inputs might lead to a completely different
state upon re-execution.
Rather than aiming to generate \emph{static} sequences of inputs, an
alternative therefore lies in \emph{dynamic} test cases in the form of
neural networks that learn to master the game, thus enabling them to
reliably reach advanced program states and coverage
goals~\cite{Feldmeier2022Neuroevolution}.

\input{images/Introduction}

The \Neatest approach~\cite{Feldmeier2022Neuroevolution} combines
search-based testing principles with neuroevolution in order to create
such dynamic test cases automatically. \Neatest targets one coverage
goal at a time and tries to optimise a neural network that reliably
reaches that goal. Once the goal is reached or \Neatest has spent a
pre-defined amount of the search budget trying to reach the current
goal, it proceeds with optimising networks for the next coverage goal
not yet covered. This process continues until all objectives are
covered or a stopping criterion is met.

Since \Neatest optimises only one coverage objective at a time, it may
fail to focus the search towards promising objectives or get stuck
trying to reach hard or even impossible-to-reach program
states. Consider the simple \emph{CreateYourWorld} game depicted in
\cref{fig:CYW}, which is a maze game where the player (displayed as a
blue square) has to reach the orange door on the right in order to
proceed to the next level, as indicated by the code example in
\cref{fig:CYW-Scripts}. A fitness function for guiding the test
generator to evaluating the corresponding if-condition to true might
use the distance between the player and the
door~\cite{Feldmeier2022Neuroevolution}. However, the fitness value is
deceiving as the player cannot walk through the wall painted in
grey~\cite{Feldmeier2024Combining}, resulting in a lot of the
allocated search budget spent on trying to cover this hard-to-reach
branch.
At the same time, easier objectives that are within reach are
ignored. For example, the fitness for the branch that increases the
score after touching the yellow coin would be based on the distance
between the player and the coin, but even though the player may get
close to the coin while the search algorithm is focusing on reaching
the door, in the end the search does not benefit from this.

A possible approach to avoid such problems is by using
\emph{many-objective} search algorithms that try to create tests for
\emph{all} coverage goals simultaneously, such that no intermediate
tests are lost, and neither the order of goals nor the time spent on
each needs to be decided by the algorithm. However, neuroevolution
differs from traditional search algorithms used for test generation:
On the one hand, neuroevolution requires principles such as speciation
to protect the evolution of subspecies, which is not used in
traditional test generation and its many-objective search
algorithms. On the other hand, existing many-objective search
algorithms use properties of static test cases such as their length to
make decisions, thus requiring suitable alternatives when dealing with
dynamic test cases, i.e., neural networks that do not have a
traditional ``length''.
In order to overcome the challenges posed by these differences, we
extend the \Neatest algorithm to a many-objective approach.
We propose alternative criteria to replace the length property, and
investigate three different search algorithms: The first two
approaches combine \Neatest's optimisation algorithm \Neat with
\MIO~\cite{Arcuri2017Many} and
\MOSA~\cite{Panichella2015Reformulating}, which are the
state-of-the-art many-objective test generation algorithms for
traditional software systems. In addition, we adapt the
state-of-the-art many-objective neuroevolution framework
\NEWSD~\cite{Salih2021Modified} to serve as a test generator for
games.

In detail, the contributions of this paper are as follows:
\begin{itemize}
    \item We extend the \Neatest neuroevolution-based test generator for games
    to a many-objective algorithm.
    \item We propose two new many-objective neuroevolution algorithms by
    combining \Neat with \MIO and \MOSA.
    \item We adapt the many-objective testing framework \NEWSD to serve as a
    test generator for games.
    \item We implement the proposed many-objective algorithms as extensions to
    the open-source \Neatest test generator.
    \item We empirically evaluate all proposed algorithms on a diverse set of 20
    \Scratch games.
\end{itemize}

Our empirical evaluation on 20 \Scratch games demonstrates that casting the test
generation problem to a many-objective optimisation task improves the overall
coverage achieved and significantly decreases the required search
time.

%% file: images/Introduction.tex
\begin{figure}[htbp]
  \centering
    \begin{subfigure}[c]{.5\columnwidth}
      \centering
      \includegraphics[width=\linewidth]{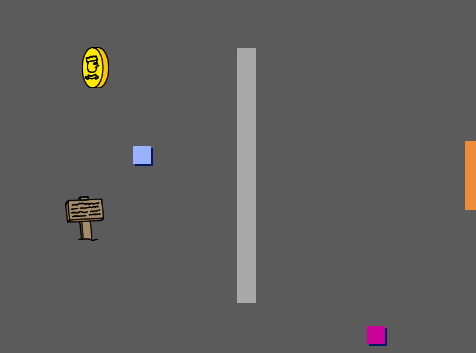}
      \caption{The \emph{CreateYourWorld} game.}
      \label{fig:CYW}
    \end{subfigure}%
   \begin{subfigure}[c]{.5\columnwidth}
      \centering
      \includegraphics[width=\columnwidth]{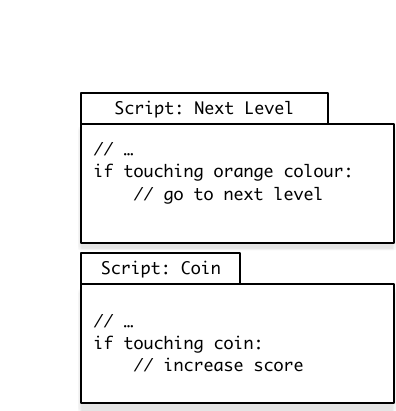}
      \caption{Example scripts of the game.}
      \label{fig:CYW-Scripts}
    \end{subfigure}

    \caption{The \emph{CreateYourWorld} game with example code scripts.}
    \label{fig:Introduction}
  \end{figure}

%% file: sections/2-Background.tex
\section{Background}
In this paper, we combine many-objective optimisation with \Neatest, a
neuroevolution-based test generator for games.

\subsection{Search-based Software Testing}
\label{sec:SBST}
Search-based Software Testing (SBST) is a technique that uses meta-heuristic optimisation search strategies, such as genetic algorithms, to automate software testing. There are various applications, such as prioritising test cases \cite{Li2007Search}, minimising test suites \cite{Habib2023Systematic} or generating test data \cite{Mcminn2004Search-based}. In this paper, we aim for the latter by generating program inputs for the program under test.
To achieve that, SBST guides the search algorithm by a problem-specific
\emph{fitness function} indicating the quality of a possible solution. If the
optimisation problem is to maximise code coverage, the fitness function states
how close an individual is to achieving full program coverage or reaching certain
statements or branches. 
A common approach to determine the distance towards covering a branch or statement is
to compute the sum of the \emph{approach level} and the normalised \emph{branch distance}
\cite{Wegener2001Evolutionary, Arcuri2010Does}. The approach level corresponds
to the distance between the control node that guards a targeted statement or
branch and the closest control node reached by a given test case. Branch
distance is measured at the control location where a path through
the program is taken that makes the execution of the coverage objective
impossible and calculates how far the predicate of the control location
is from obtaining the opposite value, which paves the way towards
reaching the objective. In traditional SBST approaches, the resulting test
suites correspond to a static sequence of test inputs for a program under test.
However, since it is difficult to evolve static sequences of inputs
that reach deep program states in games, and since static sequences are
unable to adapt to program randomisation that is common in games,
recent work~\cite{Feldmeier2022Neuroevolution} has combined SBST
principles with neuroevolution to automate the generation of tests for
games.

\subsection{Neuroevolution-based Test Generation}
\label{sec:Neatest}

\Neatest is a test generator for games that tackles the many challenges of
testing games automatically, including hard-to-reach program states and heavy
program randomisation~\cite{Feldmeier2022Neuroevolution}. The test generator is
embedded into the \Whisker~\cite{Deiner2023Automated} testing framework for
\Scratch programs and generates \emph{dynamic} test suites consisting of
neural networks. In order to handle non-determinism due to
program randomisation, these networks dynamically generate inputs to the program
under test, adapting to the current program state.  
Test suites are generated by optimising neural networks
towards reaching selected coverage objectives like program statements or
branches.

During the selection of an optimisation goal, \Neatest
prioritises branches and statements that were close to being reached in
past program executions. To this end, the game under test is modelled as an
interprocedural \emph{Control Dependence Graph}
(CDG)~\cite{Cytron1991Efficiently}, which defines a hierarchy of dependent
program statements. Based on this CDG only coverage objectives are considered
whose parents have been covered by previously generated tests. This approach
allows the test generator to employ a \emph{Curriculum
Learning}\cite{Bengio2009Curriculum} strategy such that sophisticated gameplay
can be evolved based on previously learned gameplay strategies.

Once an
optimisation goal is selected, \Neatest begins evolving neural networks using
the neuroevolution algorithm \Neat until the target is covered or a pre-defined
search budget for the targeted objective is exhausted. To ensure that generated
tests are robust against program randomisation, a \emph{robustness check}
verifies that the generated test is able to reach the coverage objective repeatedly
in a pre-defined number of randomised program executions; the robustness value is added to the fitness value. Once a test passes this
check, the objective is deemed reliably covered and \Neatest continues with selecting the
next coverage objective based on the CDG. 

\Neat~\cite{Stanley2002Evolving} employs evolutionary search strategies to co-evolve the weights and
topology of a network towards a given goal based on a specified fitness
function, which corresponds in \Neatest to a fitness function that computes the
distance towards the targeted program state using the combination of approach
level and branch distance described in \cref{sec:SBST}. 
Mutation introduces small perturbations to the network weights or modifies the
network topology by adding neurons or connections to the network. Crossover
forms a single child by combining the connections of two parent networks.
To protect novel topology 
innovations from selection pressure, \Neat splits its
population into several species and restricts the selection of parents within
species. Newly generated chromosomes are added to the species to which they
exhibit the lowest \emph{compatibility distance}, which measures the genotypic
difference between two chromosomes.

In order to avoid the threat of deceiving
fitness
landscapes~\cite{Christensen2007solving,Langdon1998ants,Lehman2008exploiting},
\Neatest has been extended with a secondary fitness function that 
promotes \emph{novel} solutions~\cite{Feldmeier2024Combining}. To this end, a
novelty score is derived from the average distance of the $k$ nearest states a
given network has reached compared to an archive of observed states in the past.
Since \Neatest iteratively selects one coverage objective after another and
because \Neat is designed as a single-objective algorithm, the test generator
focuses only on one coverage objective at a time, which may induce severe
limitations.

\subsection{Many-Objective Search-based Test Generation}
\label{sec:Many-objective}

Single-objective optimisation algorithms aim to find a single solution by maximising or minimising a certain criterion, the \emph{objective}. 
However, some problems cannot be represented by a single objective and require optimising multiple conflicting objectives simultaneously, known as \emph{multi-objective} optimisation problems. 
Optimising four or more objectives is termed \emph{many-objective} optimisation~\cite{Chand2015Evolutionary}. When applying search to generate tests based on code coverage, each coverage objective can be interpreted as one such objective.
Two popular many-objective algorithms in the SBST field are the
\emph{Many-Objective Sorting Algorithm} (\MOSA) \cite{Arcuri2017Many} and the
\emph{Many-Independent Objective algorithm} (\MIO)
\cite{Panichella2015Reformulating}. Neuroevolution research is mostly
concerned with single-objective optimisation but there exists a framework called
\emph{NeuroEvolution of Weights and Structures by Decomposition}
(\NEWSD)~\cite{Salih2021Modified} that attempts to optimise several objectives
at 
once. 

\vspace{0.2cm}
\noindent\textbf{\MOSA} maximises branch coverage by treating each branch as a
separate objective in a many-objective optimisation problem. The algorithm starts with a
random population of test cases and maintains an archive of individuals that
cover program branches. New individuals are added to the archive if they cover an
uncovered branch, while a \emph{secondary preference criterion}, favouring
shorter tests, is used to replace existing ones for already covered branches.
\MOSA generates offspring by applying mutation and crossover to selected
individuals and ranks them using \emph{preference sorting}, starting with
rank zero for individuals closest to covering uncovered branches. Remaining
individuals are ranked using non-dominated sorting, as in NSGA-II
\cite{Deb2002Fast}. Given a minimisation task, \MOSA applies the \emph{Pareto
dominance} relation, which defines that individual $x$ dominates ($x \prec y$)
individual $y$ if  
\begin{center}
	$\forall i \in \{1, \dots, m \}$ $f_i(x) \leq f_i(y)$ \\
	and \\
	$\exists j \in \{1, \dots, m \}$ such that $f_j(x) < f_j(y)$
\end{center}
with $f_i$ representing the $i$-th objective function, indicating the quality of a solution, and $m$ the number of objectives. 
The algorithm adds non-dominated fronts to the new population, starting with front $\mathbb{F}_0$, until the population size limit is reached. If the last front exceeds this limit, individuals are selected based on their \emph{crowding distances}, with more diverse individuals chosen first to maintain diversity and avoid premature convergence.
Finally, the archive is updated to store the shortest test for each covered branch. 

\vspace{0.2cm}
\noindent\textbf{\MIO} maintains an archive for each objective and either
generates a random test case or samples a random test from an archive for
mutation.
After mutating the sampled test, a heuristic score $h_k \in [0,1]$ is calculated
for each objective, with 1 indicating a covered objective. In case $h_k = 1$, the test is added to the archive
for that objective, replacing all existing individuals. Otherwise, if the
archive has not reached its maximum size yet, the test is added. If
the archive is full, the mutated test replaces the worst individual if it
performs better according to $h_k$ and the secondary preference criterion. 
To avoid sampling from unreachable objectives, \MIO uses \emph{Feedback-Directed
Sampling} by assigning each archive a counter that increases whenever a test is
sampled and is reset to zero whenever a resulting test improves the heuristic score for that objective. Sampling then prefers
archives with lowest counters. Finally, the algorithm gradually decreases the random
test generation probability and archive size while increasing mutations to focus
more on exploitation over time.

\vspace{0.2cm}
\noindent\textbf{\NEWSD} proposes a neuroevolution framework for optimising
multiple objectives simultaneously by dividing the population into $k$
single-objective subproblems. The algorithm starts with a random population of minimal
topologies and random weights, using a scalarised objective function to
aggregate objectives. Each topology has a record tracking its structure, number
of individuals within the topology, creation generation, and protected status.
First, the \emph{Score Assignment Procedure} assigns a score vector to each individual, consisting of the scalarised objective scores regarding every subproblem. 
Next, the \emph{Elite Set Procedure} selects the best individuals for each
subproblem, while the \emph{Protected Set Procedure} safeguards novice
topologies by adding them to a protected set of individuals to avoid premature
elimination of novel topologies.
In the following \emph{Selected Procedure}, the elite and protected sets are
combined and remaining population slots are filled by drawing random individuals
based on their subproblem scores. Then, the \emph{Reproduction Procedure}
evolves the weights of selected individuals from the previous step and applies
structural change if the number of topologies is below a threshold.
Finally, the \emph{Population Update Procedure} adds the best individuals of
each subproblem to the next generation, retains protected individuals and updates
the archive by adding non-dominated offspring based on Pareto dominance. This
procedure repeats until a stopping condition is met. 


%% file: sections/3-Approach.tex
\section{Extending Neatest with Many-Objective Search}
In order to extend \Neatest to a many-objective optimisation algorithm, we
adapted the three algorithms described in \cref{sec:Many-objective} to the
challenges of generating tests for games and integrated them into the \Whisker
testing framework~\cite{Deiner2023Automated}. However, in contrast to \MIO and
\MOSA, which target all coverage objectives simultaneously, we apply the same target selection strategy as in \Neatest
to facilitate \emph{Curriculum Learning}. We retain \Neat's network encoding and
\Neatest's fitness function, including the robustness check, to ensure that
generated test suites are robust against program randomisation.

\subsection{Secondary Criteria} 
\label{sec:secondary}

\MOSA and \MIO represent their test cases as input sequences, where
shorter sequences are preferable. Therefore, the algorithms use the
length as a secondary criterion. In contrast, \Neatest optimises tests
as neural networks, defined by a list of neuron and connection
genes. As smaller networks do not necessarily provide any advantages
over larger ones, we require different secondary criteria.  We
therefore propose three metrics, two based on the genetic encoding of
the network and one on the behaviour of the phenotype:
\begin{itemize}
  \item The first metric, derived from the genotype,
counts the number of peers present in the species to which a given
network belongs. Since the secondary fitness criterion is responsible for promoting
diversity, we favour networks belonging to smaller species.
\item The second metric uses the average \emph{compatibility distance} between a
given network and the remaining population.
This similarity distance between two genotypes was introduced by
\Neat~\cite{Feldmeier2022Neuroevolution} and measures how much two networks 
differ in structure. As bigger compatibility distances correspond to a greater
dissimilarity between two chromosomes, we favour networks with larger
compatibility distance averages.
\item The third preference criterion is derived from
the phenotype of a chromosome and rewards novel behaviours. To this end, we
compute a \emph{novelty score} similar to \Neatest by calculating the average
distance to the $k$ nearest program states reached by a given network compared
to an archive of observed states. Since high novelty scores indicate
distinct chromosome behaviour, we cast the novelty of a network to a
maximisation goal.
\end{itemize}

\subsection{MOSA-Neatest} 
\label{sec:MOSA-Neatest}

\SetKwFunction{generateRandomPopulation}{generateRandomPopulation}
\SetKwFunction{speciate}{speciate}
\SetKwFunction{updateArchive}{updateArchive}
\SetKwFunction{generateOffspring}{generateOffspring}
\SetKwFunction{preferenceSorting}{preferenceSorting}
\SetKwFunction{size}{size}
\SetKwFunction{add}{add}
\SetKwFunction{sort}{sort}
\SetFuncArgSty{}

\begin{algorithm}[tb]
  \Input{maximum population size~$P_{\textsf{max}}$}
  \Input{secondary preference criterion~$c_s$}
  \Output{dynamic test suite~$D$}
	\Fn{\MOSANeatest($P_{\textsf{max}}, c_s$)}{
		$P \gets \generateRandomPopulation{}$\;
		$P.\speciate{}$\;
		$D \gets$ \updateArchive{$P$}\;
	
		\While{stopping condition not reached} {
		$O \gets$ \generateOffspring{$P$}\;
		$D \gets$ \updateArchive{$O$}\;
		
		$C \gets P \cup O$\;
		$C.\speciate{}$\;
		$\mathbb{F} \gets$ \preferenceSorting{$C, c_s$}\;
		
		$P_{\textsf{next}} = \emptyset$\;
		
		\ForEach{front f $\in \mathbb{F}$}{
			\uIf{$P_{\textsf{next}}.\size + f.\size < P_{\textsf{max}}$}{$
				P_{\textsf{next}}.$\add{$f$}\;
			}
			\Else{
			\tcp{Front too large}
				$\textsf{count} \gets P_{\textsf{max}}-P_{\textsf{next}}$.size\;
				$f.$\sort{$c_s$}\;
				$f_{\textsf{best}} \gets f[0:\textsf{count}]$\;
				$P_{\textsf{next}}.$\add{$f_{\textsf{best}}$}\;
			}
		}
		$P \gets P_{\textsf{next}}$\;
		}
  		\Return{$D$}
  		}
    \caption{\MOSANeatest}
    \label{alg:MOSANeatest}
\end{algorithm}

We use \MOSA as our basic framework and adopt the new secondary
preference criteria favouring diversity (\cref{sec:secondary}).  As
the resulting test suite does not benefit from having more diverse
networks, we only check if previously uncovered coverage objectives
are newly covered during the archive update and, in contrast to the
standard \MOSA algorithm, do not evaluate the current population with
respect to already covered objectives during these updates.
Algorithm \ref{alg:MOSANeatest} shows the approach for \MOSANeatest.
In order to create a new generation of chromosomes, we adopt the offspring
generation procedure of \MOSA but use the crossover and mutation operators of
\Neat (line 6). During the selection of parents for crossover, we ensure that they
originate from the same species. To obtain the species, we speciate the initial
population in line~3 and the combined population in line~9 using the default
speciation procedure of \Neat. If a given species only contains one individual, we
select the other parent randomly from the current generation. 
After creating a new offspring generation, \MOSANeatest updates the archive in line 7 but only considers uncovered coverage objectives that are close to being reached based on \Neatest's target selection procedure, described in \cref{sec:Neatest}. 
Next, we combine the parents and offspring
and apply preference sorting based on one of the proposed
criteria. 
Similar to the traditional \MOSA algorithm, \MOSANeatest
starts with the lowest rank and gradually adds non-dominated Pareto-fronts to
the new generation in lines 12 to~21. To promote diversity, we do not sort the last front by the
crowding distance but by our secondary preference criterion, as shown in line 17.

\subsection{MIO-Neatest}

\SetKwFunction{initCounters}{initCounters}
\SetKwFunction{checkGenerateNew}{checkGenerateNew}
\SetKwFunction{getRandomObjective}{getRandomObjective}
\SetKwFunction{generateRandomNetwork}{generateRandomNetwork}
\SetKwFunction{getLowestCounterObjective}{getLowestCounterObjective}
\SetKwFunction{sampleArchive}{sampleArchive}
\SetKwFunction{randDouble}{randDouble}
\SetKwFunction{mutateStructure}{mutateStructure}
\SetKwFunction{mutateWeights}{mutateWeights}
\SetKwFunction{getBest}{getBest}
\SetKwFunction{isCovered}{isCovered}
\SetKwFunction{updateParameters}{updateParameters}
\SetFuncArgSty{}

\begin{algorithm}[tb]
  \Input{control dependence graph~$\mathit{CDG}$}
  \Input{probability for random network generation~$P_r$}
  \Input{probability for structural mutation~$P_{\textsf{st}}$}
  \Input{mutation count~$m$}
  \Input{secondary preference criterion~$c_{s}$}
    \Input{maximum archive size~$A_{m}$}
  \Output{dynamic test suite~$D$}
	\Fn{\MIONeatest($\mathit{CDG}, P_r, P_{\textsf{st}}, m, c_{s}, A_{m}$)}{
		\initCounters()\;
	
		\While{stopping condition not reached} {
			$\textsf{generateNew} \gets$ \checkGenerateNew{$P_r$}\;
			
			\uIf{$\textsf{generateNew}$}{
			$k \gets$ \getRandomObjective{$\mathit{CDG}$}\;
			$N \gets$ \generateRandomNetwork{}\;
			}
			\Else{
			$k \gets$ \getLowestCounterObjective{}\;
			$N \gets$ \sampleArchive{$k$}\;
			}
			
			\ForEach{network \( n \in N \)}{
				$n_{\textsf{best}} \gets n$\;
				$\textsf{struct} \gets \textsf{True}$\;
				
				\For{\( i \gets 1 \) \KwTo \( m \)}{
				$\textsf{mutant} \gets \emptyset$\;
				
				\uIf{$\textsf{struct} \; \textsf{\&} \; P_{\textsf{st}} < \randDouble{}$}{
					$\textsf{struct} \gets \textsf{False}$\;
					$\textsf{mutant} \gets$ \mutateStructure{$n$}\;
					$n_{\textsf{best}} \gets \textsf{mutant}$\;

				}\Else{
					$\textsf{mutant} \gets$ \mutateWeights{$n_{\textsf{best}}$}\;
					$n_{\textsf{best}} \gets$ \getBest{$\textsf{mutant}, n_{\textsf{best}}$}\;
				}
				$D \gets$ \updateArchive{$\textsf{mutant}, c_{s}, A_{m}$}\;
				
				\If{$!\textsf{generateNew} \; \textsf{\&} \;$ \isCovered{$k$}}{
					\Break network loop\;
					}
				}
			}
			
			\If{focus phase not reached}{
				\updateParameters{}\;
			}	
		}
  		\Return{$ D $}
  		}
    \caption{\MIONeatest}
    \label{alg:MIONeatest}
\end{algorithm}

Algorithm \ref{alg:MIONeatest} depicts the workflow of \MIONeatest.
We use \emph{Feedback-Directed Sampling} but initially set the
counters for all coverage objectives to $\infty$ to start with a
random chromosome instead of sampling from the archives.
When generating a random chromosome, we select a random
uncovered objective among all objectives whose parent control location in the
CDG have been covered (line 6)
based on \Neatest's target selection
procedure, described in \cref{sec:Neatest} and generate a network with minimal topology
and randomised connection weights similar to \Neatest (line 7).
In case we sample from one of the archives, as in line 10, we draw one individual per species
from an archive population to promote diversity. Like \Neatest, we assign
every novel chromosome to a species by considering all individuals stored in the
archives of uncovered goals during the speciation process.
In line 9, we set the objective from which we sampled as our target statement,
which is later used to determine whether a mutated chromosome should be mutated
further. 
If several individuals are sampled, we apply the following
steps to each of them.  

We mutate the generated or sampled individual $n$ up to $m$ times using the corresponding operator of \Neat, with $m \geq 1$ representing a user-defined
parameter. Depending on the probability $P_{\textsf{st}}$, these mutations consist of
one structural mutation (lines 18 to 20) and up to $(m-1)$ weight mutations, or a maximum of $m$ weight
mutations (lines 22 to 23). The probability $P_{\textsf{st}}$ is decreased over time alongside all
other dynamic \MIO parameters, such that with increasing search progress, more
focus is put on optimising network weights of established topologies. Within
the mutation process, we always generate one 
mutant of $n$ and execute the archive update in line 25 with the evolved
$\textsf{mutant}$ to verify whether the generated mutant reaches one of the targeted
coverage objectives. If the mutant is considered better than $n$ based on the
distance towards covering the target statement, we select the
mutant as a starting point for the next mutation. 
However, if the mutation was structural, we always take the mutant as the next
starting point to promote novel topologies. 
In case a given mutant is sampled and manages to cover its target statement, we
stop creating further mutations even if we have not yet executed $m$ mutations.
If $n$ corresponds to a newly generated chromosome, we always mutate $m$ times
as the novel network is likely to be also beneficial for other
coverage objectives.

During the archive update, described in Algorithm
\ref{alg:MIONeatestArchive}, we adapt the procedure introduced by \MIO but only
consider uncovered reachable objectives (line 3) and use the distance to the
targeted program statement as a
heuristic score (line 5).
To enable speciation in \MIONeatest, we generate a global population $P$ consisting
of all networks scattered across the archives and update the
global population whenever changes are made to an archive (lines 9 and 14). 
This allows us to assign $n$ to a global species (line 6).
In case $n$ has the same fitness value as the worst performing chromosome $w$ of
an archive, the secondary criterion decides whether $w$ should be replaced by $n$ (line 12).
If a network covers an objective, we set the sampling counters of its
direct children as derived from the CDG to zero to allow the algorithm
to start sampling from the archives of these objectives (line 18). Furthermore, to
facilitate \emph{Curriculum Learning}, we add the set of all networks that
covered an objective within the last round of $m$ mutations to the archives of
the new coverage objectives.


\SetKwFunction{getGlobalPopulation}{getGlobalPopulation}
\SetKwFunction{getArchiveForObjective}{getArchiveForObjective}
\SetKwFunction{getFitness}{getFitness}
\SetKwFunction{addAndSpeciate}{addAndSpeciate}
\SetKwFunction{getWorstNetwork}{getWorstNetwork}
\SetKwFunction{isBetterThanWorst}{isBetterThanWorst}
\SetKwFunction{remove}{remove}
\SetKwFunction{updateCounters}{updateCounters}
\SetKwFunction{getUncoveredReachableObjectives}{getUncoveredReachableObjectives}
\SetKwFunction{addIfNotExists}{addIfNotExists}
\SetKwFunction{replaceIfNotExists}{replaceIfNotExists}
\SetKwFunction{replace}{replace}

\begin{algorithm}[tb]
  \Input{network to add~$n$}
  \Input{secondary preference criterion~$c_s$}
  \Input{maximum archive size~$A_{m}$}
	\Fn{\updateArchive{$n, c_s, A_{m}$}}{
		
		$K \gets$ \getUncoveredReachableObjectives{}\;
		\ForEach{objective $k \in K$}{
			$A_k \gets$ \getArchiveForObjective{$k$}\;
			$h_k \gets$ \getFitness{$n, k$}\;
			$P.$\speciate{$n$}\;
			
			\uIf{$h_k == 1 \; \Vert \; A_k.\size < A_{m}$}{
				$A_k.$\add{$n$}\;
				$P.$\addIfNotExists{$n$}\;
			}
			\Else{
				$w \gets$ \getWorstNetwork{$A_k$}\;
				\If{\isBetterThanWorst{$n, w, c_s$}}{
				$A_k.$\replace{$w,n$}\;
				$P.$\replaceIfNotExists{$w, n$}\;				
				}
			}
  		}
  		\updateCounters{}\;
  	}
    \caption{\MIONeatest archive update}
    \label{alg:MIONeatestArchive}
\end{algorithm}

\subsection{NEWS/D-Neatest}
\label{sec:NEWSD}
Unlike \MIO and \MOSA, \NEWSD already represents a neuroevolution algorithm
that evolves chromosomes in the form of networks using  mutation and
crossover operators introduced by \Neat. Hence, we can directly adapt the proposed
processes of the \NEWSD framework~\cite{Salih2021Modified} without having to
modify the underlying algorithm. 

In the \textit{Sub-problem Generation Procedure}, we create a
sub-problem for each currently uncovered coverage objective whose
parent control location has been covered based on the target selection
procedure of \Neatest (\cref{sec:Neatest}). As we determine the
fitness values for each objective when evaluating a network, the
evaluation of a network serves as our \textit{Score Assignment
  Procedure}. If an individual covers a new objective, we add a copy
of the network to our archive that represents the test suite. The
\textit{Elite Set Procedure}, \textit{Protected Set Procedure}, and
\textit{Selection Procedure} can be adopted directly from \NEWSD, as
described in \cref{sec:NEWSD}; for the protected set we consider a
topology novice if the number of generations since its creation has
not exceeded a pre-defined number.

Once a set of chromosomes has been selected, we continue applying the
\textit{Reproduction Procedure} to each of its members.
To this end, we perform the \Neat crossover operator by randomly selecting the second
parent from the population since \NEWSD does not use speciation. If the
maximum number of evolving topologies is exceeded, we restrict the reproduction
operators to only perform weight mutations. Otherwise, if the maximum number of
topologies has not been reached yet, we also apply structural mutation. To ensure that every individual produces at least one offspring, crossover is
applied if the individual was not mutated. Once all children have 
evolved, we update the topology records with the resulting offspring networks.
Finally, we evaluate all offspring networks such that they can be compared
in the following score assignment procedure. Whenever a network covers an
uncovered objective, we store a copy of it in the archive of generated test cases. 

As a last step, we execute the \textit{Population Update Procedure},
putting all individuals of the protected set into the new generation. Then, we
combine the parent networks and the offspring obtained from the
\emph{Reproduction Procedure} and apply elitism. To this end, we fill
the generation with the $n$ best-performing individuals of every subproblem
measured by the objective fitness function, where $n = \left\lfloor
\frac{populationSize}{\#subproblems} \right\rfloor$. If the new generation has
not reached its maximum size yet, we draw a random subproblem for every free
spot and add the best individual for that problem of the remaining combined population to the new
generation. In contrast to \NEWSD, we do not add non-dominated individuals to the
archive as our Score Assignment Procedure already updates the archive with
individuals covering an uncovered objective. 


%% file: sections/4-Evaluation.tex
\section{Evaluation}

We evaluate whether many-objective search strategies can be used to improve
automated test generation for games by investigating the following two research
questions: 

\begin{itemize}
    \item \textbf{RQ1 (Secondary):} Which secondary fitness criterion is best
    suited for neuroevolution-based test generation? 
    \item \textbf{RQ2 (Performance):} Does neuroevolution-based test generation
    benefit from many-objective search strategies?
\end{itemize}

The proposed many-objective test generation algorithms are implemented as an
extension to the \Neatest algorithm and part of the open-source testing
framework \Whisker~\cite{Deiner2023Automated}. The experiment dataset, parameter configurations,
raw results of the experiments, and scripts for reproducing the results are
publicly available on Figshare:
\begin{center}
\url{https://doi.org/10.6084/m9.figshare.27135444}
\end{center}

\subsection{Dataset}

Our dataset for evaluating the proposed many-objective test
generation algorithms is based on the 25 \Scratch games that were used in 
previous work to evaluate the \Neatest
algorithm~\cite{Feldmeier2022Neuroevolution}. Since we aim to investigate
whether we can improve the test generator \Neatest, we removed games
which are too easy for the algorithm to make a difference. In
particular, we excluded games for which \Neatest is able to optimise test suites that reach 100\%
coverage within a search duration of one hour. This filtering approach led to the removal of the
following six games: \emph{BirdShooter}, \emph{BrainGame}, \emph{CatchingApples}, \emph{CatchTheGifts},
\emph{LineUp} and \emph{Pong}. 
Moreover, we removed the game \emph{RioShootout}
since achieving high block coverage for this game is purely based
on chance rather than sophisticated gameplay. 
We also added the two
games \emph{CreateYourWorld} and \emph{PokeClicker} to the dataset as they
were previously used to evaluate the effectiveness of promoting novel network
behaviour in \Neatest~\cite{Feldmeier2024Combining}.

This results in a diverse dataset of 20 \Scratch games with each one of them 
requiring the test generator to master unique gameplay challenges. As can be
seen in \cref{tab:Dataset}, the games also differ in terms of complexity
regarding the number of statements and branches to be covered, which range from
37 statements and 11 branches up to 908 and 378 coverage objectives scattered
across two and 21 game actors, respectively. Please note that \Scratch is a
domain-specific language that aims to facilitate the creation of games with a
few blocks. Hence, sophisticated games with challenging
gameplay can already be realised with a low number of blocks.

\input{tables/Dataset.tex}

\subsection{Methodology}

All experiments were conducted on a dedicated computing cluster consisting of 17
AMD EPYC 7433P CPU nodes running on 2.85GHz. To speed up test
executions, we made use of the program acceleration feature offered by
\Whisker~\cite{Deiner2023Automated}. This allows us to update the \Scratch
program as fast as the underlying machine can compute program state updates in
reaction to program inputs received without changing the underlying program
behaviour. During fitness evaluation, networks were allowed to play every game for a fixed duration
of ten seconds or until the program stopped, e.g., due to losing. 

Since a major goal of casting \Neatest to a many-objective optimisation
algorithm is to increase its efficiency, we set the global search budget for all
algorithms to two hours, which corresponds to a significantly lower search
budget compared to previous
experiments~\cite{Feldmeier2024Combining,Feldmeier2022Neuroevolution,Feldmeier2023Learning}. 
In each experiment, every algorithm was tasked with optimising branch
coverage. To ensure that the algorithms generate reliable test cases that are
robust against program randomisation, as discussed in \cref{sec:Neatest}, a
branch is only deemed covered if a generated test case manages to reach the same
branch in ten randomised program executions. We account for randomisation
inherent to games and search-based algorithms by repeating each experiment 30
times and determine statistical significance based on the
\emph{Mann-Whitney-U} test using $\alpha = 0.05$~\cite{Mann1947Test}.

\vspace{0.2cm}
\noindent\textbf{Parameter Tuning:} All proposed algorithms are based
on a set of hyperparameters that influence the performance of
the respective algorithm. To find suitable hyperparameters, we
adopted values that have shown to work well and conducted several tuning
experiments for each algorithm.  Except for the population size, which we
experimentally defined to a value of 50, we adopted the parameter of
\MOSANeatest from previous work~\cite{Deiner2023Automated}. For \MIONeatest we
adopted most parameters from related work on testing \Scratch programs and
experimentally derived a maximum archive size of 20, the avoidance of the focus
phase with $F=1$, and a probability of 50\% for the introduced
$P_{\textsf{st}}$ probability. Finally, for the \NEWSD algorithm, we reuse suggested
parameters from 
previous work~\cite{Salih2021Modified} and set the population size and the
number of generations an individual is considered novel to 50 and 3,
respectively, based on experimentation.

\input{images/RQ1.tex}

\vspace{0.2cm}
\noindent\textbf{RQ1 (Secondary):} In previous work, the secondary
preference criterion of the \MIO and \MOSA algorithms compares the
test size of two chromosomes~\cite{Arcuri2017Many}. However, this
metric is not applicable to our test cases as they correspond to evolved neural
networks for which there is no reason to prefer smaller ones. Hence, this research question
investigates alternatives to the traditional test size criterion. To this end,
we compare using (1) the compatibility distance, (2) the
species size, and (3) the novelty score
as a secondary criterion. As explained in \cref{sec:Neatest}, the novelty score is
computed based on the reached program states' similarity of the $k$-nearest
neighbours. We set the number of considered neighbours to $k=15$ and the
probability of adding new behaviours to the archive to 10\% since these values
have shown to work well in previous work~\cite{Feldmeier2024Combining}.
RQ1 is answered by comparing the branch and statement coverage achieved by \MIONeatest
and \MOSANeatest when using the compatibility distance (\emph{Compat}), the species size
(\emph{Species}) or the novelty score (\emph{Novelty}) as a preference
criterion. Moreover, we count how often each
configuration is able to cover a statement block that corresponds to winning a
game. Similar to previous work, these statements were determined through manual
analysis~\cite{Feldmeier2022Neuroevolution}.

\vspace{0.2cm}
\noindent\textbf{RQ2 (Performance):} After determining the best secondary
preference criterion for each algorithm, we continue in RQ2 with
comparing the best configurations of RQ1 and the \NEWSDNeatest
algorithm against the iterative \Neatest algorithm, which serves as a
baseline. Since the overall search duration in our experiments is
lower than in prior work, we reduced the population size of \Neatest
to 150 in order to ensure a good trade-off between exploration and
exploitation within the two-hour search budget. In line with previous
research, \Neatest changes the currently selected optimisation target
after ten generations without fitness improvements. All remaining
hyperparameters were set according to previous
work~\cite{Feldmeier2022Neuroevolution}. RQ2 is answered by comparing
the statement and branch coverage achieved by each algorithm as well
as the corresponding \emph{Vargha \& Delaney} effect
sizes~\cite{Vargha2000Critique} and by evaluating how often the
algorithms learned to master games by counting the reached winning
states.

\vspace{0.2cm}

\subsection{Threats to Validity}

\noindent \textbf{External Validity:} Although our dataset of 20 \Scratch
programs offers a wide range of gameplay challenges, we cannot
guarantee that the results generalise. 

\vspace{0.1cm}

\noindent \textbf{Internal Validity:} The large degree of randomisation inherent
to games and the evaluated algorithms will lead to varying results
each time the experiments are repeated. Thus, we ensure robust results by
repeating every experiment 30 times. 

\vspace{0.1cm}

\noindent \textbf{Construct Validity:} Both RQs evaluate the analysed algorithms
based on the achieved block and branch coverage. These metrics are similar to
statement and branch coverage in mature programming languages. However, since
\Scratch allows the creation of sophisticated games with relatively few
statement blocks, already small changes in coverage have a significant impact on the set
of tested program states.

\subsection{RQ1 (Secondary): Which secondary fitness criterion is best
suited for neuroevolution-based test generation?}

Our first research question investigates which secondary preference criterion is
best suited for comparing test cases in the form of neural networks. To this
end, we experimented with three different criteria, including the average
compatibility distance (\emph{Compat}), the species size (\emph{Species}) and a
novelty score (\emph{Novelty}). \cref{fig:RQ1} gives an overview of the average
branch and statement coverage achieved across all 20 \Scratch games of the
experiment dataset. 

For \MIONeatest, we can see in \cref{fig:RQ1-MIO} that there are only
minor differences in the achieved coverage values, ranging from a median of
83.86\% branch coverage to 84.30\%. This observation is supported
by the average number of reached winning states as they differ only slightly
across all three algorithm configurations, with \emph{Compat}, \emph{Novelty}
and \emph{Species} winning games on average \MeanWinsCompatMIO,
\MeanWinsNoveltyMIO\xspace and \MeanWinsSpeciesMIO\xspace times. We explain the
small impact on the secondary preference criterion for \MIONeatest due to the
criterion being only used for determining whether a given chromosome should
replace the worst-performing chromosome if both were assigned the same objective
fitness value. However, all games are randomised, and many fitness values in our
game testing scenario correspond to distances between actors of a game measured
at pixel resolution. Hence, two individuals rarely achieve the exact same
fitness value, leading to the secondary preference criterion being scarcely
used as a selection criterion.

While changes to the secondary preference criterion have only a small impact on
\MIONeatest, we observe a much higher impact in \cref{fig:RQ1-MOSA} for 
\MOSANeatest. Using the species size as a secondary criterion,
\emph{\MOSANeatest} achieves a median statement coverage of 93.24\% and
branch coverage of 85.35\%. Compared to the other two
configurations, this corresponds to an increase of 2.6 and 1.69 percentage points
in statement and branch coverage, respectively. The advantage in statement
coverage of the \emph{Species} configuration is also reflected in the
average number of reached winning states since this configuration
manages to master games an average of \MeanWinsSpeciesMOSA\xspace times, while the
\emph{Compat} and \emph{Novelty} configurations win games only an average of
\MeanWinsCompatMOSA\xspace and \MeanWinsNoveltyMOSA\xspace times. We explain the
higher impact of the secondary preference criterion for \MOSANeatest compared to
\MIONeatest due to its use in forming the final \emph{Pareto} front in the \MOSA
algorithm. Thus, the secondary criterion is an important factor in deciding which
individuals are allowed to reproduce.

Statistically significant differences in statement coverage can be observed for
the \emph{FruitCatching} game, with the \emph{Species} configuration achieving
\MeanBranchCoverageFruitCatchingSpeciesMOSA\% coverage compared to
\MeanBranchCoverageFruitCatchingCompatMOSA\% and
\MeanBranchCoverageFruitCatchingNoveltyMOSA\% when employing the compatibility
distance or the novelty score, respectively. In terms of
branch coverage, we identify a statistically significant advantage of using the species size
as a secondary criterion for \emph{Frogger} and
\emph{FruitCatching}. In these games, the \emph{Species} configuration improves the achieved
coverage by 0.5 and 5.24 percentage points compared to the \emph{Novelty}
or \emph{Compat} configuration. No statistically significant results were
observed in the remaining 17 games.

RQ1 reveals that while \MIONeatest is hardly impacted by a change in the
secondary preference criterion, greater differences may be observed for the
\MOSA extension of \Neatest. The good results of promoting small species sizes as a
secondary criterion indicate that \MOSANeatest benefits from maintaining a
species distribution, although the traditional \MOSA algorithm does not split
its population into species. These results raise the question of
whether the traditional \MOSA algorithm may also benefit from maintaining a
species distribution to use the size of species as a secondary
preference criterion. Since the results demonstrate that \MIONeatest and
\MOSANeatest benefit the most from using the novelty score, or respectively, the
species size as a secondary fitness criterion, we proceed with these
configurations in RQ2 to determine their performance compared to \Neatest.

\vspace{0.1em}
 \begin{tcolorbox}
	\textbf{RQ1 (Secondary)}: While the secondary preference criterion has hardly an
	impact on \MIONeatest, \MOSA-based \Neatest benefits the most from using the
	species size as a preference criterion.
 \end{tcolorbox}
 \vspace{0.1em}

\subsection{RQ2 (Performance): Does neuroevolution-based test generation
benefit from many-objective search strategies?}

\input{tables/RQ2.tex}

After identifying the best secondary preference criteria for \MIONeatest and
\MOSANeatest, we continue in RQ2 with evaluating the performance of the proposed
many-objective test generators against the \Neatest baseline. In the
following, we will refer to \MIONeatest as \MIO, \MOSANeatest
as \MOSA, and \NEWSDNeatest as \NEWSD.

\cref{tab:RQ2} shows that among the many-objective algorithms, \MOSA reaches the
highest average branch coverage value of \MeanBranchCoverageTotalMOSA\% compared
to \MeanBranchCoverageTotalMIO\% (\MIO) and \MeanBranchCoverageTotalNEWSD\%
(\NEWSD). This advantage is a consequence of \MOSA's ability to reach the most 
winning states with an average of \MeanWinsMOSA\xspace compared to
\MeanWinsMIO\xspace (\MIO) and \MeanWinsNEWSD\xspace (\NEWSD) won games. The
biggest differences in the reached winning states can be observed for the games
\emph{Frogger}, \emph{FruitCatching}, and \emph{HackAttack}. Since all three
games have in common that most branches are covered fairly easily and the
remaining branches depend on mastering the game, we conclude that \MOSA is best
suited to focus the search on winning games. We hypothesise that this effect
results from \MOSA being able to focus most of the population on a few branches
that correlate with winning the game while protecting novel innovations through
speciation, which is incorporated via the secondary preference criterion.

All proposed many-objective algorithms reach significantly more
branches than the \Neatest baseline. While \Neatest reaches an average branch
coverage of \MeanBranchCoverageTotalNeatest\%, the many-objective algorithms
improve the achieved coverage by 4.74 (\MIO), 5.45 (\MOSA) and 5.12 (\NEWSD)
percentage points. However, the baseline performs significantly better than all
many-objective algorithms for \emph{CatchTheDots}. This game
contains many branches that can be reached accidentally but are
much harder to reach consistently as they require sophisticated gameplay. Since
the many-objective algorithms aim to cover all these branches simultaneously,
they tend to execute robustness checks involving 10 program executions whenever
one of these branches is covered accidentally. Even though the
robustness check is done only once regardless of how many goals are
 checked, repeated robustness check executions are costly, as seen in the number of fitness evaluations performed by each
algorithm. While \Neatest evaluates on average,
\MeanFitnessEvaluationsCatchTheDotsNeatest\xspace networks within two hours of
search time, the many-objective algorithms evaluate less than half as
many, with averages of \MeanFitnessEvaluationsCatchTheDotsMIO\xspace
(\MIO), \MeanFitnessEvaluationsCatchTheDotsMOSA\xspace (\MOSA) and
\MeanFitnessEvaluationsCatchTheDotsNEWSD\xspace (\NEWSD) networks evaluations.
A similar issue can be observed for \emph{Dragons}, where all many-objective
algorithms, except \NEWSD, perform significantly worse than \Neatest. In the
\emph{FruitCatching}, \emph{HackAttack} and \emph{SpaceOdyssey} games, 
most, if not all, branches are dependent on winning the game,  which benefits the
single-objective algorithm as it is able to put more resources into covering
these goals.

Although all proposed many-objective algorithms perform worse in the
\emph{CatchTheDots} game,
they manage to achieve a statistically significant improvement of the achieved
branch coverage in 9 (\MIO) or 10 (\MOSA \& \NEWSD) games of all 20 dataset programs. We attribute this
significant advantage to two major reasons: The many-objective
algorithms effectively optimise tests for several branches
simultaneously and avoid getting stuck trying to cover hard or even
impossible-to-reach coverage objectives. We can observe this effect in the
\emph{CreateYourWorld} game shown in 
\cref{fig:CYW}, for which \Neatest does not cover a single branch in 3/30
experiment repetitions within two hours due to getting stuck at optimising
towards an impossible-to-reach coverage objective. In contrast, the many-objective algorithms
spread their resources on several branches simultaneously, which avoids getting
stuck and helps reach easier objectives such as touching the coin, leading to
a significant effect size of at least 0.88. However, not focusing on
hard-to-reach branches does not imply that these branches do not get covered
eventually, as all algorithms are able to reach the next level
in \emph{CreateYourWorld} three times.

Comparing the average program coverage achieved over time across all projects of
the dataset in \cref{fig:RQ2} reveals that the many-objective algorithms reach
significantly more branch and statement coverage than the baseline at each stage
of the search. \MIO, \MOSA, and \NEWSD reach the same average branch coverage
that \Neatest achieves after two hours already after
\SameBranchCoverageMIO\xspace(\MIO) and \SameBranchCoverageMOSA\xspace(\MOSA \&
\NEWSD) minutes, which corresponds to a significant speedup of
\SpeedUpBranchMIO\% and \SpeedUpBranchNEWSD\%. In light of this tremendous
speedup, we want to emphasise that whenever \Neatest covers a statement or
branch, it checks if the optimised network also covers other
objectives that depend on the covered goal. Hence, this speedup is
indeed the effect of \emph{optimising} towards several coverage goals at once
and not due to \Neatest aiming to generate a separate test for every coverage
objective.

\input{images/RQ2.tex}

The results of RQ2 demonstrate that the neuroevolution-based generation of tests
for games benefits from casting the test generation problem to a many
objective optimisation goal. While \MOSA achieves the best results, all proposed
many-objective algorithms demonstrate clear improvements over the single
objective \Neatest baseline. Due to the good performance of \MOSA, especially if
games have to be mastered to make progress, the proposed combination of
neuroevolution and \MOSA may also be beneficial for other problem
scenarios that are unrelated to generating tests for games.

\vspace{0.1em}
\begin{tcolorbox}
	\textbf{RQ2 (Performance)}: All many-objective algorithms significantly
	improve over the iterative baseline and reduce the required search time by at least
	\SpeedUpBranchMIO\%. 
 \end{tcolorbox}
 \vspace{0.1em}


%% file: tables/Dataset.tex
\begin{table}[tb]
    \centering
    \caption{Games used for the evaluation.}
    \label{tab:Dataset}
    \resizebox{\columnwidth}{!}{
        \begin{tabular}{lrrrlrrr}
            \toprule
            Project & \rotatebox{90}{\texttt{\#} Sprites} & \rotatebox{90}{\texttt{\#} Statements} & \rotatebox{90}{\texttt{\#} Branches} &
            Project & \rotatebox{90}{\texttt{\#} Sprites} & \rotatebox{90}{\texttt{\#} Statements} & \rotatebox{90}{\texttt{\#} Branches} \\
            \midrule
            CatchTheDots        & 4  & 82 & 35  & FlappyParrot  & 2   & 37    & 11    \\
            CityDefender        & 10 & 97 & 37  & Frogger       & 8   & 104   & 40 \\
            CreateYourWorld     & 12 & 165 & 91 & FruitCatching & 3   & 55    & 28\\
            DessertRally        & 10 & 213 & 104 & HackAttack   & 6   & 93    & 38\\
            Zauberlehrlinge     & 4  & 87  & 31 & OceanCleanup  & 11 & 156 & 62\\
            Dodgeball           & 4 & 78 & 43 & PokeClicker & 21 & 908 & 378 \\
            Dragons             & 6 & 377 & 188 & Snake & 3 & 60 & 14 \\
            EndlessRunner       & 8 & 163 & 50 & SnowballFight & 3 & 39 & 23 \\
            FallingStars        & 4 & 91 & 44 & SpaceOdyssey & 4 & 116 & 57 \\
            FinalFight          & 13 & 286 & 130 & WhackAMole & 10 & 391 & 148 \\
            \bottomrule
        \end{tabular}
    }
\end{table}

%% file: images/RQ1.tex
\begin{figure*}[!tbp]

    \begin{subfigure}[b]{\columnwidth}
      \centering
      \includegraphics[width=\columnwidth]{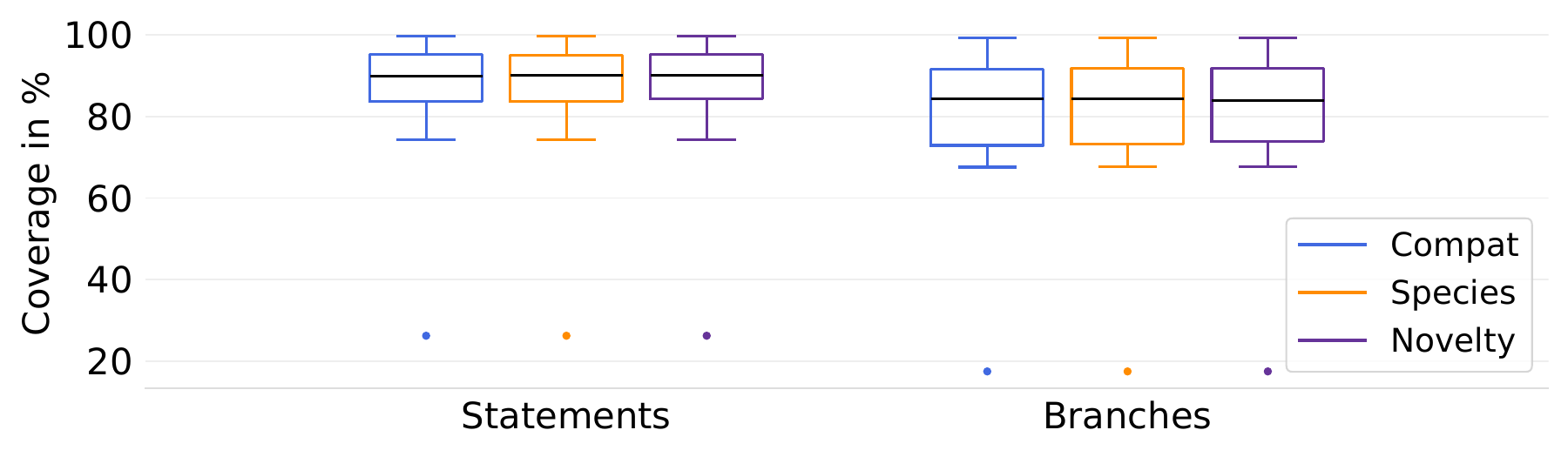}
      \caption{\MIONeatest}
      \label{fig:RQ1-MIO}
    \end{subfigure}%
   \begin{subfigure}[b]{\columnwidth}
      \centering
      \includegraphics[width=\columnwidth]{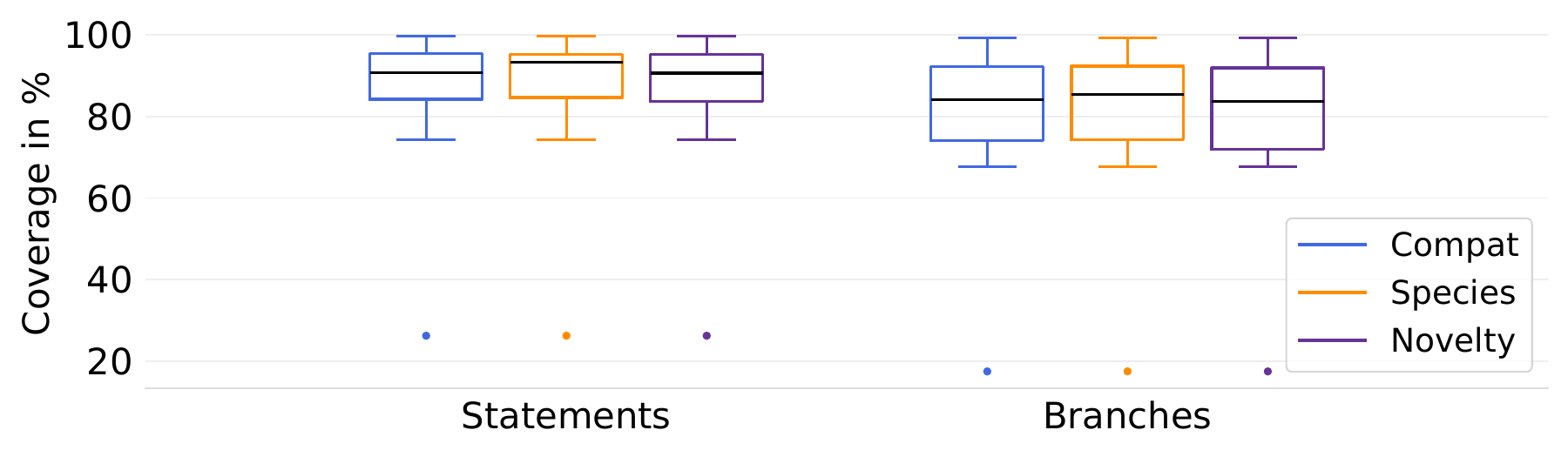}
      \caption{\MOSANeatest}
      \label{fig:RQ1-MOSA}
    \end{subfigure}%

    \caption{Achieved coverages over all projects when varying the secondary fitness criteria.}
    \label{fig:RQ1}
    
  \end{figure*}

%% file: tables/RQ2.tex
\begin{table*}[t!]
    \center
    \caption{Average branch coverage achieved (respective $\hat{A}_{12}$
    effect size) and reached winning states of \Neatest and the proposed test generators out of 30 experiment repetitions. Values in bold represent
    statistically significant results with $p < 0.05$.}
    \label{tab:RQ2}
    \resizebox{\linewidth}{!}{
        \begin{tabular}{lrrrrrrrrr}
            \toprule
            \multicolumn{1}{r}{} & \multicolumn{4}{c}{Branch Coverage} &
            \multicolumn{1}{r}{} & \multicolumn{4}{c}{Wins} \\
            \cline{2-5} \cline{7-10}
            Project & 
            \Neatest & \MIO & \MOSA & \NEWSD & & 
            \Neatest & \MIO & \MOSA & \NEWSD \\
            \midrule
            CatchTheDots 
            & \MeanBranchCoverageCatchTheDotsNeatest 
            & \MeanBranchCoverageCatchTheDotsMIO~(\EffectSizeBranchCoverageCatchTheDotsNeatestMIO)
            & \MeanBranchCoverageCatchTheDotsMOSA~(\EffectSizeBranchCoverageCatchTheDotsNeatestMOSA)
            & \MeanBranchCoverageCatchTheDotsNEWSD~(\EffectSizeBranchCoverageCatchTheDotsNeatestNEWSD)
            &
            & \WinCountCatchTheDotsNeatest
            & \WinCountCatchTheDotsNoveltyMIO
            & \WinCountCatchTheDotsMOSA
            & \WinCountCatchTheDotsNEWSD
            \\
            CityDefender 
            & \MeanBranchCoverageCityDefenderNeatest 
            & \MeanBranchCoverageCityDefenderMIO~(\EffectSizeBranchCoverageCityDefenderNeatestMIO)
            & \MeanBranchCoverageCityDefenderMOSA~(\EffectSizeBranchCoverageCityDefenderNeatestMOSA)
            & \MeanBranchCoverageCityDefenderNEWSD~(\EffectSizeBranchCoverageCityDefenderNeatestNEWSD)
            &
            & \WinCountCityDefenderNeatest
            & \WinCountCityDefenderNoveltyMIO
            & \WinCountCityDefenderMOSA
            & \WinCountCityDefenderNEWSD
            \\
            CreateYourWorld 
            & \MeanBranchCoverageCreateYourWorldNeatest 
            & \MeanBranchCoverageCreateYourWorldMIO~(\EffectSizeBranchCoverageCreateYourWorldNeatestMIO)
            & \MeanBranchCoverageCreateYourWorldMOSA~(\EffectSizeBranchCoverageCreateYourWorldNeatestMOSA)
            & \MeanBranchCoverageCreateYourWorldNEWSD~(\EffectSizeBranchCoverageCreateYourWorldNeatestNEWSD)
            &
            & \WinCountCreateYourWorldNeatest
            & \WinCountCreateYourWorldNoveltyMIO
            & \WinCountCreateYourWorldMOSA
            & \WinCountCreateYourWorldNEWSD
            \\
            DessertRally
            & \MeanBranchCoverageDessertRallyNeatest 
            & \MeanBranchCoverageDessertRallyMIO~(\EffectSizeBranchCoverageDessertRallyNeatestMIO)
            & \MeanBranchCoverageDessertRallyMOSA~(\EffectSizeBranchCoverageDessertRallyNeatestMOSA)
            & \MeanBranchCoverageDessertRallyNEWSD~(\EffectSizeBranchCoverageDessertRallyNeatestNEWSD)
            &
            & \WinCountDessertRallyNeatest
            & \WinCountDessertRallyNoveltyMIO
            & \WinCountDessertRallyMOSA
            & \WinCountDessertRallyNEWSD
            \\
            Zauberlehrlinge 
            & \MeanBranchCoverageDieZauberlehrlingeNeatest 
            & \MeanBranchCoverageDieZauberlehrlingeMIO~(\EffectSizeBranchCoverageDieZauberlehrlingeNeatestMIO)
            & \MeanBranchCoverageDieZauberlehrlingeMOSA~(\EffectSizeBranchCoverageDieZauberlehrlingeNeatestMOSA)
            & \MeanBranchCoverageDieZauberlehrlingeNEWSD~(\EffectSizeBranchCoverageDieZauberlehrlingeNeatestNEWSD)
            &
            & \WinCountDieZauberlehrlingeNeatest
            & \WinCountDieZauberlehrlingeNoveltyMIO
            & \WinCountDieZauberlehrlingeMOSA
            & \WinCountDieZauberlehrlingeNEWSD
            \\
            Dodgeball 
            & \MeanBranchCoverageDodgeballNeatest 
            & \MeanBranchCoverageDodgeballMIO~(\EffectSizeBranchCoverageDodgeballNeatestMIO)
            & \MeanBranchCoverageDodgeballMOSA~(\EffectSizeBranchCoverageDodgeballNeatestMOSA)
            & \MeanBranchCoverageDodgeballNEWSD~(\EffectSizeBranchCoverageDodgeballNeatestNEWSD)
            &
            & \WinCountDodgeballNeatest
            & \WinCountDodgeballNoveltyMIO
            & \WinCountDodgeballMOSA
            & \WinCountDodgeballNEWSD
            \\
            Dragons
            & \MeanBranchCoverageDragonsNeatest 
            & \MeanBranchCoverageDragonsMIO~(\EffectSizeBranchCoverageDragonsNeatestMIO)
            & \MeanBranchCoverageDragonsMOSA~(\EffectSizeBranchCoverageDragonsNeatestMOSA)
            & \MeanBranchCoverageDragonsNEWSD~(\EffectSizeBranchCoverageDragonsNeatestNEWSD)
            &
            & \WinCountDragonsNeatest
            & \WinCountDragonsNoveltyMIO
            & \WinCountDragonsMOSA
            & \WinCountDragonsNEWSD
            \\
            EndlessRunner
            & \MeanBranchCoverageEndlessRunnerNeatest 
            & \MeanBranchCoverageEndlessRunnerMIO~(\EffectSizeBranchCoverageEndlessRunnerNeatestMIO)
            & \MeanBranchCoverageEndlessRunnerMOSA~(\EffectSizeBranchCoverageEndlessRunnerNeatestMOSA)
            & \MeanBranchCoverageEndlessRunnerNEWSD~(\EffectSizeBranchCoverageEndlessRunnerNeatestNEWSD)
            &
            & \WinCountEndlessRunnerNeatest
            & \WinCountEndlessRunnerNoveltyMIO
            & \WinCountEndlessRunnerMOSA
            & \WinCountEndlessRunnerNEWSD
            \\
            FallingStars
            & \MeanBranchCoverageFallingStarsNeatest 
            & \MeanBranchCoverageFallingStarsMIO~(\EffectSizeBranchCoverageFallingStarsNeatestMIO)
            & \MeanBranchCoverageFallingStarsMOSA~(\EffectSizeBranchCoverageFallingStarsNeatestMOSA)
            & \MeanBranchCoverageFallingStarsNEWSD~(\EffectSizeBranchCoverageFallingStarsNeatestNEWSD)
            &
            & \WinCountFallingStarsNeatest
            & \WinCountFallingStarsNoveltyMIO
            & \WinCountFallingStarsMOSA
            & \WinCountFallingStarsNEWSD
            \\
            FinalFight
            & \MeanBranchCoverageFinalFightNeatest 
            & \MeanBranchCoverageFinalFightMIO~(\EffectSizeBranchCoverageFinalFightNeatestMIO)
            & \MeanBranchCoverageFinalFightMOSA~(\EffectSizeBranchCoverageFinalFightNeatestMOSA)
            & \MeanBranchCoverageFinalFightNEWSD~(\EffectSizeBranchCoverageFinalFightNeatestNEWSD)
            &
            & \WinCountFinalFightNeatest
            & \WinCountFinalFightNoveltyMIO
            & \WinCountFinalFightMOSA
            & \WinCountFinalFightNEWSD
            \\
            FlappyParrot
            & \MeanBranchCoverageFlappyParrotNeatest 
            & \MeanBranchCoverageFlappyParrotMIO~(\EffectSizeBranchCoverageFlappyParrotNeatestMIO)
            & \MeanBranchCoverageFlappyParrotMOSA~(\EffectSizeBranchCoverageFlappyParrotNeatestMOSA)
            & \MeanBranchCoverageFlappyParrotNEWSD~(\EffectSizeBranchCoverageFlappyParrotNeatestNEWSD)
            &
            & \WinCountFlappyParrotNeatest
            & \WinCountFlappyParrotNoveltyMIO
            & \WinCountFlappyParrotMOSA
            & \WinCountFlappyParrotNEWSD
            \\
            Frogger
            & \MeanBranchCoverageFroggerNeatest 
            & \MeanBranchCoverageFroggerMIO~(\EffectSizeBranchCoverageFroggerNeatestMIO)
            & \MeanBranchCoverageFroggerMOSA~(\EffectSizeBranchCoverageFroggerNeatestMOSA)
            & \MeanBranchCoverageFroggerNEWSD~(\EffectSizeBranchCoverageFroggerNeatestNEWSD)
            &
            & \WinCountFroggerNeatest
            & \WinCountFroggerNoveltyMIO
            & \WinCountFroggerMOSA
            & \WinCountFroggerNEWSD
            \\
            FruitCatching
            & \MeanBranchCoverageFruitCatchingNeatest 
            & \MeanBranchCoverageFruitCatchingMIO~(\EffectSizeBranchCoverageFruitCatchingNeatestMIO)
            & \MeanBranchCoverageFruitCatchingMOSA~(\EffectSizeBranchCoverageFruitCatchingNeatestMOSA)
            & \MeanBranchCoverageFruitCatchingNEWSD~(\EffectSizeBranchCoverageFruitCatchingNeatestNEWSD)
            &
            & \WinCountFruitCatchingNeatest
            & \WinCountFruitCatchingNoveltyMIO
            & \WinCountFruitCatchingMOSA
            & \WinCountFruitCatchingNEWSD
            \\
            HackAttack
            & \MeanBranchCoverageHackAttackNeatest 
            & \MeanBranchCoverageHackAttackMIO~(\EffectSizeBranchCoverageHackAttackNeatestMIO)
            & \MeanBranchCoverageHackAttackMOSA~(\EffectSizeBranchCoverageHackAttackNeatestMOSA)
            & \MeanBranchCoverageHackAttackNEWSD~(\EffectSizeBranchCoverageHackAttackNeatestNEWSD)
            &
            & \WinCountHackAttackNeatest
            & \WinCountHackAttackNoveltyMIO
            & \WinCountHackAttackMOSA
            & \WinCountHackAttackNEWSD
            \\
            OceanCleanup
            & \MeanBranchCoverageOceanCleanupNeatest 
            & \MeanBranchCoverageOceanCleanupMIO~(\EffectSizeBranchCoverageOceanCleanupNeatestMIO)
            & \MeanBranchCoverageOceanCleanupMOSA~(\EffectSizeBranchCoverageOceanCleanupNeatestMOSA)
            & \MeanBranchCoverageOceanCleanupNEWSD~(\EffectSizeBranchCoverageOceanCleanupNeatestNEWSD)
            &
            & \WinCountOceanCleanupNeatest
            & \WinCountOceanCleanupNoveltyMIO
            & \WinCountOceanCleanupMOSA
            & \WinCountOceanCleanupNEWSD
            \\
            PokeClicker
            & \MeanBranchCoveragePokemonClickerNeatest 
            & \MeanBranchCoveragePokemonClickerMIO~(\EffectSizeBranchCoveragePokemonClickerNeatestMIO)
            & \MeanBranchCoveragePokemonClickerMOSA~(\EffectSizeBranchCoveragePokemonClickerNeatestMOSA)
            & \MeanBranchCoveragePokemonClickerNEWSD~(\EffectSizeBranchCoveragePokemonClickerNeatestNEWSD)
            &
            & \WinCountPokemonClickerNeatest
            & \WinCountPokemonClickerNoveltyMIO
            & \WinCountPokemonClickerMOSA
            & \WinCountPokemonClickerNEWSD
            \\
            Snake
            & \MeanBranchCoverageSnakeNeatest 
            & \MeanBranchCoverageSnakeMIO~(\EffectSizeBranchCoverageSnakeNeatestMIO)
            & \MeanBranchCoverageSnakeMOSA~(\EffectSizeBranchCoverageSnakeNeatestMOSA)
            & \MeanBranchCoverageSnakeNEWSD~(\EffectSizeBranchCoverageSnakeNeatestNEWSD)
            &
            & \WinCountSnakeNeatest
            & \WinCountSnakeNoveltyMIO
            & \WinCountSnakeMOSA
            & \WinCountSnakeNEWSD
            \\
            SnowballFight
            & \MeanBranchCoverageSnowballFightNeatest 
            & \MeanBranchCoverageSnowballFightMIO~(\EffectSizeBranchCoverageSnowballFightNeatestMIO)
            & \MeanBranchCoverageSnowballFightMOSA~(\EffectSizeBranchCoverageSnowballFightNeatestMOSA)
            & \MeanBranchCoverageSnowballFightNEWSD~(\EffectSizeBranchCoverageSnowballFightNeatestNEWSD)
            &
            & \WinCountSnowballFightNeatest
            & \WinCountSnowballFightNoveltyMIO
            & \WinCountSnowballFightMOSA
            & \WinCountSnowballFightNEWSD
            \\
            SpaceOdyssey
            & \MeanBranchCoverageSpaceOdysseyNeatest 
            & \MeanBranchCoverageSpaceOdysseyMIO~(\EffectSizeBranchCoverageSpaceOdysseyNeatestMIO)
            & \MeanBranchCoverageSpaceOdysseyMOSA~(\EffectSizeBranchCoverageSpaceOdysseyNeatestMOSA)
            & \MeanBranchCoverageSpaceOdysseyNEWSD~(\EffectSizeBranchCoverageSpaceOdysseyNeatestNEWSD)
            &
            & \WinCountSpaceOdysseyNeatest
            & \WinCountSpaceOdysseyNoveltyMIO
            & \WinCountSpaceOdysseyMOSA
            & \WinCountSpaceOdysseyNEWSD
            \\
            WhackAMole
            & \MeanBranchCoverageWhackAMoleNeatest 
            & \MeanBranchCoverageWhackAMoleMIO~(\EffectSizeBranchCoverageWhackAMoleNeatestMIO)
            & \MeanBranchCoverageWhackAMoleMOSA~(\EffectSizeBranchCoverageWhackAMoleNeatestMOSA)
            & \MeanBranchCoverageWhackAMoleNEWSD~(\EffectSizeBranchCoverageWhackAMoleNeatestNEWSD)
            &
            & \WinCountWhackAMoleNeatest
            & \WinCountWhackAMoleNoveltyMIO
            & \WinCountWhackAMoleMOSA
            & \WinCountWhackAMoleNEWSD
            \\
            \midrule
            Mean
            & \MeanBranchCoverageTotalNeatest
            & \MeanBranchCoverageTotalMIO~(\MeanEffectSizeBranchCoverageNeatestMIO)
            & \MeanBranchCoverageTotalMOSA~(\MeanEffectSizeBranchCoverageNeatestMOSA)
            & \MeanBranchCoverageTotalNEWSD~(\MeanEffectSizeBranchCoverageNeatestNEWSD)
            &
            & \MeanWinsNeatest
            & \MeanWinsMIO
            & \MeanWinsMOSA
            & \MeanWinsNEWSD
            \\
            \bottomrule
        \end{tabular}
    }
\end{table*}

%% file: images/RQ2.tex
\begin{figure}[!tbp]

    \begin{subfigure}[b]{.5\columnwidth}
      \centering
      \includegraphics[width=\columnwidth]{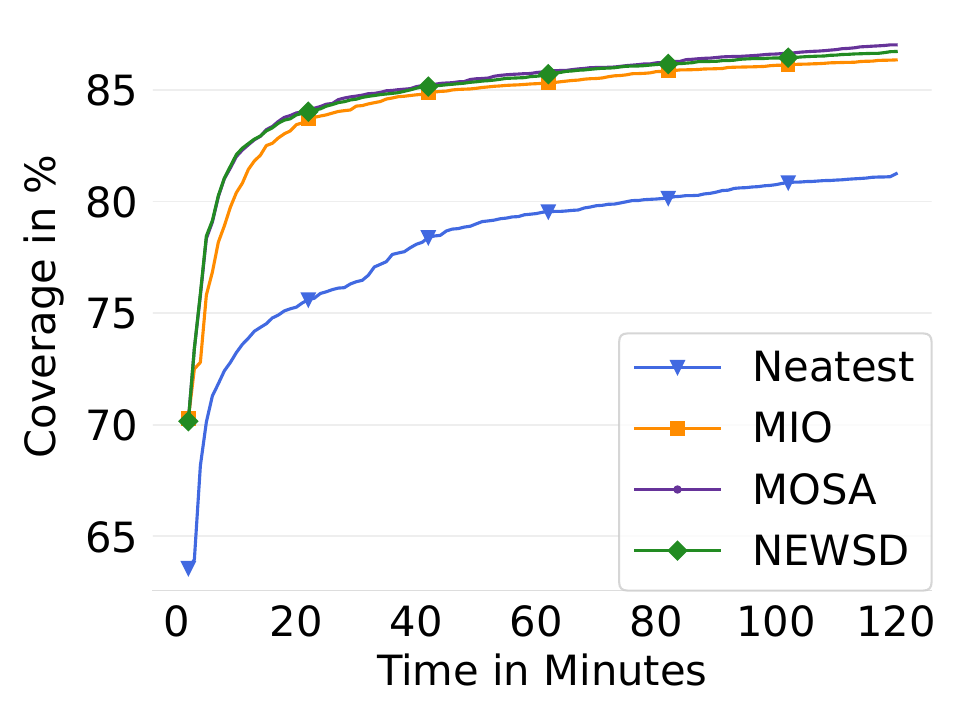}
      \caption{Statement coverage.}
      \label{fig:RQ2-Statement}
    \end{subfigure}%
   \begin{subfigure}[b]{.5\columnwidth}
      \centering
      \includegraphics[width=\columnwidth]{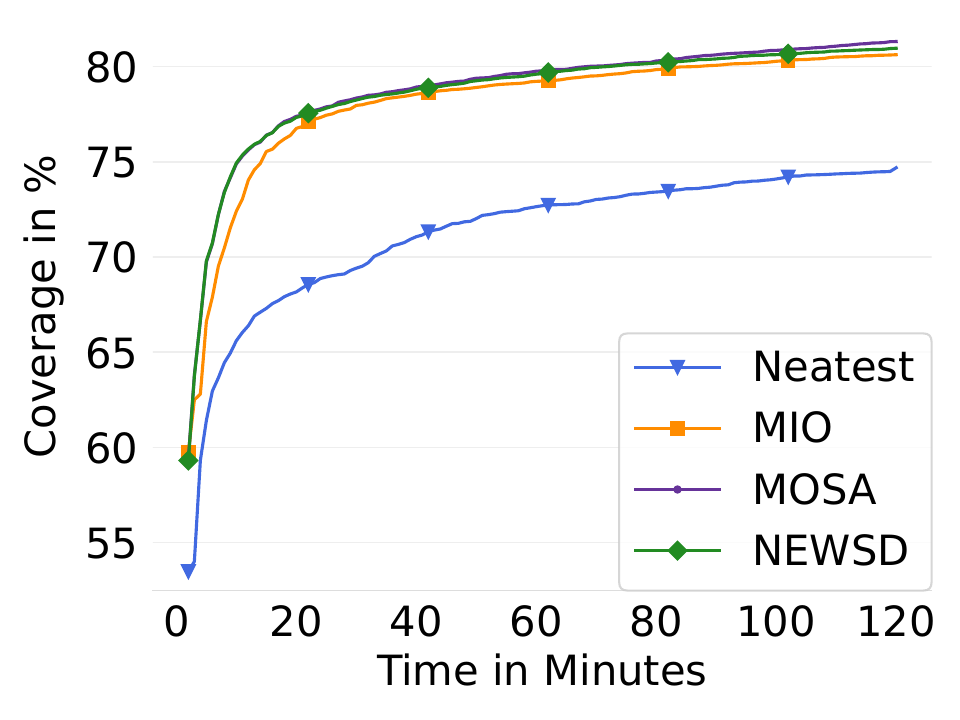}
      \caption{Branch coverage}
      \label{fig:RQ2-Branch}
    \end{subfigure}

    \caption{Average program coverage achieved over time.}
    \label{fig:RQ2}
  \end{figure}

%% file: sections/5-RelatedWork.tex
\section{Related Work}

Challenges such as randomisation and retrospective changes, as well as
the high costs and time required to write individual tests for each
game, limit the amount of automation when testing computer
games~\cite{Politowski2021Survey, Murphy-hill2014Cowboys}. Several
approaches have been introduced to address some of these issues, such
as using reinforcement learning \cite{Bergdahl2020Augmenting,
  Tufano2022Using, Ariyurek2021Automated, Zheng2019Wuji,
  Pfau2017Automated} to train smart agents that generate inputs to the
game under test. While all these approaches are similar to the
\Neatest testing framework in the regard that they evolve agents to
generate test inputs, only \Neatest optimises towards reaching
specific coverage objectives, such as program statements or branches.

To the best of the authors' knowledge, \Neatest is the only test generator that
aims to cover targeted coverage objectives by optimising test cases in the form
of neural networks using neuroevolution~\cite{Feldmeier2022Neuroevolution}.
While previous work has explored extending \Neatest with novelty
search~\cite{Feldmeier2024Combining} or with gradient-descent-based
backpropagation using traces of human gameplay as a training
dataset~\cite{Feldmeier2023Learning}, the test generator has never been cast to
a many-objective algorithm before. However, neuroevolution has already
been used in context with games before in order to simulate NPC behaviour 
\cite{Stanley2005Real-time}, master games \cite{Hausknecht2014Neuroevolution},
create game content \cite{Hastings2009Automatic} or react to user
behaviour \cite{Jallov2017Evocommander}. 

The neuroevolution-based algorithm \Neat has already been combined
with multi- and many-objective optimisation, in contrast to its
descendant \Neatest.  However, this combination was not applied for
testing purposes but for other tasks, including creating NPC behaviour
\cite{Schrum2021Constructing} or constructing controllers for
autonomous vehicles \cite{Willigen2013Evolving,
  Willigen2013Multi-objective}.


%% file: sections/6-Conclusions.tex
\section{Conclusions}

Generating tests for games is a daunting challenge due to
hard-to-reach program states and heavy program randomisation. The
neuroevolution-based test generator \Neatest tackles these challenges
by evolving neural networks that serve as test cases for games. Since
\Neatest targets one coverage objective after another, its efficiency
is reduced because it wastes valuable search time trying to cover
unreachable coverage objectives, redundantly evolving similar network
structures, and potentially missing individuals relevant for one
objective while searching for another. Thus, we combine the
single-objective \Neatest test generation algorithm with
many-objective search algorithms and propose three variations of
\Neatest that target several coverage objectives simultaneously, based
on the test generation algorithms \MIO and \MOSA, as well as the
many-objective neuroevolution framework \NEWSD. Our empirical
evaluation on 20 \Scratch programs demonstrates that extending
\Neatest to a many-objective test generator increases the average
branch coverage achieved while significantly reducing the required
search time.

Since our empirical evaluation has shown that the \MOSA-based test
generator benefits the most from using speciation as a preference
criterion, future work may investigate whether the traditional \MOSA
algorithm could also benefit from splitting its population into
several species. Although our implementation as part of the \Whisker
framework targets \Scratch programs, the algorithms could easily be
implemented for other testing domains, such as testing mobile apps or
reactive systems. In fact, the algorithms could even serve as a basis
for novel many-objective neuroevolution algorithms in other problem
scenarios unrelated to software testing.

A central issue in applying neuroevolution to testing games lies in
the high costs of fitness evaluations, particularly caused by
excessive robustness checks. We expect that surrogate models could
help reduce these costs in the future. Finally, in this paper we
focused on the problem of test input generation, but dynamic test
suites have also shown promise as test
oracles~\cite{Feldmeier2022Neuroevolution}, meriting further research.

\newpage